\documentclass[11pt, a4paper]{article}
\usepackage{jheparxiv}
\usepackage[latin1]{inputenc}
\usepackage{amsmath}
\usepackage{amsfonts}
\usepackage{amssymb}
\usepackage{latexsym}
\usepackage{mathrsfs}
\usepackage{twistor}

\newcommand{\V}{\, V}
\newcommand{\sA}{\mathscr{A}}

\title{MHV diagrams in twistor space and the twistor action}
\author{Tim Adamo}
\author{and Lionel Mason}

\affiliation{The Mathematical Institute,\\University of Oxford,\\24-29 St. Giles', Oxford, OX1 3LB, United Kingdom}
\emailAdd{adamo@maths.ox.ac.uk}
\emailAdd{lmason@maths.ox.ac.uk}

\abstract{MHV diagrams give an efficient Feynman diagram-like formalism for calculating gauge theory scattering amplitudes on momentum space.  Although they arise as the Feynman diagrams from an action on twistor space in an axial gauge, the main ingredients were previously expressed only in momentum space and momentum twistor space.  Here we show how the formalism can be elegantly derived and expressed entirely in twistor space.  This brings out the underlying superconformal invariance of the framework (up to the choice of a reference twistor used to define the axial gauge) and makes the twistor support transparent.  Our treatment is largely independent of signature, although we focus on Lorentz signature.

Starting from the $\cN=4$ super-Yang-Mills twistor action, we obtain
the propagator for the anti-holomorphic Dolbeault-operator as a delta
function imposing collinear support with the reference twistor
defining the axial gauge.  The MHV vertices are also expressed in
terms of similar delta functions.  We obtain concrete formulae for
tree-level $\mathrm{N}^{k}$MHV diagrams as a product of MHV amplitudes
with an R-invariant for each propagator; here the R-invariant
manifests superconformal as opposed to dual-superconformal invariance.
This gives the expected explicit support on $k+1$ lines linked by $k$
further lines associated to the propagators.  The R-invariants arising
correspond to those obtained in the dual conformal invariant momentum
twistor version of the formalism, but differences arise in the
specification of the boundary terms.  Surprisingly, in this framework,
some finite loop integrals can be performed as simply as those for
tree diagrams. }


\begin{document}
\maketitle

\section{Introduction}
An important output of Witten's twistor string theory
\cite{Witten:2003nn} was the MHV diagram formalism
\cite{Cachazo:2004kj, Cachazo:2004zb, Cachazo:2004by}.  This is a
momentum space Feynman-diagram formalism for gauge
theories that is much more efficient than standard ones.  In this
formalism, the propagators are the standard massless scalar
propagators, $1/p^2$, and the vertices are obtained from a simple
off-shell extension of the Parke-Taylor formula \cite{PT-86, BG-88}
for the `MHV' amplitudes (these being, in our conventions, the tree
amplitudes that have the maximal number of negative helicity external
gluons but nevertheless being nontrivial, leaving just two of positive
helicity).  This formalism was shown to give the correct amplitudes at
tree-level \cite{Risager:2005vk, Elvang:2008na, Elvang:2008vz} and for
1-loop MHV \cite{Brandhuber:2004yw}.  It has also recently been
expressed in momentum twistor space \cite{Bullimore:2010pj} where it
was shown to give the correct planar momentum space loop integrand to
all orders for supersymmetric theories which are
cut-constructible \cite{Bullimore:2010dz}.  We emphasize that this
recent work in {\em momentum} twistor space \cite{Hodges:2009hk} is
distinct to that in ordinary twistor space as described in this paper.
Momentum twistor space is essentially a new rational coordinatization
of momentum space that brings out dual superconformal invariance and
only applies to planar gauge theories (really one is computing a
Wilson-loop there, not the S-matrix \cite{Mason:2010yk}).  Amplitudes
on twistor space are not locally related to those on momentum twistor
space and have quite different analytic properties as we shall see.
We remark also that the loop integrands that these studies concern are
canonical and finite, but lead to infrared divergences in
four-dimensions when integrated and then require regularization, though
we will not address this major issue in this paper.

MHV diagrams were originally motivated from twistor-string theory.
Twistor-string theory had already led to formulae for tree-level
scattering amplitudes in $\cN=4$ Super-Yang-Mills (SYM) as a
path-integral over curves in super twistor space $\CP^{3|4}$
\cite{Roiban:2004yf,Skinner:2010cz}.  In these formulae, N$^k$MHV
amplitudes (i.e., those involving $(k+2)$ positive helicity gluons
with the rest negative) correspond to the part of the amplitude
supported on curves in twistor space of degree $k+1$.  Although it has
not been possible to extend these ideas to loop amplitudes (conformal
super-gravity corrupts the calculations beyond tree-level
\cite{Berkovits:2004jj}) the MHV formalism is not obstructed in the
same way and as noted above works to all loop orders at the level of the four-dimensional integrand.  It was based on
the idea that, instead of a connected degree $k+1$ curve, one could
consider $k+1$ lines that are geometrically disconnected, but are
joined by propagators \cite{Gukov:2004ei}.  This was expressed only
loosely in twistor space, but has a well-defined momentum space diagram
formalism.

The connection between classical Yang-Mills theory, twistor-string
theory and the MHV formalism was subsequently understood in terms of
twistor actions for supersymmetric gauge theories \cite{Mason-05,
  Boels:2006ir,Mason:2007zv,Jiang:2008xw} on twistor space (see
\cite{Mansfield:2005yd,Lovelace:2010ev} for other approaches).  These actions have greater gauge freedom than that on space-time; on 
one hand they reduce to the space-time actions non-peturbatively in
one gauge.  On the other, twistor actions can be gauge fixed in an
axial gauge on twistor space that is inaccessible from space-time; the
corresponding Feynman diagrams are then precisely the MHV diagrams
\cite{Boels:2007qn}.  Although this gave a field theory
explanation of the origins of the MHV formalism in momentum space,
it was not able to exploit the advantages that could have been hoped for from a twistorial formulation, such as making contact
with ideas from twistor-string theory and being able to exploit the
superconformal invariance of the twistor formulation to obtain a
simpler and more natural formalism.

More recently however, there has been an emphasis on understanding
amplitudes directly in twistor space rather than via a momentum space
representation; in effect we consider the scattering of particles that
are supported at a single twistor, an `elemental state'.  The first
systematic works \cite{Mason:2009sa, ArkaniHamed:2009si} (following on
from some earlier works, particularly \cite{Bena:2004xu}) were based
on   Witten's half-Fourier transform from momentum space to twistor space
using the transform of the BCFW recursion relations
\cite{Britto:2004ap, Britto:2005fq}.  This required analytic
continuation to split signature, which is unphysical, but avoided the
need to worry about the cohomological nature of twistor wave
functions.  It led to a superconformally invariant formulation (up to
some symmetry breaking signs) that clearly brought out the twistor
support of the BCFW representations of the amplitudes
\cite{Korchemsky:2010ut, Bullimore:2009cb}.  For example, terms in the
BCFW decomposition of an N$^k$MHV tree diagram are supported on
configurations of $2k+1$ lines containing $k$ loops (triangles for
NMHV).  These ideas led to a Grassmannian representation for the tree
amplitudes and leading singularities
\cite{ArkaniHamed:2009dn,Mason:2009qx,Bullimore:2009cb}.

In this paper we continue this investigation for the twistor
representation of amplitudes arising from the MHV formalism.  We obtain
amplitudes directly in twistor space starting from the twistor action
without referring to space-time or momentum space.  We have also
developed the technology further, using ideas from
\cite{Mason:2009qx, Skinner:2010cz, Mason:2010yk} to obtain a
signature independent formulation that incorporates the
cohomological nature of twistor wave functions whilst maintaining an
explicit and superconformally invariant formulation up to the choices
required for the axial gauge fixing. This is based on the use of
distributional $(0,p)$-forms in the multiple copies of twistor
space. Although at first sight these might seem to go against the
general holomorphic philosophy of twistor theory, there is no
obstruction to basing the calculus on \v{C}ech cohomology that uses holomorphic
functions as its representatives.  However, there would then be more gauge
freedom and complicated combinatorics associated with the choice of
cover, whereas these distributional 
forms provide a more efficient calculus for the corresponding
cohomological residue calculations in many complex variables.  In
particular the
delta function support reduces many of the integrations to algebra
(much as the Cauchy residue theorem would in a \v{C}ech approach).
Although we do not do this here, there is a direct translation from
the formulae obtained in split signature in \cite{Mason:2009sa} to
formulae that are valid in any signature and make better cohomological
sense. Indeed the formulae simplify as the conformal symmetry breaking
signs of \cite{Mason:2009sa, ArkaniHamed:2009si} are simply ommitted
under this translation.

In more detail, the twistor actions consist of a holomorphic
Chern-Simons action supplemented by a non-local term that generates
the MHV vertex contributions. The axial gauge arises from a choice of
a reference twistor denoted $Z_\rf$ (or simply $\rf$) and is
implemented by requiring that the component of any Dolbeault form
should vanish in the direction of the lines through $Z_\rf$.  In this
gauge, the MHV vertices are the only vertices as the Chern-Simons
cubic vertex vanishes.  The propagator is the Green's function for the
anti-holomorphic Dolbeault operator ($\dbar$) on twistor space.  A key
tool is a simple expression for this propagator that first appeared in
this form in \cite{Mason:2010yk} but which built on calculations from
space-time representatives appearing in appendix \ref{Prop} (see also
the last section of \cite{Cachazo:2004kj}).  The propagator
$\Delta(Z,Z')$ is essentially a superconformally invariant delta
function $\bar\delta^{2|4}(Z,\rf,Z')$ which is a Dolbeault
$(0,2)$-form current that imposes the condition that $Z_\rf$, $Z$ and
$Z'$ should be collinear.  The MHV vertices also have a description as
products of the same superconformally invariant delta functions that
enforce collinearity of the field insertion points.

Each N$^k$MHV tree diagram yields an integral of a product of $k+1$
MHV vertices supported on $k+1$ lines and $k$ propagators with ends
inserted on different lines.  The propagators are delta-functions that
restrict the insertion points on the MHV vertices to lie on a line
through $Z_\rf$.  The solution for the insertion points is unique and so
the integrals over the insertion points can be performed explicitly
against the delta functions in the propagator and vertices.  We are
left with the product of the MHV vertices but now with only external
twistors inserted, multiplied by a certain standard superconformal invariant
for each propagator.  These are the twistor R-invariants of
\cite{Mason:2009qx}, but are now invariants of the standard
superconformal group rather than the dual superconformal group.
The R-invariants that arise are quite similar to those that arise in
the momentum twistor version of the MHV formalism
\cite{Bullimore:2010pj} (i.e., one for each propagator), but there are
a number of differences: the R-invariants are those built out of
ordinary twistors rather than momentum twistors, the geometry of the
shifts for the boundary terms is different and in the momentum twistor
formulation there are no vertex contributions.

We therefore obtain a straightforward calculus in which it is possible
to perform the integrals arising for generic diagrams (both trees and
loops).  Here generic is meant in the sense of fixed NMHV degree $k$
and large particle number.  In this generic case, we have at least two
external particles on each vertex and the location of its
corresponding line in twistor space is fixed by its external
particles.  The formulae make manifest the expected support of a given
MHV diagram contribution in twistor space: N$^k$MHV tree diagrams are
supported on $k+1$ lines that correspond to the MHV vertices,
connected by $k$ propagators.  A propagator corresponds to the unique
line that passes through the fixed `reference twistor' and is
transversal to the two lines corresponding to the MHV vertices at each
end.  Thus, as in the case of a BCFW decomposition, we obtain support
on $2k+1$ lines for a N$^k$MHV tree amplitude, but here arranged as a
tree with $k$ of the lines clearly playing a distinct role as
propagators.

We obtain a similar support picture for loop diagrams with lines for
each vertex and for each propagator.  Remarkably, it is as easy to
perform the integrals for a loop diagram as it is for a tree diagram,
at least when the diagram is finite.  For divergent diagrams, it is also possible to perform these integrals, but the results require regulation; there are
many loop diagrams that do not lead to divergences though, and
these can be evaluated as simply as tree diagrams in the generic case.
These have the same structure as tree amplitudes, being a product of
MHV vertices evaluated only on external twistors with R-invariants.

We will discuss a number of further ramifications of these ideas, including
infrared divergencies and their possible regularisation, crossing symmetry,
connections with momentum twistors and the Grassmannian formulation in
\S\ref{discuss}.

The paper is structured as follows.  After some brief preliminaries to
establish notation and conventions, \S \ref{TA} reviews the twistor
action for SYM and develops the theory of the superconformally
invariant delta functions that provide the basic building blocks of
the formalism.  Section \ref{MHV} then provides a derivation of the
Feynman rules for this action in the CSW axial gauge.  We obtain
formulae for the propagator and for the MHV amplitudes (which are
also the vertices in this formalism) on twistor space using the
superconformal invariant delta functions developed earlier.  \S \ref{DMomsp} outlines the proof that these Feynman rules are equivalent to the momentum space MHV rules of \cite{Cachazo:2004kj}.  Next, we
demonstrate how the twistor space MHV formalism works at tree-level by
giving explicit computations for the various classes of
$\mathrm{N}^{k}$MHV tree diagrams in \S \ref{Tree}.  We go on to show
how this twistor formalism extends to loop diagrams.  We
consider the class of finite 1-loop non-planar MHV and planar NMHV diagrams, and the simplest case of the planar 1-loop MHV diagram in \S \ref{Loop} and
explain in general how to
identify the divergences.  Although we leave a full discussion of
regulation of divergences to another
paper, we dicuss this and a number of other key issues in \S \ref{discuss}.

The appendices contain discussions of the 2-point vertex on twistor
space (\ref{2pt}); the particulars of twistor theory for Euclidean
signature space-time (\ref{A}); the details of the proof deriving the
momentum space MHV formalisms from that in twistor space
(\ref{Momsp}); and the calculation of the twistor propagator from
space-time representatives (\ref{Prop}) which also demonstrates that
the propagator we use is the Feynman propagator.

%string theory\cite{Gukov:2004ei}.

\section{Background, notation and conventions}
\label{NC}

We adhere to the conventions of \cite{HT-85,WW-90} for bosonic
twistors but twistor space will be $\cN=4$ super-twistor space,
denoted $\PT$, the Calabi-Yau supermanifold $\CP^{3|4}$, with
homogeneous coordinates: \be{eqn: tsc} Z^{I}=(Z^{\alpha},\chi^{i
})=(\omega^{A},\pi_{A'},\chi^{i }), \ee where $\omega^{A}$ and
$\pi_{A'}$ are respectively negative and positive chirality Weyl
spinors, and $\chi^{i }$ (for $i =1\ldots 4$) are anti-commuting
Grassmann coordinates.

Points $(x^{AA'},\theta^{iA'})$ in
complexified chiral super Minkowski space-time $\M^{4|8}$ correspond to lines
$L_{(x,\theta)}$ in twistor space by 
the incidence relation
\be{eqn: inc}
\omega^A=ix^{AA'}\pi_{A'}\, , \quad \chi^i=\theta^{iA'}\pi_{A'}\, .
\ee
These lines $L_{(x,\theta)}$ are Riemann spheres ($\CP^1$s) and will be
parametrized with the homogeneous coordinates $\pi_{A'}$.

The Penrose transform relates helicity $n/2$ solutions to the
zero-rest-mass (z.r.m.) equations on a region $U'$ in complexified
Minkowski space to the first cohomology group of functions of homogeneity
degree $-n-2$ over the corresponding region $U$ in bosonic twistor
space ($\PT_{b}\cong\CP^{3}$); $U$ is the region swept out by lines
corresponding to points of $U'$.  We have \be{eqn: ptp}
H^{1}(U,\cO(-n-2))\cong \left\{
  \mathrm{z.r.m.}\;\mbox{fields on $U'$ of helicity}\;
  n/2\right\}, \ee \be{eqn: ptm} H^{1}(U,\cO(n-2))\cong \left\{
  \mathrm{z.r.m.}\;\mbox{fields on $U'$ of helicity}\;
  -n/2\right\}, \ee see for example \cite{WW-90} for a proof.  Here
$H^{1}$ denotes analytic cohomology.  We will use the Dolbeault
representation for the cohomology in which the cohomology classes are
represented as $\dbar$-closed $(0,1)$-forms modulo $\dbar$-exact ones.
(See \S\ref{discuss} for further discussion of other representations.)
This transform is most easily realized by an integral formula
\begin{align}
 \phi_{A_1' A_2' \ldots A_k'} & = \frac{1}{2 \pi i} \int_{L_x}
\pi_{A_1'} \pi_{A_2'} \ldots \pi_{A_k'} f (ix^{AA'}\pi_{A'},
   \pi_{A'} ) \D\pi \, , \label{Pintdirect} \\ 
\phi _{A_1 A_2 \ldots A_k} & = \frac{1}{2 \pi i} \int_{L_x}
\frac\p{\p\omega ^{A_1}} \frac \p{\p\omega ^{A_2}} \ldots \frac\p{\p\omega
    ^{A_k}} g (ix^{AA'}\pi_{A'}, \pi_{A'} ) \D\pi
\, , \label{Pintpot} 
\end{align}
where $\D\pi=\pi_{C'} \rd \pi^{C'} $.  The fact that these integral
formulae yield solutions to the field equations can easily be seen by
differentiating under the integral sign.

The standard choices of $U$ for positive/negative frequency fields are
the sets \be{future-tube} \PT^\pm=\{Z|\pm Z\cdot \bar Z\geq 0\}\quad
\mbox{ where }\quad Z\cdot\bar Z=\omega^A\bar\pi_A
+\pi_{A'}\bar\omega^{A'}\, .  \ee These are the sets that correspond
in space-time to the future/past tubes $\M^{\pm}$, i.e., the sets on
which the imaginary part of $x^{AA'}$ is past or future pointing
time-like respectively.  This follows from the fact that if we take
$x=u+iv$ and substitute into the incidence relation, then $Z\cdot \bar
Z= -v^{AA'}\bar\pi_A\pi_{A'}$ which has a definite sign when $v$ is
timelike depending on whether $v$ is future or past pointing.
The significance of this is that a field of positive frequency, whose
Fourier transform of a field is supported on the future lightcone in
momentum space, automatically extends over the future tube because
$\e^{ip\cdot x}$ is rapidly decreasing there, bounded by its values on
the real slice.

Another frequently used set is $U=\PT'$ on which $\pi_{A'}\neq 0$;
this corresponds to excluding the lightcone of the `point at infinity'
in complex space-time.

\subsection{The supersymmetric extension}
The transform has
a straightforward supersymmetrization to give an action for the superfield 
\be{superfield}
\cA=a+\chi^{i }\psi_{i }+\frac{1}{2}\chi^{i }\chi^{j }\phi_{i j
}+\frac{1}{3!}\epsilon_{i j k l }\chi^{i }\chi^{j }\chi^{ k
}\tilde\psi^{ l }+\chi^{1 }\chi^{2 }\chi^{ 3 }\chi^{ 4} g\, .  \ee
where $a$, $\tilde{\psi}$,
$\phi$, $\psi$, and $g$ are of weights $0$ $-1$, $-2$, $-3$, and $-4$
respectively corresponding respectively to zero-rest mass fields $(F_{AB}, \Psi_{iA}, 
 \Phi_{ij}, \tilde\Psi^{i}_{A'}, G_{A'B'})$.  

The formulae \eqref{Pintpot} extend directly to this supersymmetric
context to give superfields on space-times incorporating derivatives
on \eqref{Pintdirect}
\begin{eqnarray}
\cF_{AB}&:=&F_{AB}+\theta^{iA'}\p_{AA'}
\left(\Psi_{Bi}+ \theta^{jB'}\p_{BB'}\left(\Phi_{ij}+
    \theta^{kC'}\varepsilon_{ijkl}\left(\frac{\widetilde\Psi_{C'}^l}{3!}+
    \theta^{lD'}\frac{G_{C'D'}}{4!}  \right) \right)\right) \nonumber\\
&=&\int _{L_{(x,\theta)}} \frac\p{\p\omega ^{A}} \frac \p{\p\omega ^{B}}
 \cA (ix^{AA'}\pi_{A'}, \pi_{A'},\theta^{iA'}\pi_{A'} ) \D\pi \nonumber\\
\cF_{ij}&:=& \Phi_{ij} +
\theta^{kC'}\varepsilon_{ijkl}(\widetilde\Psi_{C'}^l+\theta^{lD'}\frac{G_{C'D'}}2)
\nonumber \\ 
&=&\int _{L(x,\theta)}  \frac\p{\p\chi^i} \frac \p{\p\chi^j}
 \cA (ix^{AA'}, \pi_{A'},\theta^{iA'}\pi_{A'} ) \D\pi 
\label{Susy-Pint}
\end{eqnarray}
These fields have the interpretation as being the non-zero parts of
the curvature
\be{susy-curv}
\cF=\cF_{AB}\varepsilon_{A'B'}\rd x^{AA'}\wedge \rd
x^{BB'}+\cF_{ij}\varepsilon_{A'B'}\rd \theta^{iA'}\wedge \rd
\theta^{jB'}
\ee
of the superconnection
\begin{multline}
\sA=\left(A_{AA'} +\theta^i_{A'}\left(\tilde\Psi_A+
    \theta^{jB'}\p_{AB'}\left(\frac{\Phi_{ij}}{2!} +\varepsilon_{ijkl}
\theta^{kC'}\left(\frac{\Psi_{C'}^l}{3!}+\theta^{lD'}\frac{G_{C'D'}}{4!}
\right)\right)\right) \right) \rd x^{AA'}
\\ +
\left(\Phi_{ij}+\varepsilon_{ijkl}\theta^{kB'}\left(\frac{\Psi^l_{B'}}{2} +
  \theta^{lC'}\frac{G_{B'C'}}{3!}\right)\right)  \theta^i_{A'}\rd\theta^{jA'}\, .
\label{superconn}
\end{multline}
Indeed, this superconnection can be obtained directly from $\cA$ via
the Ward transform, which treats $\cA$ geometrically as a deformation
of the $\dbar$ operator on  a line bundle and obtains $\sA$ as a
(super)-conection on a corresponding line bundle on space-time.

%%%%%%%%%%%%%%%%%%%%%%%%%%%%%%%%%%%%%%%%%%%%%%%%%%%%%%%%%%%%%%%%%%%%%
%%%%%%%%%%%%%%%%%%%%%%%%%%%%%%%%%%%%%%%%%%%%%%%%%%%%%%%%%%%%%%%%%%%%%

\subsection{The Twistor Yang-Mills Action}
\label{TA}

Here we give a brief review of the twistor
action on $\PT'$ for $\cN=4$ SYM~\cite{Mason-05,Boels:2006ir,Jiang:2008xw} (these
papers also discuss different amounts of supersymmetry, but we will
stick to $\cN=4$ here).

The space-time version of this action is an extension to $\cN=4$ SYM
of one introduced by Chalmers-Siegel for 
ordinary Yang-Mills.   This action is a 
reformulation of the standard one designed 
in such a way as to expand around the
anti-self-dual (ASD) sector.  In addition to the
connection 1-form $A(x)$ on
a bundle $\tilde{E}\rightarrow\M$, they 
introduce an auxiliary SD 2-form $G\in\Omega^{2+}(\M,
\End(\tilde{E}))$, and 
action \cite{CS-96}:   
\be{eqn: CS}
S[A,G]=\int_{\M}\tr(G\wedge F)-\frac{\varepsilon}{2}\int_{\M} \tr(G\wedge G),
\ee
where $\varepsilon$ is the expansion parameter.  Splitting the
curvature into its SD and ASD parts, $F=F^{+}+F^{-}$, this action 
gives the field equations
\begin{equation*}
F^{+}=\varepsilon G, \qquad \nabla\wedge G=0,
\end{equation*}
with $\nabla=\d +A$ the connection corresponding to $A$. 
These equations are easily seen to be
equivalent to the full Yang-Mills equations ($\nabla \wedge F^{*}=0$),
but for
$\varepsilon=0$, they reduce to the ASD Yang-Mills equations with a
background coupled SD field $G$.

%Introducing spinor indices, we encode the Yang Mills field into the
%helicity $+1$ field $G_{A'B'}(x)$ satisfying
%$\nabla^{A'}_AG_{A'B'}=0$  and the gauge potential $A_{AA'}(x)$
%with curvature 
%\begin{equation*}
%F_{AA'BB'}=F_{AB}\epsilon_{A'B'}+F_{A'B'}\epsilon_{AB}.
%\end{equation*}

The full $\cN=4$ SYM action can be similarly written as a
sum of two terms: 
\be{eqn: syma}
S[A,\Psi,\Phi,\widetilde\Psi,
G]=S_{\mathrm{ASD}}[A,\Psi,\Phi, \widetilde\Psi,
G]-\frac{\varepsilon}{2}I[A,\Psi,\Phi, \widetilde\Psi,G], 
\ee
where $S_{\mathrm{ASD}}$ accounts for the purely ASD sector and $I$
accounts for the remaining interactions which couple via the parameter
$\varepsilon$, $\Psi_{iA}$ and $\widetilde \Psi^i_{A'}$ are
respectively the ASD and SD spinor parts of the multiplet and
$\Phi_{ij}=\frac12\varepsilon_{ijkl}\Phi^{kl}$ the scalars.  Explicitly:
\begin{eqnarray}\label{eqn: symasd}
S_{\mathrm{ASD}}[A,
\Psi,\Phi,\widetilde\Psi,
G]&=&\int_{\M}\tr\left(\frac{\scriptstyle 1}{\scriptstyle 2}\, G\cdot
  F+\widetilde\Psi^i_{A'}\nabla^{AA'}\Psi_{i\, A} %\right. \\ \left. 
-\frac{\scriptstyle 1}{\scriptstyle 8}\, \nabla^a\Phi^{i j }\nabla_a\Phi_{ij }+\Phi^{i j
  }\Psi_i^{A}\Psi_{j\, A}\right)\d^{4}x, \nonumber \\
%\end{multline}
%\be{eqn: symsd}
I[A,\Psi,\Phi,
\widetilde\Psi,G]&=& \frac{
    1}{2}\int_{\M}\tr\left(
\, G\cdot G+ \Phi_{i j }\widetilde\Psi^i_{A'}\widetilde\Psi^{j\,
    A'}+\frac{\scriptstyle
    1}{\scriptstyle 4}\, \Phi^{i
    k}\Phi_{ij}\Phi^{jl}\Phi_{kl}\right)\d^{4}x. 
\end{eqnarray}

\subsubsection*{\textit{The Twistor Action}}

We now consider a topologically trivial vector bundle
$E\rightarrow\PT$ with $\dbar$-operator $\dbar_\cA=\dbar + \cA$ for
$\cA\in\Omega^{0,1}(\PT',\End(E))$.  If the $\dbar$-operator is
integrable, $\dbar_\cA^2=0$, the supersymmetric Ward transform
\cite{WW-90, Manin:1997ds} gives a correspondence between such
holomorphic vector bundles on twistor space and solutions to the
anti-self-dual sector of $\cN=4$ SYM.  The integrability conditions
$\dbar_\cA^2=0$ are the field equations of holomorphic Chern-Simons
theory with action \be{eqn: taasd}
S_{\mathrm{ASD}}[\cA]=\frac{i}{2\pi}\int_{\PT'}\D^{3|4}Z\wedge\tr\left(\cA\wedge\dbar\cA+\frac{2}{3}\cA\wedge\cA\wedge\cA\right)\,
.  \ee Here $\cA$ depends holomorphically on the fermionic coordinates
$\chi^{i }$ and has no components in the $\bar{\chi}$-directions.  We
can expand $\cA$ in terms of the $\chi^i$ to get 
\be{eqn: A}
\cA=a+\chi^{i }\psi_{i }+\frac{1}{2}\chi^{i }\chi^{j }\phi_{i j
}+\frac{1}{3!}\epsilon_{i j k l }\chi^{i }\chi^{j }\chi^{ k
}\tilde\psi^{ l }+\chi^{1 }\chi^{2 }\chi^{ 3 }\chi^{ 4} g\, .  \ee
Since $\cA$ has weight 0 and $\chi^{i }$ has weight 1, we find that $a$ has weight
0, $\psi$ weight $-1$, $\phi$ weight $-2$, $\tilde\psi$ weight $-3$
and $g$ weight $-4$. When taken to be cohomology classes, these give
the multiplet appropriate to $\cN=4$ SYM under the
Penrose transform with the lower case quantity on twistor space
corresponding to its upper case counterpart on space-time.

To introduce the remaining interactions of the theory, we add the term
\be{eqn: tasd}
I[\cA]=\int_{\M^{4|8}_\R}\d^{4|8}x\log \det\left(\dbar_{\cA}|_{L_{(x,\theta)}}\right),
\ee
where $L_{(x,\theta)}$ is the line or $\CP^{1}$ corresponding to 
$(x,\theta)\in\M^{4|8}_\R$ in $\PT'$ and $\M^{4|8}_\R$ is a real
4-dimensional contour in the complexified Minkowski space $\M^{4|8}$; $\dbar^{\cA}|_{L_{(x,\theta)}}$ is
the restriction of the deformed complex structure $\dbar_{\cA}$ to
this $L(x,\theta)$; and $\d^{4|8}x$ is the natural holomorphic volume
form on chiral superspace: 
\begin{equation*}
\d^{4|8}x=\frac1{4!}\epsilon_{abcd}\d x^{a}\wedge\d x^{b}\wedge\d x^{c}\wedge\d x^{d}\wedge\d^{8}\theta.
\end{equation*}
Although $\det(\dbar_{\cA}|_{L_{(x,\theta)}})$ might seem rather 
intimidating at first sight, we will see that it is easy to
understand perturbatively and indeed this leads both to the
finite set of terms in the space-time action in one gauge and the
infinite set of MHV vertices in another.  Although it is a section of
a line bundle $\cL$ over $\M^{4|8}$, it can be checked that the
integral of its log is independent of the choice of gauge as a
consequence of the fermionic integration in this
context \cite{Boels:2006ir}.

Hence, the twistor action for $\cN=4$ SYM is: \be{eqn: tact}
S_{\PT}[\cA]=S_{\mathrm{ASD}}[\cA]-\frac{\varepsilon}{2}I[\cA].  \ee
This action has gauge freedom \be{eqn: gauge} \dbar_\cA \rightarrow
h\dbar_\cA h^{-1}, \qquad h\in\Gamma(\PT',\End(E)), \ee and since
$\PT'$ has six real bosonic dimensions, $S_{\PT}$ has much more gauge
freedom than the Yang-Mills action in space-time.  In order to prove
that \eqref{eqn: tact} is equivalent to $\cN=4$ SYM on space-time, we
must make a gauge choice which reduces \eqref{eqn: gauge} to the
freedom of ordinary space-time gauge transformations.  One
particularly useful choice is a harmonic gauge up the fibres of a Euclidean
fibration first introduced by Woodhouse \cite{Wood-85} and this leads to the
reduction to the space-time action as described in \cite{Boels:2006ir,Jiang:2008xw}.

%%%%%%%%%%%%%%%%%%%%%%%%%%%%%%%%%%%%%%%%%%%%%%%%%%%%%%%%%%%%%%%%%%%%%
%%%%%%%%%%%%%%%%%%%%%%%%%%%%%%%%%%%%%%%%%%%%%%%%%%%%%%%%%%%%%%%%%%%%%

\section{The Twistor Space MHV Formalism}
\label{MHV}

In \cite{Boels:2007qn}, the MHV formalism on momentum space was
recovered as the Feynman diagrams of the twistor action \eqref{eqn:
  tact} in an axial gauge.  This was done by using twistor cohomology
classes that correspond to momentum eigenstates as the basic
scattering states. Although this provides an explanation of the origin
of the MHV formalism, it does not exploit the advantages that one
might hope to gain from a twistorial formulation such as making
contact with ideas from twistor-string theory and being able to
exploit the superconformal invariance of the twistor formulation.  The
novelty of the following treatment is that the presentation will be
self-contained in twistor space, using twistor cohomology classes that
are supported at points of twistor space.  This will bring out the
underlying superconformal invariance up to the choices that are
required to impose the gauge condition, and also make explicit the
support of the various contributions to the amplitude.  

We must first introduce the distributional wave functions that we will
use as the asymptotic states for scattering processes: the elemental
states supported at points of twistor space (these will in fact be the
twistor transform of those originally introduced by Andrew Hodges
\cite{Hodges:1990}).  These distributions turn 
out to be part of a framework of distributions supported at points,
lines, planes and the bosonic `body' in supertwistor space.  These
allow us to give a more careful treatment of the propagator and
to better understand the MHV vertices.

\subsection{Amplitudes, cohomology and distributional forms}

As has already been mentioned, the asymptotic states for the particles
in a scattering process are given by cohomology classes on
twistor space in $H^1_{\dbar}(U,\cO(n))$ for $U=\PT^\pm$.  We will
represent these as
$(0,1)$-forms 
$\phi$ that are $\dbar$-closed, $\dbar \phi=0$, defined modulo the
gauge freedom $\phi\rightarrow \phi+\dbar f$ on some domain $U
\subset \CP^{3|4}$.  Amplitudes are functionals of such asymptotic
states.  The kernel of an $n$-particle amplitude will therefore be in
the $n$-fold product of the dual to such $H^1$s. Although we could use
the Hilbert space structure on such $H^1$s, this turns out to be
complicated in our context; we obtain the best formalism by
representing the kernel of an amplitude using a local duality between
$(0,1)$-forms and distributional $(0,2)$-forms that are compactly
supported.  This is simply given by
\begin{equation*}
\Omega^{0,2}_{c,}(\PT,\cO)\otimes\Omega^{0,1}(U,\cO)\rightarrow\C,
\qquad (\alpha,\phi)\mapsto
\int_{\PT}\D^{3|4}Z\wedge\alpha\wedge\phi. 
\end{equation*}
For manifest crossing symmetry, we must be able to
take our asymptotic states to be of both positive and negative
frequency, and so we must be able to take both $U=\PT^+$ or $\PT^-$.
This will be possible if the compact support of the amplitude is
within $\PN=\PT^+\cap \PT^-=\{Z|Z\cdot\bar Z=0\}$.  See below
\eqref{eqn: TMHV} for the example of the MHV amplitude.and   
\S\ref{discuss} for further discussion.

Tree-level amplitudes in $\mathcal{N}=4$ SYM can be decomposed in terms of color-sector subamplitudes; a $n$-particle tree-amplitude $\mathcal{A}_{n}^{0}$ can be written as:
\begin{equation*}
\mathcal{A}^{0}_{n}=\sum_{\sigma\in S_{n}/\mathbb{Z}_{n}}\mathrm{tr}\left(T^{a_{\sigma(1)}}T^{a_{\sigma(2)}}\cdots T^{a_{\sigma(n)}}\right)A^{0}_{n}(\sigma(1),\ldots, \sigma(n)),
\end{equation*}
where the sum runs over all non-cyclic permutations of the $n$ particles and the $T^{a}$s are the generators of the gauge group.  In this paper, we will be interested in the color-stripped amplitudes $A^{0}_{n}$; due to the color trace, these objects obey a cyclic symmetry in their arguments, and this will extend to the twistorial amplitude as a function of twistor wave functions or distributional forms.  We make use of this cyclic symmetry both explicitly and implicitly often for the remainder of this work.

The amplitude will be defined modulo $\dbar$-exact forms with compact
support as these will give zero by integration by parts.  In an ideal
world, an $n$-particle amplitude would take values in
$H^{2n}_c(\times^{n}\PT,\cO)$.  However, we will see that our
amplitudes, including in particular the MHV amplitude fail to be
$\dbar$-closed due to anomalies arising from infrared divergences, see
\eqref{dbar-MHV} below.  This failure of the amplitude to be
$\dbar$-closed will lead to anomalies in gauge invariance.  This is
mitigated by the fact that throughout we will fix a gauge and, if we
were to change the gauge fixing condition, quantum field theory would
lead to very different formulae for the amplitudes.  It is
nevertheless a feature that should be understood better.  See
\S\ref{discuss} for further discussion.

In order to obtain explicit formulae, 
we need to introduce some natural distributions on twistor space.
We first note that on $\C$ with coordinate $z=x+iy$, the delta
function supported at the origin 
is naturally a $(0,1)$-form which we denote
\be{delta-bar}
\bar\delta^1(z)=\delta(x)\delta(y)\, \d \bar z =\frac1{2\pi i}\dbar \frac1z
\ee
the second equality being a consequence of the standard Cauchy kernel
for the $\dbar$-operator.  This second representation makes clear the
homogeneity property $\bar\delta(\lambda
z)=\lambda^{-1}\bar\delta(z)$.  

The fermionic delta function in the fermionic
variable $\eta$ is
$$
\delta^{0|1}(\eta)=\eta\, .
$$
This follows from the Berezinian integration rule $\int
\eta\rd\eta=1$ so that $\int f(\eta)\eta\rd\eta=f(0)$.

Following  \cite{Mason:2009qx}, to obtain delta functions on projective
space we first introduce the Dolbeault delta functions on $\C^{4|4}$: 
\be{delta44}
\bar\delta^{4|4}(Z)=\prod_{\alpha=0}^3 \, \bar \delta(Z^\alpha)
\prod_{i=1}^4 \chi^i. 
\ee
This is a $(0,4)$-form on $\C^{4|4}$ of weight zero. We then define projective
delta functions by 
\begin{equation*}
\bar{\delta}^{3|4}(Z_{1},Z_{2})
=\int_{\C}\frac{\d
  s}{s}\,\bar{\delta}^{4|4}(Z_{1}+sZ_{2})
\end{equation*}  
These can easily be seen to be antisymmetric and to satisfy the
obvious delta function relation
\begin{equation*}
f(Z')=\int_{\PT} f(Z) \bar{\delta}^{3|4}(Z,Z') \D^{3|4}Z \, .
\end{equation*}

By integrating against further parameters, we can obtain the following
superconformally invariant delta functions
\begin{eqnarray}
\bar{\delta}^{2|4}(Z_{1},Z_{2},Z_3)
&:=&\int_{\CP^{2}}\frac{\D^{2} c}{c_{1}c_{2}c_3
}\bar{\delta}^{4|4}(c_1Z_{1}+c_2Z_{2}+c_3Z_3) \nonumber\\
&=&\int_{\C\times\C}\frac{\d s}{s}\frac{\d t}{t}
\bar{\delta}^{4|4}(Z_3+sZ_{1}+tZ_{2})\nonumber \\
&=& \int_\C \frac{\d s}{s} \bar{\delta}^{3|4}(Z_{1},Z_{2}+s Z_3)
\label{delta-line} 
\end{eqnarray}
where $\D^{2} c=c_{1}\d c_{2}\wedge\d c_{3}+$cyclic permutations.  The
delta function $\bar{\delta}^{2|4}(Z_{1},Z_{2},Z_3)$ will play a large
role in what follows, giving both a representation of the propagator
and being an ingredient of the MHV amplitude.  It is
antisymmetric in its arguments and has support where the three points
$Z_1$, $Z_2,$ and $Z_3$ are collinear and has simple poles where two
of them coincide.

We can similarly define a coplanarity delta function
\begin{eqnarray}
\bar{\delta}^{1|4}(Z_{1},Z_{2},Z_3, Z_4)
&:=&\int_{\CP^{3}}\frac{\D^{3} c}{c_{1}c_{2}c_3c_4
}\bar{\delta}^{4|4}(c_1Z_{1}+c_2Z_{2}+c_3Z_3+c_4Z_4) \nonumber\\
&=&\int_{\C^{3}}\frac{\d s}{s}\frac{\d t}{t}\frac{\d u}{u}
\bar{\delta}^{4|4}(Z_4+sZ_3+tZ_{2}+uZ_{1})\nonumber \\
&=& \int_\C \frac{\d s}{s} \bar{\delta}^{2|4}(Z_{1},Z_{2},Z_3+s Z_4).
\label{delta-plane} 
\end{eqnarray}
Finally we can define the rational `R-invariant' 
\begin{eqnarray}
[Z_1,Z_2,Z_3,Z_4,Z_5 ]
&:=&\bar{\delta}^{0|4}(Z_{1},Z_{2},Z_3,Z_4,Z_5)\nonumber \\
&=&\int_{\CP^4}\frac{\D^4
  c}{c_1c_2c_3c_4c_5}\bar{\delta}^{4|4}\left(\sum_{i=1}^5c_iZ_i\right) 
\nonumber \\
&=&\frac{\delta^{0|4}\left( (1234)\chi_5 + \mbox{cyclic}\right)}
{(1234)(2345)(3451)(4512)(5123) }
\label{superconf}
\end{eqnarray}
where $\D^4c=c_1\d c_2\d c_3\d c_4\d c_5 +$cyclic and
$(1234)=\epsilon_{\alpha\beta\gamma\delta}Z_1^\alpha Z_2^\beta
Z_3^\gamma Z_4^\delta$.  We will also abbreviate
$[Z_1,Z_2,Z_3,Z_4,Z_5]$ by $[1,2,3,4,5]$.  Although there are no
longer bosonic delta functions, on the support of this fermionic delta
function, the five twistors span a four dimensional space inside
$\C^{4|4}$ so that this can be thought of as a delta function
supported on a choice of bosonic `body' of supertwistor space.  In the
context of momentum twistors this is the standard dual superconformal
invariant of \cite{Drummond:2008vq}.  The second formula is obtained
by integration against the delta functions, see \cite{Mason:2009qx}
for full details.  This will also play a significant role in this
story here, but as an invariant of the usual superconformal group as
opposed to the dual superconformal group.

It will be useful to know how these delta functions behave under the
$\dbar$-operator.  In general we have relations of the form 
$$
\dbar
\bar\delta^{r|4}(Z_1,\cdots,Z_{5-r})= (2\pi i) \sum_{i=1}^{5-r}
(-1)^{i+1}\bar\delta^{r+1|4}(Z_1,\cdots,\widehat{Z_i}
, \cdots, Z_{5-r})
$$ 
where $\widehat{Z_i}$ is ommitted.  The right
hand side necessarily vanishes for $r=3$. We will have frequent use
for the case of $r=2$ \be{dbar-delta-line}
\dbar_Z\bar{\delta}^{2|4}(Z_{1},Z_{2},Z_3 )=2\pi
i\left(\bar{\delta}^{3|4}(Z_{1},Z_{2})+\bar{\delta}^{3|4}(Z_{2},Z_3 )
  + \bar{\delta}^{3|4}(Z_3,Z_{1})\right).  \ee so we give the
derivation in full detail and leave the remaining relations as an
exercise.

Since
$\bar\delta^{4|4} (Z)$ is a top degree form, it is $\dbar$-closed.
Thus $$\dbar_{\mathrm{T}}\, \bar\delta^{4|4}(c_1Z_1 + c_2 Z_2 + c_3
Z_3 )=0\, ,$$
where $\dbar_{\mathrm{T}}=\dbar_c+\dbar_Z$ is the total
$\dbar$-operator on the space of parameters $(c_0,c_1,c_2)$ together
with the twistors $Z_i$, where $\dbar_c$ is that on 
the $c$s alone and $\dbar_Z$ being that on the $Z$s alone.  

We can use this to 
calculate $\dbar_Z \bar\delta^{2|4}(Z_1,Z_2,Z_3)$ as follows
\begin{eqnarray*}
\dbar_Z\bar{\delta}^{2|4}(Z_{1},Z_{2},Z_3 )&=&\int_{\CP^{2}}
\frac{\D^{2}c}{c_{1}c_{2}c_{3}}  
\wedge\dbar_Z \;\bar\delta^{4|4}(c_1Z_1 + c_2 Z_2 + c_3 Z_3 )
\\ 
&=&\int_{\CP^{2}}\frac{\D^{2}c}{c_{1}c_{2}c_{3}}\wedge(-\dbar_{c})
\, \bar\delta^{4|4}(c_1Z_1 + c_2 Z_2 + c_3 Z_3 )
\\
&=&\int_{\CP^{2}}\dbar_c\left(\frac{\D^{2}c}{c_{1}c_{2}c_{3}}\right)\wedge
\bar\delta^{4|4}(c_1Z_1 + c_2 Z_2 + c_3 Z_3 ) 
\\
&=&\int_{\CP^{2}}\D^{2}c
\left(\sum_{i=1}^{3}\frac{1}{c_{i+1}c_{i+2}}\dbar_{c} \frac{1}{c_{i}}\right)
\, \bar\delta^{4|4}(c_1Z_1 + c_2 Z_2 + c_3 Z_3 ) \\
&=&2\pi i\int_{\C}\frac{\d s}{s} \left( \bar\delta^{4|4}(Z_1+sZ_2) 
+ \bar\delta^{4|4}( Z_2 + s Z_3 )  + \bar\delta^{4|4}(Z_1 + s Z_3 ) \right)
\\
&=& 2\pi i\left(\bar{\delta}^{3|4}(Z_{1},Z_{2})+\bar{\delta}^{3|4}(Z_{2},Z_3 ) +
\bar{\delta}^{3|4}(Z_3,Z_{1})\right). 
\end{eqnarray*}
In the first equality we have taken $\dbar_Z$ under the integral, and
in the second, we have used $\dbar_Z=\dbar_{\mathrm{T}}-\dbar_c$ and
used the fact that $\dbar_{\mathrm{T}}^2=0$, in the third we have
integrated by parts, and the fourth we have expanded out $\dbar_c$
using finally $\dbar_c 1/c_i=2\pi i \, \bar \delta(c_i)$ to perform one of the
$c$-integrals to reduce it down to just one parameter.

We finally remark that with these distributional delta functions, many
integrals can be performed essentially algebraically.  Examples
that we will frequently use are
\begin{eqnarray}
\int \bar\delta^{2|4}(Z_1,Z_2,Z)\bar\delta^{2|4}(Z,Z_3,Z_4) \D^{3|4}Z=
\bar\delta^{1|4}(Z_1,Z_2, Z_3,Z_4)  \nonumber \\
\int \bar\delta^{2|4}(Z_1,Z_2,Z)\bar\delta^{1|4}(Z,Z_3,Z_4,Z_5) \D^{3|4}Z=
\bar\delta^{0|4}(Z_1,Z_2, Z_3,Z_4,Z_5)  \, . \label{composition}
\end{eqnarray}

\subsection{The CSW Gauge and Twistor Space Feynman Rules}
In order to obtain the dual form of the amplitude described above,
instead of inserting $H^1$ wave functions into colour stripped
amplitudes or vertices to obtain a number, we will insert external
fields $\cA_{a}$ (for $a=1,\ldots,n$) \be{eqn: insert}
\cA_{a}(Z)=\bar{\delta}^{3|4}(Z_{a},Z), \ee to obtain an expression
for the amplitude taking values in the $n$-fold tensor product of
$H^{2}(\PT,\cO)$ (one for each external particle).

To recover the MHV formalism on twistor space, we impose an axial gauge.
The choice of reference spinor $\iota^{A}$ in the MHV formalism corresponds to
the choice of a twistor `at infinity' 
denoted $Z_\rf=(\iota^{A},0,0)\in\PT$; this induces a foliation of $\PT-\{\rf\}$
by the lines that pass through $Z_\rf$.  We require that $\cA$ should vanish when
restricted to the leaves of this foliation: 
\be{eqn: CSWg}
\overline{Z_\rf\cdot\frac{\partial}{\partial{Z}}}\, \lrcorner\, \cA =0.  
\ee 
This gauge explicitly breaks conformal invariance due to the
choice of $\rf$, but we will obtain a formalism that
is invariant up to this choice.  We will often refer to this as the
CSW gauge as it was first introduced in \cite{Cachazo:2004kj}.

The main benefit is that it reduces the number of components of $\cA$
from three to two, so the cubic Chern-Simons vertex in
$S_{\mathrm{ASD}}[\cA]$ will vanish.  Since this cubic vertex corresponds to the anti-MHV three-point amplitude, the choice of CSW gauge eliminates this vertex; anti-MHV amplitudes will of course still exist, but are now constructed from the remaining vertices of the theory.  The twistor action becomes: 
\be{eqn: CSWg2}
S_{\PT}[\cA]=\frac{i}{2\pi}\int_{\PT'}\D^{3|4}Z\wedge\tr\left(\cA\wedge\dbar\cA
  \right)-\frac{\varepsilon}{2}\int_{\E^{4|8}}\d^{4|8}x\,\log\det\left(\dbar_\cA
    |_{L_{(x,\theta)}}\right).  
\ee 
We now determine the Feynman rules of this action in twistor space.

\subsubsection*{\textit{Propagator}}

Usually the propagator is determined by the quadratic part of the
action.  However, 
there are two such contributions in \eqref{eqn: CSWg2}: one from the
kinetic Chern-Simons portion and another from the perturbative
expansion of the $\log\det$ (see \eqref{eqn: exp} below).  Since it
occurs as part of a generating functional of vertices, we choose to
treat this latter contribution perturbatively, so it will not enter
into our definition of the propagator.  However, this means that our
formalism will include a two-point vertex.  We discuss this issue in
great detail later, and in appendix \ref{2pt}, but the main point is
that the two-point vertex itself vanishes as a conseqeuence of
momentum conservation, and so never appears as a vertex in the diagram
formalism.  However,  it does also play a role as a constituent of the
higher point MHV vertices where it is no longer forced to vanish.

Hence, the propagator is fixed by the kinetic part of the action 
\begin{equation*}
\int_{\PT'}\D^{3|4}Z\wedge\tr\left(\cA\wedge\dbar\cA\right),
\end{equation*}  
to be the inverse of the $\dbar$-operator on
$\PT'$ acting on $(0,1)$-forms in the CSW gauge \eqref{eqn: CSWg}:
\begin{equation*}
\dbar \Delta(Z_{1},Z_{2})=\bar{\delta}^{3|4}(Z_{1},Z_{2}), \qquad \overline {(*\cdot\p_1)}\, \lrcorner
\,\Delta=\overline {(*\cdot\p_2)}\, \lrcorner \,\Delta=0.
\end{equation*} 
The final answer is simply one of our superconformal delta functions 
\be{eqn: prop2}
\Delta
=\bar{\delta}^{2|4}(Z_{1},\rf ,Z_{2}):=
%\int_{\C^{*}\times\C^{*}}\frac{\d s}{s}\frac{\d t}{t}\bar{\delta}^{4|4}(N+sZ_{1}+tZ_{2}) \\
\int_{\C\times \C}\frac{\rd s\rd t}{st}\, \bar{\delta}^{4|4}(Z_{1}+
sZ_\rf +tZ_{2} )\, .  \ee 

In order to check this, we need to see that it
is indeed a Green's function for $\dbar$ and is also in the CSW gauge.
The gauge condition $\overline {(*\cdot\p_1)}\, \lrcorner
\,\Delta=\overline {(*\cdot\p_2)}\, \lrcorner \,\Delta=0$ follows from the
fact that to obtain a non-trivial integral we must take the
coefficient of $\rd \bar s$ in the expansion of the form part of
$\bar{\delta}^{4|4}(Z_{1}+ sZ_\rf +tZ_{2} )$ and since this is
accompanied by the constant $\bar Z_\rf$, the remaining form indices are
skew symmetrized with $\bar Z_\rf$.  To see that $\Delta$ indeed defines a
Green's function, we have from \eqref{dbar-delta-line} that, taking
$Z_{1},Z_{2}\in\PT'$ and $Z_3=Z_\rf$ as the reference twistor we have
$$
\dbar\Delta=2\pi i\left( \bar{\delta}^{3|4}(Z_{1},Z_{2})+
\bar{\delta}^{3|4}(Z_{1},\rf)+ \bar{\delta}^{3|4}(Z_{2},\rf)\right) .
$$ 
The first term is the delta-function that we would like to have,
whereas the last two terms essentially vanish in the degrees relevant
to the inversion of the $\dbar$-operator on 1-forms; $\Delta$ should be a
$(0,1)$-form in each variable, $Z_1$ and $Z_2$, whereas the error
terms arise from $(0,2)$-components in $Z_1$ and $Z_2$.  Such minor
`errors' (or at least unphysical poles in momentum space) in the
propagator are a familiar feature of axial gauges and are not
problematic. Indeed, if we restrict ourselves to the open set in $\PT$ that excludes the `point at infinity' (i.e., $U=\PT'$), then the error terms do not have support and the Green's function equation is satisfied exactly.

We need to be sure that we have found the
correct Feynman propagator with the appropriate $i\epsilon$
prescription.  This is addressed in the remarks in
\S\ref{discuss} and appendices\S\ref{A}--\ref{Prop}.

%Hence, if $\beta=\phi\wedge\D^{3|4}Z$ is an arbitrary test form on $\PT'$, we have the following action of $\dbar\Delta$ as a current:
%\begin{equation*}
%\langle\Delta, \beta\rangle=\int_{\PT'\times\PT'}\dbar\Delta\wedge\beta =\int_{\PT'\times\PT'}\bar{\delta}^{3|4}(Z_{1},Z_{2})\wedge\beta =\int_{\Delta}\frac{\beta|_{\Delta}}{\D^{3|4}Z} =\int_{\Delta}\phi|_{\Delta}.
%\end{equation*}
%But this is precisely the action of the anti-holomorphic Dirac
%current, so $\dbar\Delta=\bar{\delta}_{\Delta}$, as required.  

%With \eqref{eqn: prop2} defining the propagator, we now have the full
%set of Feynman rules for the twistor action $S_{\PT}[\cA]$ in the CSW
%gauge on twistor space; furthermore, these rules correspond precisely
%with the MHV formalism, but on twistor space rather than momentum
%space as in \cite{Cachazo:2004kj}.

\subsubsection*{\textit{Vertices}}

In the CSW gauge the vertices all come from the logarithm of the determinant in \eqref{eqn: CSWg2}  (the
interactions of the full theory that we added to the ASD action).
These vertices can be made explicit by perturbatively expanding out
the logarithm of the determinant which gives \cite{Boels:2006ir, Boels:2007qn}
\be{eqn: exp} \log\det\left(\dbar_\cA
  |_{L}\right)=\tr\left(\log\dbar|_{L}\right)+\sum_{n=2}^{\infty}\frac{1}{n}\int_{L^{n}}\tr\left(\dbar
  |_{L}^{-1}\cA_{1}\dbar |_{L}^{-1}\cA_{2}\cdots\dbar
  |_{L}^{-1}\cA_{n}\right).  \ee Here, $\dbar|_{L}$ is the restriction
of the $\dbar$-operator from $\PT$ to $L\cong\CP^{1}$, and $\cA_{a}$
is a field inserted a point $Z_{a}\in L$.  The $\dbar|_L^{-1}$ are the
Green's functions for the $\dbar$-operator restricted to $L$.  If we
suppose that the line $L$ is that joining twistors $Z_A$ and $Z_B$, we
can introduce the coordinate $\sigma$ on $L$ by
\be{sigma-def}
Z(\sigma)=Z_A+\sigma Z_B\,  .
\ee
In terms of this coordinate, $\dbar|_L^{-1} $ 
is just integration against the Cauchy kernel 
\begin{equation*}
\left(\dbar|_L^{-1}\cA\right) (\sigma_{a-1})=\frac{1}{2\pi i}\int
\frac{\cA(Z(\sigma_a))\wedge \d\sigma_a }{\sigma_{a}-\sigma_{a-1}},
\end{equation*}
Thus, the $n$th term in our expansion yields the vertex \be{eqn: vert}
\frac{1}{n}\left(\frac{1}{2\pi i}\right)^{n}\int_\M \d^{4|8}x
\int_{L^{n}}\tr\left(\prod _{a=1}^n \frac{\cA_a(Z(\sigma_a))\wedge
    \d\sigma_a }{\sigma_a-\sigma_{a-1}} \right) \, .  \ee 
Here the index $a$ is understood cyclically with
$\sigma_a=\sigma_{n+a}$ and $\M$ denotes a real slice of complexified
space-time.     In the
action of course all the $\cA_a=\cA$, but as a vertex in the Feynman
rules, all the $\cA_a$ are allowed to be different.  When the $\cA_a$
are the twistor wave functions that correspond to momentum
eigenstates, we will see that this  reduces to the standard $n$-particle
Parke-Taylor formula \eqref{eqn: PT} for the MHV (recall that scattering
  amplitudes for $n$ incoming gluons 
in which $k+2$ gluons have positive helicity with the rest negative
will be the N$^k$MHV amplitudes in our conventions). amplitude
\cite{Boels:2007qn}.  This form, as an
integral over the space of lines in twistor space, is a Dolbeault
analogue of 
Nair's original twistor formulation \cite{Nair:1988}.

A key point in this formula is
that the bosonic part of the $\d^{4|8}x$
integral is a contour integral in the space of complex $x$, performed
over the $4$-dimensional real slice $\M$ which can be taken to be some
real slice of complex Minkowski space.  The choice of signature of
this slice will determine the support of the vertices that we obtain;
if it is taken to be the standard Minkowski slice, our vertices will
clearly be supported in $\PN$ as all the lines in the integration will
lie in $\PN$.
\begin{figure}
\centering
\includegraphics[width=5 in, height=1.5 in]{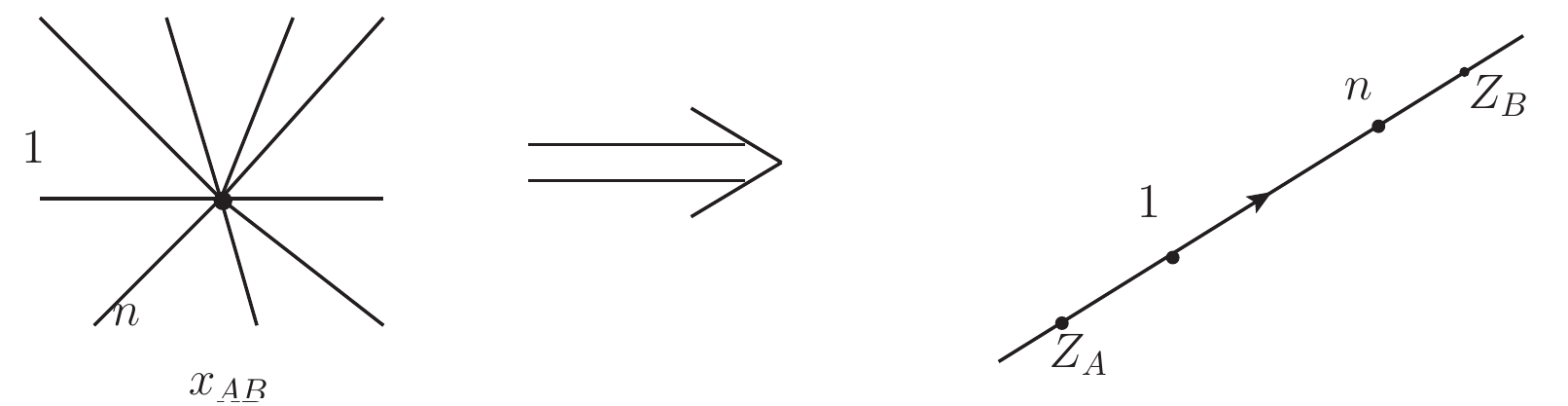}\caption{MHV Amplitude in Twistor Space}
\end{figure}

In order to obtain a manifestly conformally invariant formulation, we
represent the volume form $\d^{4|8}x$ as 
\be{vol-form} 
\d^{4|8}x=\frac{\D^{4|4}Z_{A}\wedge\D^{4|4}Z_{B}}{\vol(\GL(2,\C))}\, 
\ee 
i.e., rather than represent the line $L$ as $L_x$ via \eqref{eqn: inc} we
use \eqref{sigma-def} and quotient by $\GL(2)$ corresponding to the
choice of $Z_A$ and $Z_B$ on $L$. 

Defining $\cA_{a}$ as in \eqref{eqn: insert} will now give us the
superconformally invariant formula \be{eqn: TMHV}
\V
(Z_{1},\ldots,Z_{n})=\int
\frac{\D^{4|4}Z_{A}\wedge\D^{4|4}Z_{B}}{\vol(\GL(2,\C))}\int_{(L_{AB})^{n}}
\prod_{a=1}^{n}\frac{\bar{\delta}^{3|4}(Z_{a},Z(\sigma_{a}))\d
  \sigma_{a}}{(\sigma_{a}-\sigma_{a-1})},  \ee where we suppress color indices and the implicit trace.  This form most fully
manifests the symmetry of the amplitude, including the cyclic symmetry mentioned earlier.  It is the twistor-string
formulation \cite{Witten:2003nn,Roiban:2004yf} given as a `path-integral'
over the space of lines. Having defined our external fields $\cA_{a}$
as $(0,3)$-forms on $\PT'$, in \eqref{eqn: TMHV}, the integration over
$\d\sigma _{a}$ reduces the $(0,3)$-form to a $(0,2)$-form in each
$Z_a$ variable.

We can re-express higher point MHV vertices in terms of lower point
ones multiplied by delta functions by the relation
\be{inverse-soft}
\V(Z_{1},\ldots,Z_{n+1})
=\V
(Z_{1},\ldots,Z_{n})\bar\delta^{2|4}(Z_n,Z_{n+1},Z_1)\, .
\ee
To see this, 
observe that if we replace the $\sigma_{n+1}$ variable by
$$
s=\frac {\sigma_{n+1}-\sigma_1}{\sigma_n-\sigma_{n+1} }\, .
$$
then 
$$ 
(1+s) Z(\sigma_{n+1}) = Z(\sigma_n) +s Z(\sigma_1) \quad 
\mbox{ and }\quad 
\frac{\rd  \sigma_{n+1}}{(\sigma_{n+1}-\sigma_n)(\sigma_1-
  \sigma_{n+1})} = \frac {\rd s}{s(\sigma_n-\sigma_1)}\, .
$$
Using this in the defining formula \eqref{eqn: TMHV} for
$\V(Z_{1},\ldots,Z_{n+1})$, we can separate 
out $\V(Z_{1},\ldots,Z_{n})$ and an $s$-integral
$$
\V(Z_{1},\ldots,Z_{n+1})
=\V
(Z_{1},\ldots,Z_{n})\int \frac {\rd s}s \, \bar\delta^{3|4}(Z_{n+1},Z_{n}+sZ_1)\, 
$$
which leads to \eqref{inverse-soft} as desired.  This relationship
between the $n-1$ point MHV amplitude and the $n$-point amplitude
appeared in a totally real version of \eqref{inverse-soft} in
\cite{Mason:2009sa} and has become known as an inverse soft limit
\cite{ArkaniHamed:2009si}.

This can be used to reduce the general MHV vertex to a product of
delta functions and the two point vertex in many different ways.  A
typical such formula is
\be{eqn: TMHV3}
\V(Z_1,\ldots,Z_n)
= \V (Z_1,Z_2) \prod_{i=2}^n
\bar{\delta}^{2|4}(Z_1,Z_{i-1},Z_i)\, .
\ee   
This formula exhibits explicit superconformal invariance and has a
minimal number of residual integrations, but at the expense of the
cyclic symmetry which is manifest in \eqref{eqn: TMHV}. 
We can obtain many such formulae for the MHV
amplitude.  The different versions are generated by the identity
arising from the cyclic symmetry of the four point amplitude
\be{eqn: cyclic}
\V (Z_1,Z_{2},Z_3)\,
\bar{\delta}^{2|4}(Z_1,Z_{3},Z_4)=
\V (Z_2,Z_{3},Z_4)\,
\bar{\delta}^{2|4}(Z_2,Z_{4},Z_1)\, .
\ee
\begin{figure}
\centering
\includegraphics[width=2 in, height= 1.5 in]{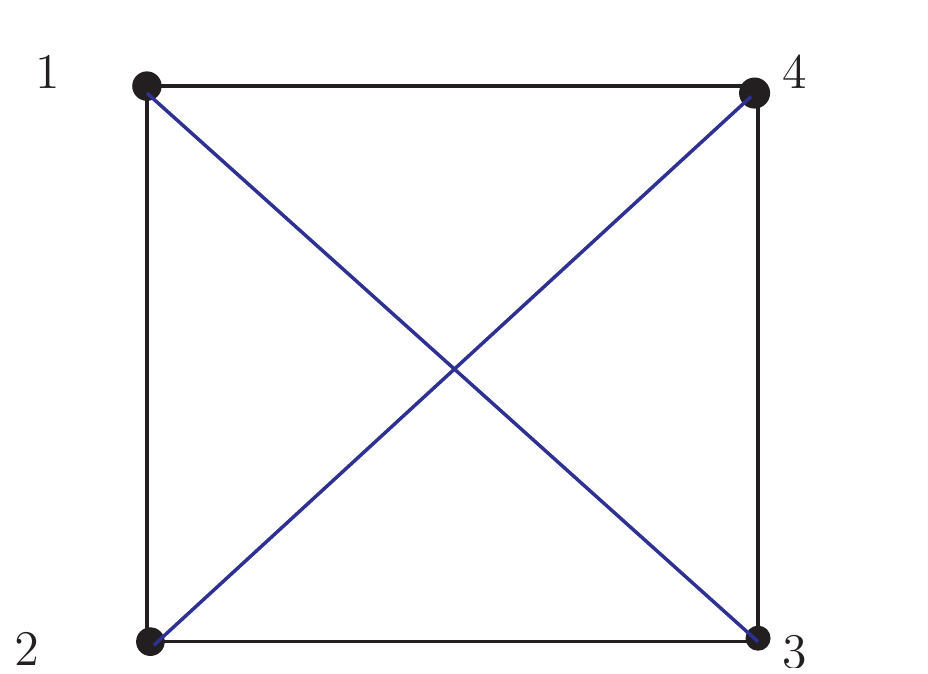}\caption{Geometric realization of \eqref{eqn: cyclic}}\label{DI1}
\end{figure}
This can be understood pictorially via Figure \ref{DI1}, where 
\eqref{eqn: cyclic} is the statement that the two triangulations of
the square
represented by blue lines are equivalent and that the location of $\V(i,j,k)$ is
interchangable with that of any $\bar\delta^{2,4}(\ldots)$ . 
For an MHV vertex with $n$ points, we can give a polygon representation
of formula \eqref{eqn: TMHV3} in which the $\bar
\delta^{2|4}(Z_1,Z_{i-1},Z_i)$ factors corresponds to a triangulation
of the polygon using just the triangles $(1,i-1,i)$ (with their being
one special line $(1,2)$ representing
the residual two-point MHV amplitude).  Then \eqref{eqn: cyclic}
allows us to change the given triangulation to an essentially an
arbitrary triangulation leading to many alternate formulae for the MHV
vertices.  In general we will just use equation \eqref{inverse-soft}  and its cyclic
permutations as necessary to pull out the dependence on twistors that
need to be integrated and leave the remaining MHV vertices as residual
factors in the answer. 

In  appendix \ref{2pt} the
two-point vertex $\V(Z_1,Z_2)$ is reduced to the integral
\be{eqn: 2pt1b-text} 
\V(Z_{1},Z_{2})
=\int_{\M\times (\CP^1)^2}
{\D^{3}Z_{A}\D^{3}Z_{B}}\, 
\bar{\delta}^{3}_{0,-4} (Z_{1},Z_A)\, \bar{\delta}^{3}_{0,-4}(Z_{2},Z_B),
\ee
where the $\CP^1$ factors in the contour are now understood as arising from
integrating $Z_A$ and $Z_B$ over the $\CP^1$ corresponding to $x\in\M$
and then integrating over the real slice $\M$ of complexified
Minkowski space (some other more concrete formulae are also given there but
this will be sufficient for our purpose here).  This is an integral of
a $12$-form over an $8$-dimensional contour so that we are left with a
$4$-form in $Z_1$ and $Z_2$ (a $(0,2)$-form in each factor).  In
particular if $\M$ is the Minkowski real slice, $Z_1$ and $Z_2$ are
pinned to lying on a line in $\PN$ and since the remaining
collinearity delta functions in \eqref{eqn: TMHV3} force the remaining
points to lie on this line, the general MHV vertex is supported for
$Z_i\in \PN$.

The 2-point vertex does not vanish, and indeed is a non-trivial factor
in each of our vertices.  Moreover it plays a nontrivial role in the
calculation of correlation functions in the context of Wilson loops
\cite{Mason:2010yk}.  However, it is shown in appendix \ref{2pt}
that it is $\dbar$-exact in the sense of compactly supported
cohomology, so if it appears on the exterior of a diagram, then it will
give a vanishing contribution because it will be integrated against a
$\dbar$-closed form.  A more subtle argument should obtain when it appears in the interior of a diagram (see \S \ref{2pt0} and appendix \ref{2pt} for discussion).  Thus it never
appears in the Feynman diagram calculus.  In \cite{Boels:2007qn} it is
shown that its evaluation on momentum eigenstates vanishes explicitly.
However, because our amplitudes are cohomological they don't need to
vanish explicitly to be trivial in cohomology.

\subsection{Derivation of Momentum Space MHV Formalism}
\label{DMomsp}

As a reality check, we now show that the Feynman rules for the twistor
action in the CSW axial gauge lead directly to the momentum space MHV
formalism of \cite{Cachazo:2004kj}.  This formalism was based on the
use of the Parke-Taylor MHV amplitudes \eqref{eqn: PT} below as vertices
and scalar $1/p^2$ propagators, with the off-shell prescription for
the vertices that the primed spinor associated to an off-shell momenta
$p_{AA'}$ should be taken to be $p_{A'}=p_{AA'}\hat \iota^A$ for some
reference spinor $\hat\iota^A$ (which as the notation suggests, is the
Euclidean conjugate of the spinor part of
the reference twistor).

In this complex 
framework, it is no longer possible to use the half-Fourier transform
to convert the twistor amplitudes used here to momentum space.
Nevertheless, it is possible to transform the ingredients of the MHV
formalism on twistor space term by term into their counterparts on
momentum space using (super) momentum eigenstates with supermomenta
$P=(p_{AA'},\eta_i)$.  For the propagator we will necessarily have
$p^2\neq 0$, but the external particles will be on shell with \be{}
P_a=(p_{AA'},\eta_{i})=(\tilde{p}_{A}p_{A'},\eta_i).  \ee For such an
on-shell momentum eigenstate we have the twistor cohomology class
\be{momentum-eigenstate} \cA_{P}=\int\frac{\rd s}{s}
\e^{\left(s(\omega^A\tilde p_A + \chi^i\eta_i)\right)}
\bar\delta^2(s\pi_{A'}-p_{A'})\, .  \ee That this gives the space-time
momentum eigenstates can be verified directly for the component
fields using \eqref{Pintdirect} and \eqref{Pintpot}; the integral is
performed algebraically against the delta function enforcing
$s\pi_{A'}=p_{A'}$ (see \cite{Witten:2004cp} for a discussion of such
individual momentum eigenstates).

To obtain the momentum space formula corresponding to a final
integrated diagram on twistor space, we integrate out the $(0,2)$ form
in each external twistor variable against the above Dolbeault
$(0,1)$-forms representing momentum eigenstates.  However, we can go
further and 
show that when expressed in momentum space, the vertices and propagators yield the appropriate CSW counterparts. 
We start with the vertices.

Momentum space representations break conformal invariance.  So there
is no loss in using a version of the MHV vertex in which the $\GL(2)$
symmetry has been fixed by coordinatizing the line $L_{(x,\theta)}$ by
the $\pi_{A'}$ coordinate.  This reduces the volume form on
$\M$ to $\rd ^{4|8}x$, as in \eqref{eqn: inc} yielding the
formula \be{vert-spinb} 
%\frac{1}{n}\left(\frac{1}{2\pi i}\right)^{n}
\int_{\M_{\R}} \d^{4|8}x\int_{L^{n}}\tr\left(\prod
  _{a=1}^n
  \frac{\cA_a(ix^{AA'}\pi_{A'a},\pi_{A'a},\theta_i^{A'}\pi_{aA'})\wedge
    D\pi_a }{[\pi_{a-1}\, \pi_{a}] } \right) \, , \ee where as usual
$[\cdot\;\cdot ]$ denotes the spinor inner product and we have ignored
normalization factors.  The first check is to show
that these MHV vertices give the standard momentum space MHV
amplitudes.  This can be done by taking the $\cA_{a}$ to be momentum
eigenstates as above \eqref{momentum-eigenstate}. It is now easily seen that the delta functions
allow the $\pi$-integrals to be done directly, simply enforcing
$s\pi_{A'a}=p_{A'a}$.  The remaining integral of the product of
exponential factors over $\rd^{4|8}x$ now 
gives the super-momentum conserving delta function to end up with the
Parke-Taylor \cite{PT-86, BG-88,Nair:1988} formula for the MHV
tree amplitude extended to $\cN=4$ SYM: \be{eqn: PT}
A_{\mathrm{MHV}}^{0}(P_{1},\ldots,P_{n})=%i(2\pi)^{4}(-\kappa)^{n-2}
\frac{\delta^{4|8}\left(\sum_{a=1}^{n}P_{a}\right)}{\prod_{a=1}^{n}
  [p_{a-1}\, p_{a}]}, \ee where %$\kappa$ is the gauge coupling, 
we have stripped off a normalization and an overall color trace factor, and the
super\-momentum conserving delta-function is
$$
\int \exp{i\left(\sum_a p_a\cdot x + \eta_{ai}p_{aA'}\theta^{iA'} \right)}\rd^{4|8}x=\delta^{4|8}\left(\sum P_{a}\right)=\delta^{4|0}\left(\sum
p_a\right)\delta^{0|8}\left(\sum_a \eta_a^ip_{a\, A'}\right), 
$$
with
$$
\delta^{0|8}\left(\sum_a \eta_ap_{a\, A'}\right)
=\prod_{i,\,A'}\left(\sum_a \eta_a^ip_{a\, A'}\right)\, .
$$

These Parke-Taylor MHV amplitudes \eqref{eqn: PT} are the vertices in
the momentum space MHV formalism extended off-shell by associating the
primed spinor $p_{A'}=p_{AA'}\hat\iota^A$ to an off-shell momentum
$p_{AA'}$.  To see how this arises from our twistor space formalism,
we first remark that the integrals in \eqref{vert-spinb} are over (a
contour in) the spin bunde $\PS$ coordinatized by
$(x^{AA'},\theta^{iA'},\pi_{A'})$ where $(x,\theta)$ is real.  This has a
natural projection to twistor space following from \eqref{eqn: inc}
given by:
\begin{equation*}
q:\PS\rightarrow\PT, \qquad (x^{AA'},\theta^{iA'},\pi_{A'})\mapsto
(ix^{AA'}\pi_{A'},\theta^{iA'}\pi_{A'},\pi_{A'})\, .
\end{equation*}
The MHV vertex is evaluated by first pulling back the cohomology
classes for the external fields and for the propagators
$\Delta(Z,Z')$ to the spin bundle $\PS$, and then integrating using
\eqref{vert-spinb}.  In order to obtain a momentum space
representative, we wish to Fourier transform the ingredients so as to
replace the $\rd^{4}x$ integral by a corresponding momentum space
integral; this is a conventional Fourier transform over a real slice.
We pull $\Delta(Z,Z')=\bar{\delta}^{2|4}(Z,\rf,Z')$ back to $\PS\times
\PS$ using $q$ and Fourier transform in the $x$ and $x'$ variables to
obtain the Fourier representation \be{eqn: momp1*}
\Delta(x,\theta,\pi,x'\theta',\pi')= \int\rd^4p\, \rd^4\eta\,
\e^{i(x-x')\cdot p+ \eta\cdot (\theta|\pi]-\theta'|\pi'])} \,
\widetilde \Delta(p,\eta,\pi,\pi') \ee After some calculation we
obtain \be{eqn: momp2*} \widetilde\Delta(p)=\frac{
  \bar{\delta}^{1}(\langle\hat{\iota}|p|\pi])\,
  \wedge\bar{\delta}^{1}(\langle\hat{\iota}|p|\pi'])}{p^{2}} \ee where
$\hat{\iota}^{A}$ is related to the original constant spinor
$\iota^{A}$ (the primary part of $Z_\rf$) by means of a quaternionic complex
conjugation induced by the choice of Euclidean real slice (see
appendix \ref{A}).  Appendix \ref{Momsp} contains the details of these
calculations; in order to obtain the correct answer here, it was
necessary to perform the Fourier transform on a Euclidean real slice.

If we now substitute this expression for the propagator into the MHV
vertex \eqref{vert-spinb}, then the delta functions in $\pi$ and
$\pi'$ again allow these $\pi$ integrals to be done algebraically with
the effect of substituting them with the primed spinor
$p_{A'}=\hat{\iota}^{A}p_{AA'}$. The integral over $(x,\theta)$ then
incorporates the super\-momentum $(p_{AA'},p_{A'}\eta_i)$ into the
super\-momentum conserving delta function.  This corresponds exactly
with the prescription given by \cite{Cachazo:2004kj} for the momentum
space MHV formalism as required.

\subsubsection{The vanishing of the 2-point vertex}\label{2pt0}
It is now straightforward to show, via this transform to
momentum space, that the two point vertex does not play a role in the
formalism: if it is present in a diagram, the whole diagram will
vanish.  The most nontrivial case is when the vertex is in the middle
of the diagram with propagators attached to each leg with supermomenta
$(P,\eta)$ and $(P',\eta')$.  The fermionic part of the momentum
conserving delta function in \eqref{eqn: PT} then reduces to $[p\,
p']^4\delta^{0|4}(\eta)\delta^{0|4}(\eta')$ and so the spinor products
cancel those in the denominator, yielding an overall $[p\, p']^2$ in
the numerator.  The bosonic delta function then forces $P+P'=0$ so
that $p=-p'$, and the numerator factor then forces the vertex to vanish.

\section{Tree Diagrams}
\label{Tree}

Having demonstrated that the twistor action in CSW gauge produces a perturbative expansion equivalent to the momentum space MHV formalism, we now endeavour to calculate amplitudes in a manner which is self-contained on twistor space.  The Feynman rules using the propagator and vertices we have just
obtained lead to formulae for amplitudes in terms of integrals over
intermediate twistors in the standard way.  We will see that for
generic diagrams all these integrals can be performed explicitly.  We
will find that each N$^k$MHV diagram yields a product of $k+1$ MHV
amplitudes/vertices multiplied by $k$ R-invariants.  The MHV vertices
are those corresponding to the external legs of each of the vertices
and there is an R-invariant for each propagator; the R-invariant has
five arguments, one of which is always the reference twistor and the
other four are the external twistors adjacent to the propagator when
propagators are not inserted adjacent to each other at a vertex.  Such
a picture holds when no propagators are adjacent at a vertex and we
will refer to these as generic diagrams (which is the case for fixed
$k$ and large $n$, but will not be the case when $k$ approaches $n$).
When propagators are adjacent we call the diagram a boundary term and
either the nearest external twistor, or a shifted version thereof is
used to determine the R-invariant.  There are also boundary-boundary
terms in which some vertex has fewer than two external vertices.  Here
there are not sufficient delta-functions to integrate out all the
internal twistors and some integrals remain.

\subsection{Tree-level NMHV Amplitudes}

\begin{figure}
\centering
\includegraphics[width=5.5 in, height=1.5 in]{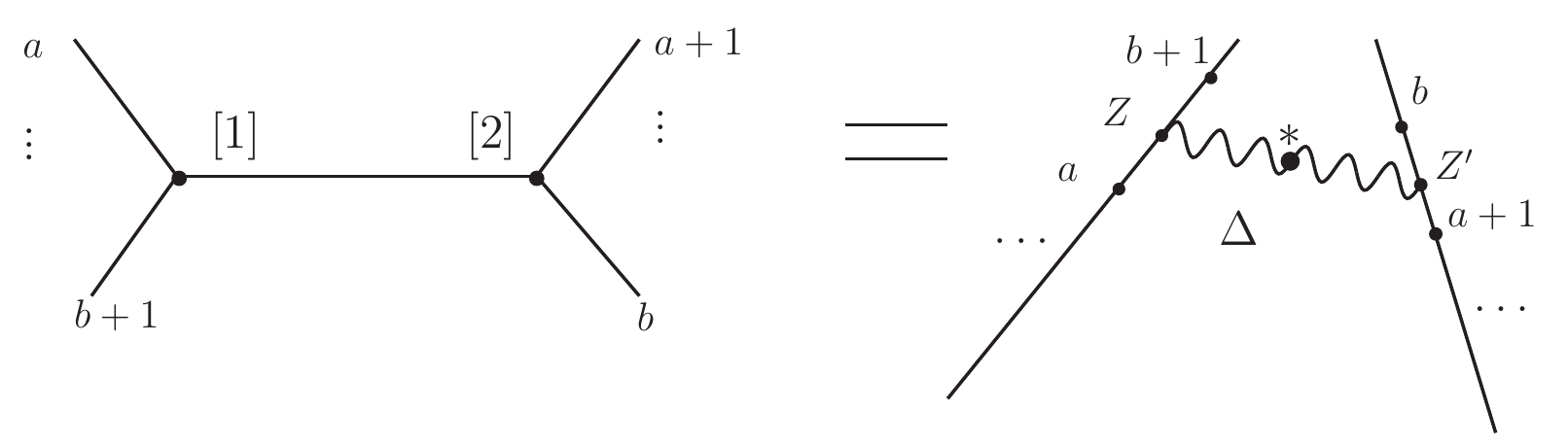}\caption{Twistor support of a typical NMHV tree diagram}\label{NMHV}
\end{figure}

A tree-level NMHV amplitude is expressed in the MHV formalism by a sum
over tree diagrams, each with two MHV vertices joined by a single
propagator.  The corresponding picture in twistor space, given in Figure
\ref{NMHV}, has a line corresponding to each of the two MHV
vertices of \eqref{eqn: TMHV3} connected by a
propagator as given by \eqref{eqn: prop2}.   Thus, the contribution of
such a term to the NMHV amplitude is: 
$$
\int_{\PT\times\PT} \D^{3|4}Z\D^{3|4}Z'
\V ({b+1},\ldots, a, Z)
\, \bar{\delta}^{2|4}(Z,\rf ,Z') 
\V ({a+1},\ldots, b, Z') \, .
$$
We can simplify this using \eqref{inverse-soft} to obtain
\begin{multline*}
\V ({b+1},\ldots, a)\V ({a+1},\ldots, b)
\times  \hfill\\
\hfill 
\int_{\PT\times\PT}\D^{3|4}Z\,\D^{3|4}Z'\,
\bar{\delta}^{2|4}({a},{b+1},Z)\, \bar{\delta}^{2|4}(Z,\rf ,Z'
)\, \bar{\delta}^{2|4}({b},{a+1},Z')\, .
\end{multline*}
The first two factors here are MHV amplitudes.  The remaining
factor can be integrated explicitly against the delta functions using
\eqref{delta-line}
as follows
\begin{multline*}
%\int_{\PT\times\PT}\D^{3|4}Z\,\D^{3|4}Z'\, \bar{\delta}^{2|4}({j},{i+1},Z)\, \bar{\delta}^{2|4}(Z,\rf,Z')\, \bar{\delta}^{2|4}({i},{j+1},Z')\, =\\
\int_{\PT\times\PT}\D^{3|4}Z\D^{3|4}Z'\int_{\C\times \C} 
\frac{\d  s\d t}{st}\, 
\bar{\delta}^{3|4}(Z, Z_{a}+sZ_{b+1}) 
\, \bar{\delta}^{2|4}(Z,Z',\rf )
\, \bar{\delta}^{3|4}(Z', Z_{b}+tZ_{a+1})\\
= \int_{\C\times\C}\frac{\d  s\d t}{st} \,
\bar{\delta}^{2|4}(Z_{a}+sZ_{b+1},\rf , Z_{b}+tZ_{a+1}
)
\\
\hfill = [{b+1},{a},\rf ,{b},{a+1}] 
\hfill
\end{multline*}
Hence, we see that such a contribution to the NMHV amplitude 
is given by: 
\begin{equation*}
\V ({b+1},\ldots, a)\V ({a+1},\ldots, b)
\, [{b+1},{a},\rf ,{b},{a+1}] 
\end{equation*}
The sum over tree diagrams gives the NMHV amplitude as
\be{eqn: TNMHV}
A_{\mathrm{NMHV}}^{0}=\sum_{a<b}\V (b+1,\ldots,
a)\, [b,b+1,\rf , a,a+1]\V (a+1,\ldots, b). 
\ee
We remark that the corresponding formula in momentum twistor space
is $A^{0}_{\mathrm{NMHV}}=\sum_{a<b} [{a},{a+1},\rf, b, {b+1} ] $,
which is the formula above stripped of the MHV factors.

\subsection{Tree-level $\mathrm{N}^{2}$MHV Amplitudes}

\begin{figure}
\centering
\includegraphics[width=6 in, height=1.5 in]{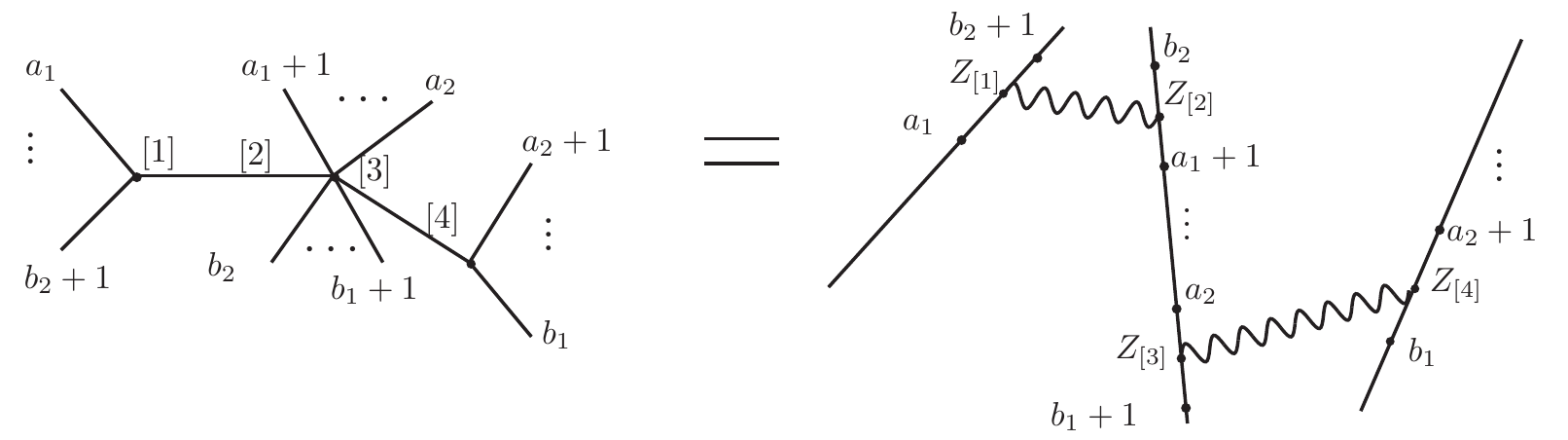}\caption{Twistor support of a non-boundary $\mathrm{N}^{2}$MHV tree diagram}\label{TN2MHV}
\end{figure}

At $\mathrm{N}^{2}$MHV, there is still essentially one family of
diagrams, Figure \ref{TN2MHV}, with the external legs distributed
around it in all possible ways.  However, our treatment will be
different in the two cases either where the two propagators are not or are adjacent (i.e., not separated by external particles) Figure
\ref{TN2MHV} or \ref{TN2C} respectively.  We refer to these as
`generic' and `boundary' diagrams.  The twistor space
support of these diagrams is also shown in Figures \ref{TN2MHV} and
\ref{TN2C}.

\subsubsection*{\textit{Generic terms}}
Applying our twistor space Feynman rules for the MHV formalism, a diagram of this sort gives:
\begin{multline}
\int \D^{3|4}Z_{[1]}\D^{3|4}Z_{[2]}\D^{3|4}Z_{[3]}\D^{3|4}Z_{[4]}
\bar{\delta}^{2|4}([1],\rf ,[2])  \bar{\delta}^{2|4}([3],\rf ,[4]
) \times 
\nonumber \\
\V(b_{2}+1,\ldots, a_1,[1]) \V([2],a_{1}+1, \ldots, a_2, [3],
b_1+1,\ldots,b_2) \V([4],a_2+1,\ldots,b_1)
\end{multline}
We can use \eqref{inverse-soft} four times (twice
on the middle vertex and once each on the others) to replace an MHV
vertex by one with one fewer arguments multiplied by a
$\bar\delta^{2|4}(\ldots)$ to isolate the dependence on the propagator
variables $Z_{[i]}$ to get
\begin{multline*}
\int \D^{3|4}Z_{[1]}\D^{3|4}Z_{[2]}
\bar{\delta}^{2|4}([1],\rf ,[2])  \bar{\delta}^{2|4}(a_{2},b_{1}+1,[3])
\bar{\delta}^{2|4}(b_{1},a_{2}+1,[4]) 
\times \\
\D^{3|4}Z_{[3]}\D^{3|4}Z_{[4]}  \bar{\delta}^{2|4}([3],\rf ,[4])
\bar{\delta}^{2|4}(a_{1},b_{2}+1,[1]) \bar{\delta}^{2|4}(b_{2},a_{1}+1,[2])\times
\\
\V(b_{2}+1,\ldots, a_1)\V(a_1+1,\ldots, a_2, b_1+1,\ldots,b_2)
\V(b_1,\ldots,a_{2}+1) 
\, .
\end{multline*}
These integrations can be done against the delta functions just as in
the NMHV case to obtain R-invariants. This gives an R-invariant for
each propagator multiplied by an MHV amplitude for each vertex to yield
\begin{multline}\label{eqn: TNC} 
[a_1,b_2+1,\rf,a_1+1,b_2] \,[a_2,b_1+1,\rf,a_2+1, b_1]\, 
  \times   \\ 
\V(b_{2}+1,\ldots, a_1)   \V(a_1+1,\ldots, a_2, b_1+1,\ldots,b_2) 
\V(b_1,\ldots,a_{2}+1)
\, , 
\end{multline}
for the general contribution to the N$^2$MHV amplitude with
non-consecutive propagators.  

%\FIGURE[t]{\includegraphics[width=2 in, height=1.5
%  in]{DI2.epsi}\caption{Geometry of non-consecutive propagator
%    insertions along a MHV amplitude}\label{DI2}} 

%The inverse soft limits of \eqref{inverse-soft} have a nice pictorial
%interpretation as in Figure \ref{DI2}.  Consider an MHV amplitude with
%$n$ external fields and two propagator insertions along it (such as
%the middle line in Figure \ref{TN2MHV}). The vertex labeled 1
%corresponds to the marked insertion appearing in every
%$\bar{\delta}^{2|4}$ of representation \eqref{eqn: TMHV3} of the MHV
%amplitude, coming from when we fixed the scale invariance of
%\eqref{eqn: TMHV} on the way to \eqref{eqn: TMHV3}.  Geometrically,
%this corresponds to a triangulation of the polygon in Figure \ref{DI2}
%by the blue lines from this fixed point.  However, the equivalence of
%(local) triangulations implied by \eqref{eqn: cyclic} tells us that
%the points corresponding to propagator insertions can be totally
%removed via the triangulations indicated by the red dashed lines.
Hence, after the integral over propagator insertions has been
performed, the remaining external legs on the middle line of Figure
\ref{TN2MHV} can be treated as a single MHV vertex, with 
the propagator insertions removed. 

%This leads to the following expression for any
%non-consecutive tree diagram with the leg-numbering structure of
%Figure \ref{TN2MHV}: 
%\begin{multline}\label{eqn: TN2NC}
%\V(b_{2}+1, \ldots, a_{1})\bar{\delta}^{0|4}(a_{1},a_{1}+1, b_{2},b_{2}+1,\rf ) \\
%\times \V(a_{1}+1,\ldots, b_{2})\bar{\delta}^{0|4}(a_{2},a_{2}+1,b_{1},b_{1}+1,\rf ) \\
%\times \V(a_{2}+1,\ldots, b_{1}).
%\end{multline}

\subsubsection*{\textit{Boundary terms}}
\begin{figure}
\centering
\includegraphics[width=6 in, height=1.5 in]{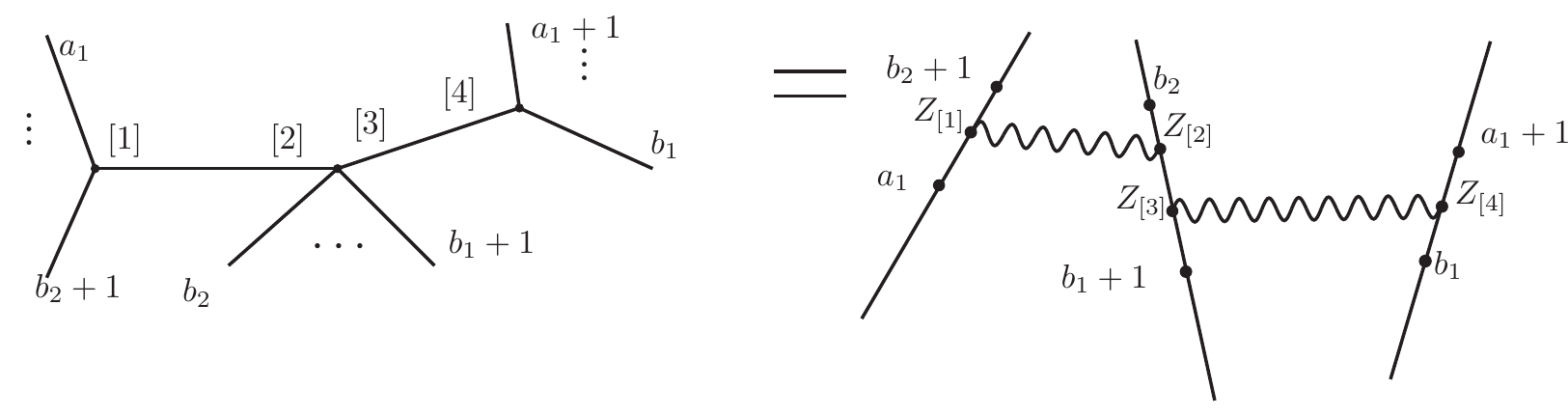}\caption{Twistor support of a boundary $\mathrm{N}^{2}$MHV tree diagram}\label{TN2C}
\end{figure}

A boundary diagram, on the other hand, is one in which the
propagator insertions are adjacent on the middle vertex (see Figure
\ref{TN2C}).  We obtain
\begin{multline}
\int \D^{3|4}Z_{[1]}\D^{3|4}Z_{[2]}\D^{3|4}Z_{[3]}\D^{3|4}Z_{[4]}
\, \bar{\delta}^{2|4}([1],\rf,[2]
)  \, \bar{\delta}^{2|4}([3],\rf,[4]) \times 
\nonumber \\
\V(b_{2}+1,\ldots, a_1,[1]) \V([2], [3],
b_1+1,\ldots,b_2) \V([4],a_1+1,\ldots,b_1) 
\label{N2MHVFNC}
\end{multline}
As before we can use \eqref{inverse-soft} to factor out three MHV
amplitudes/vertices, one for each vertex, depending only on the
external twistors. Because of the adjacency of $Z_{[2]}$ and $Z_{[3]}$
on the middle vertex there are two ways to do this depending on which
of these propagator insertions 
we use \eqref{inverse-soft} on first.  Taking $Z_{[2]}$ first we
obtain
\begin{multline}
\int \D^{3|4}Z_{[1]}\D^{3|4}Z_{[2]}\D^{3|4}Z_{[3]}\D^{3|4}Z_{[4]}
\, \bar{\delta}^{2|4}([1],\rf,[2])  \, \bar{\delta}^{2|4}([3],\rf,[4]) \times 
\nonumber \\
\bar{\delta}^{2|4}(a_1,[1],b_2+1)  \, \bar{\delta}^{2|4}(b_2, [2],[3])
\,
\bar{\delta}^{2|4}(b_2,[3],b_1+1)  \, \bar{\delta}^{2|4}(b_1,[4],a_1+1) \times \\
\V(b_{2}+1,\ldots, a_1) \V(
b_1+1,\ldots,b_2) \V(a_1+1,\ldots,b_1) 
\end{multline}
proceeding to do the $Z_{[1]}$ and $Z_{[2]}$ integrals as before, we
obtain 
\begin{multline*}
\int [a_{1},b_{2}+1,\rf,b_{2},[3]]\, \bar{\delta}^{2|4}([3],\rf,[4]
) \, \bar{\delta}^{2|4}(b_{1},a_{1}+1,[4])\, \bar{\delta}^{2|4}(b_{1}+1,b_2,[3])\, 
\D^{3|4}Z_{[3]}\D^{3|4}Z_{[4]}
\\
\V(b_{2}+1,\ldots, a_1) \V(
b_1+1,\ldots,b_2) \V(a_1+1,\ldots,b_1) 
\end{multline*}
Here 
we see that $Z_{[3]}$ is inserted into the first R-invariant but will
be fixed by integration against the remaining delta-functions. 
It is clear that $Z_{[3]}$ is uniquely determined to be 
at the intersection between the line $L_{b_1+1,b_2}$ joining
$Z_{b_1+1}$ to $Z_{b_2}$ and the plane spanned by $\langle
  \rf,b_{1},a_1+1\rangle$.  We therefore define
\begin{equation*}
Z_{\widehat{b_1+1}}=L_{b_{2},b_{1}+1}\cap\langle
\rf,b_{1},a_{1}+1\rangle\, .
\end{equation*}
which will be the final value of $Z_{[3]}$.  Now, integrating out
$Z_{[3]}$ and $Z_{[4]}$ against the delta functions we obtain
\begin{multline}\label{eqn: TN2C}
 [a_{1}, b_{2}+1 ,\rf,\widehat{b_1+1}, b_{2}] \, [b_2,b_1+1,\rf,
 a_1+1,b_1] \times  \\
\V(b_{2}+1,\ldots, a_1) \V(
b_1+1,\ldots,b_2) \V(a_1+1,\ldots,b_1) 
\end{multline}

If we had decomposed the middle MHV vertex using \eqref{inverse-soft}
in a different order, removing $Z_{[3]}$ first and then $Z_{[2]}$, we would
have obtained a different, albeit equivalent, formula.  
Following the above procedure, we obtain
\begin{multline}\label{eqn: TN2C-opp}
 [a_{1},b_{2}+1,\rf,{b_1+1},b_{2}]
\, [\widehat{b_2},b_1+1,\rf,a_1+1,b_1] \times
\\
\V(b_{2}+1,\ldots, a_1) \V(
b_1+1,\ldots,b_2) \V(a_1+1,\ldots,b_1) 
\end{multline}
where 
$$
Z_{\widehat{b_2}}= L_{b_2,b_1+1}\cap \la a_1,b_2+1,\rf\ra\, .
$$
As in the momentum twistor case, these two shifts are equivalent in
the sense that \eqref{eqn: TN2C} and \eqref{eqn: TN2C-opp} are equal.   

\subsubsection*{\textit{Boundary-Boundary terms}}
There is a final class of N$^2$MHV diagrams that doesn't quite fit into the
above framework; those in which there is only one external
particle on the  middle vertex (see the first diagram of Figure
\ref{boundary}.  This yields 
\begin{multline*}
V(b_{1}+1,\ldots , a_{1})V(a_{1}+2,\ldots ,b_{1}) \times \\
\int
\D^{3|4}Z_{[2]}\, \D^{3|4}Z_{[3]}\, V([2],a_{1}+1,[3])\, 
\bar{\delta}^{1|4}(b_{1}+1,a_{1},\rf,[2])\bar{\delta}^{1|4}(a_{1}+2,b_{1},\rf,[3]
)
, 
\end{multline*}
and by pulling out a delta function from the middle vertex reducing it
to a two-vertex, we can integrate out $Z_{[3]}$ to reduce to
\begin{multline}\label{eqn: bb-term}
V(b_{1}+1,\ldots , a_{1})V(a_{1}+2,\ldots ,b_{1}) \times \\
\int \D^{3|4}Z_{[2]} \,  V([2],a_{1}+1)\, 
[a_{1}+2,b_1,\rf ,a_1+1,[2] ]  
\, \bar{\delta}^{1|4}(b_{1}+1,a_{1},\rf,[2])\, . 
\end{multline}
At this point the remaining intergrations can be performed in various
ways, for example one can use the remaining explicit delta function or
one of those implicit in the 2-point vertex to perform (some of) the
remaining $Z_{[2]}$ integration. (If one were working in
  Euclidean signature, we could use \eqref{euc-2pt} to obtain a
  formula as a product of R-invariants but involving a complex
  conjugate twistor.)  However, the integral over the space of lines
through the given fixed point $Z_{a+1}$ on the middle vertex is
essential.  If these lines are to correspond to points of real Minkowski
space, then this is a one-dimensional integral, but in Euclidean
signature this would be zero-dimensional.

The full $\mathrm{N}^{2}$MHV amplitude is a sum over generic, boundary and
boundary-boundary diagrams using \eqref{eqn: TNC}, \eqref{eqn: TN2C}
and \eqref{eqn: bb-term}.

\subsection{$\mathrm{N}^{k}$MHV Tree Amplitudes}

We now extend this computational strategy to general
$\mathrm{N}^{k}$MHV tree amplitudes.  For arbitrary $k$, the amplitude
is built from a sum of diagrams, the building blocks of which we have
already encountered at the $\mathrm{N}^{2}$MHV level: generic terms
(i.e., diagrams with no adjacent propagator insertions); boundary
terms (i.e., diagrams in which one or more vertices have two or more
adjacent propagator insertions); and boundary-boundary terms (i.e.,
boundary terms in which a vertex with adjacent propagator insertions
has fewer than two external legs).  We deal with each type of term
separately in what follows. 

\subsubsection{Generic Terms}

\begin{figure}
\centering
\includegraphics[width=2 in, height= 1 in]{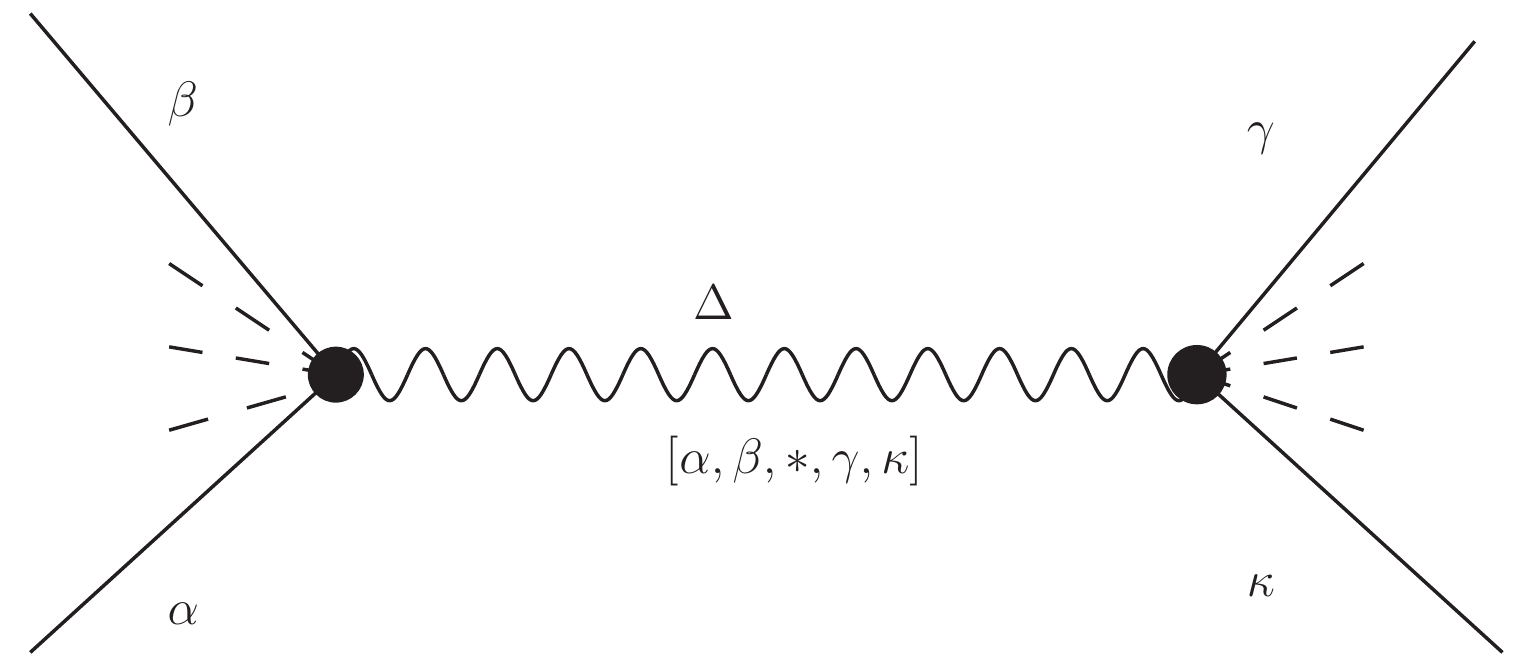}\caption{Propagator contributions}\label{PropR} 
\end{figure}
For diagrams in which there are no adjacent propagator insertions at any
vertex, our twistor space Feynman rules generalize directy from our
prior investigations.  At each vertex, using \eqref{inverse-soft}, we
can strip off a $\bar{\delta}^{2|4}$ at each propagator insertion
leaving an MHV vertex that depends only on the external particles at
that vertex.  Each propagator $\bar \delta^{2|4}(Z_{1},\rf,Z_{2})$
is then multiplied by the $\bar{\delta}^{2|4}(Z_1,\alpha, \beta)$ and 
$\bar{\delta}^{2|4}(Z_2,\gamma, \delta)$ that have been stripped off
from the MHV vertices at each end.  Here $\alpha$ and $\beta$ are the
two nearest external particles on one side of the progagator, while
$\gamma$ and $\kappa$ are the closest on the other side (see Figure
\ref{PropR}).  
As before, the integrations over $Z_1$ and $Z_2$ can be done
algebraically against the delta functions to yield the $R$-invariant 
$[\alpha,\beta,\rf,\gamma,\kappa]$.

\begin{figure}
\centering
\includegraphics[width=5.5 in, height=2 in]{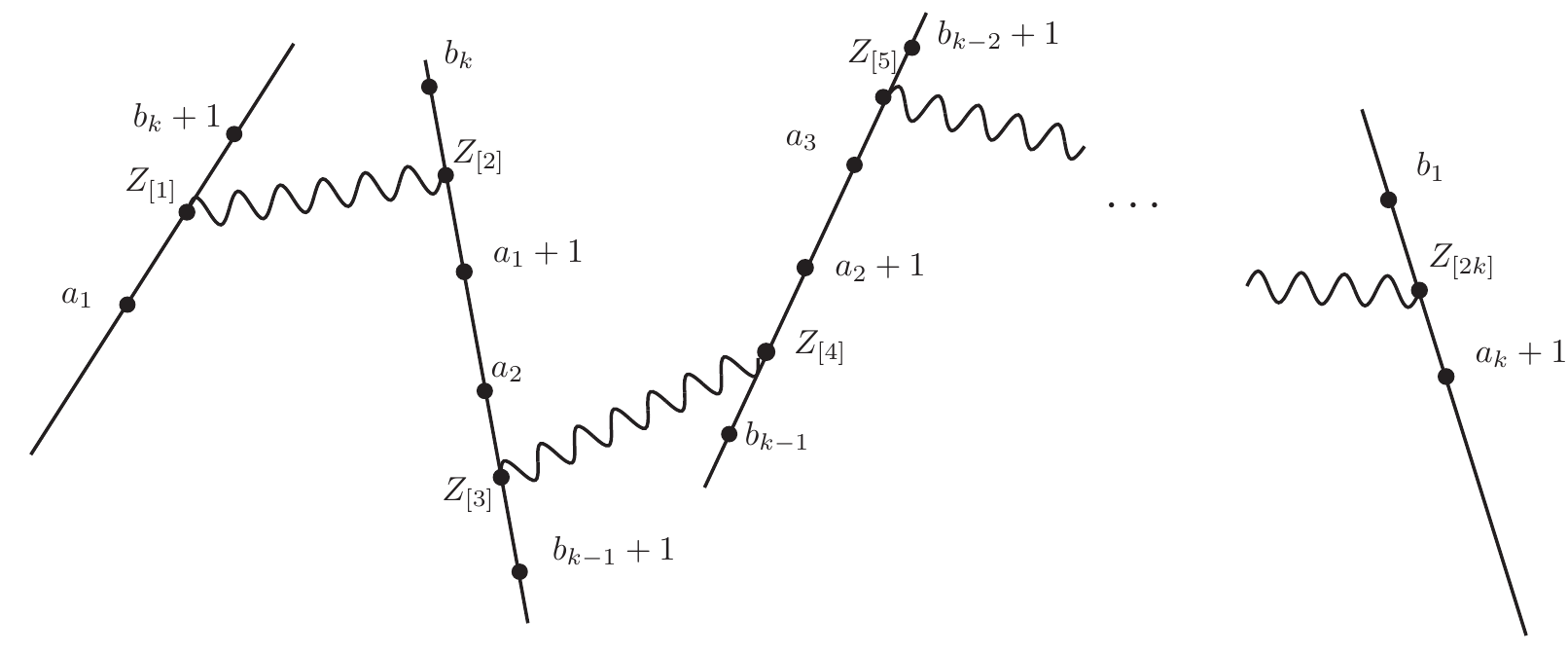}\caption{Twistor support of an example non-boundary $\mathrm{N}^{k}$MHV term}\label{TNkMHV}
\end{figure}
As an example, consider the generic term with diagram in twistor space given by Figure \ref{TNkMHV}.  Stripping off all external MHV amplitudes and then integrating over propagator insertions gives, in the numbering scheme for the external particles used by the diagram:
\begin{multline*}
V(b_{k}+1,\ldots, a_{1})V(a_{k}+1,\ldots, b_{1})\prod_{j=1}^{k-1}V(a_{j}+1,\ldots, a_{j+1}, b_{k-j}+1,\ldots, b_{k+1-j}) \\
\times \prod_{j=1}^{k}[b_{k+1-j}+1,a_{j},\rf,a_{j}+1,b_{k+1-j}].
\end{multline*}

Of course, for generic non-boundary tree diagrams, the precise form of
the contribution to the overall amplitude will depend on the
diagram's topology.  However, the formula is constructed
algorithmically as described above with 
$k+1$ MHV amplitudes built from external particles at each vertex, and
$k$ R-invariants built by connecting the external particles closest to
the two ends of each propagator and the reference twistor.

\subsubsection{Boundary Terms}
A boundary term will be a diagram for which some propagators are
inserted next to each other on some vertices, although we will for the
time being require that there are at least two external particles on
each vertex.

For boundary terms, the formulae are similar to the non-boundary case:
we obtain a product of $k+1$ MHV amplitudes, one for each vertex
containing only the twistors for the external particles at that vertex, and $k$
R-invariants, one for each propagator.  However, because of
adjacent propagator insertions, some of the entries in the
R-invariants associated to the propagators are now shifted.

The rule for the shifts can be obtained by studying each end of the
propagator separately; the R-invariant for a given propagator will
still have two pairs of twistors inserted into it, each pair lying on
one of the two lines associated to the vertices into which the ends of
the propagator are 
inserted.

\begin{figure}
\centering
\includegraphics[width=4 in, height=2 in]{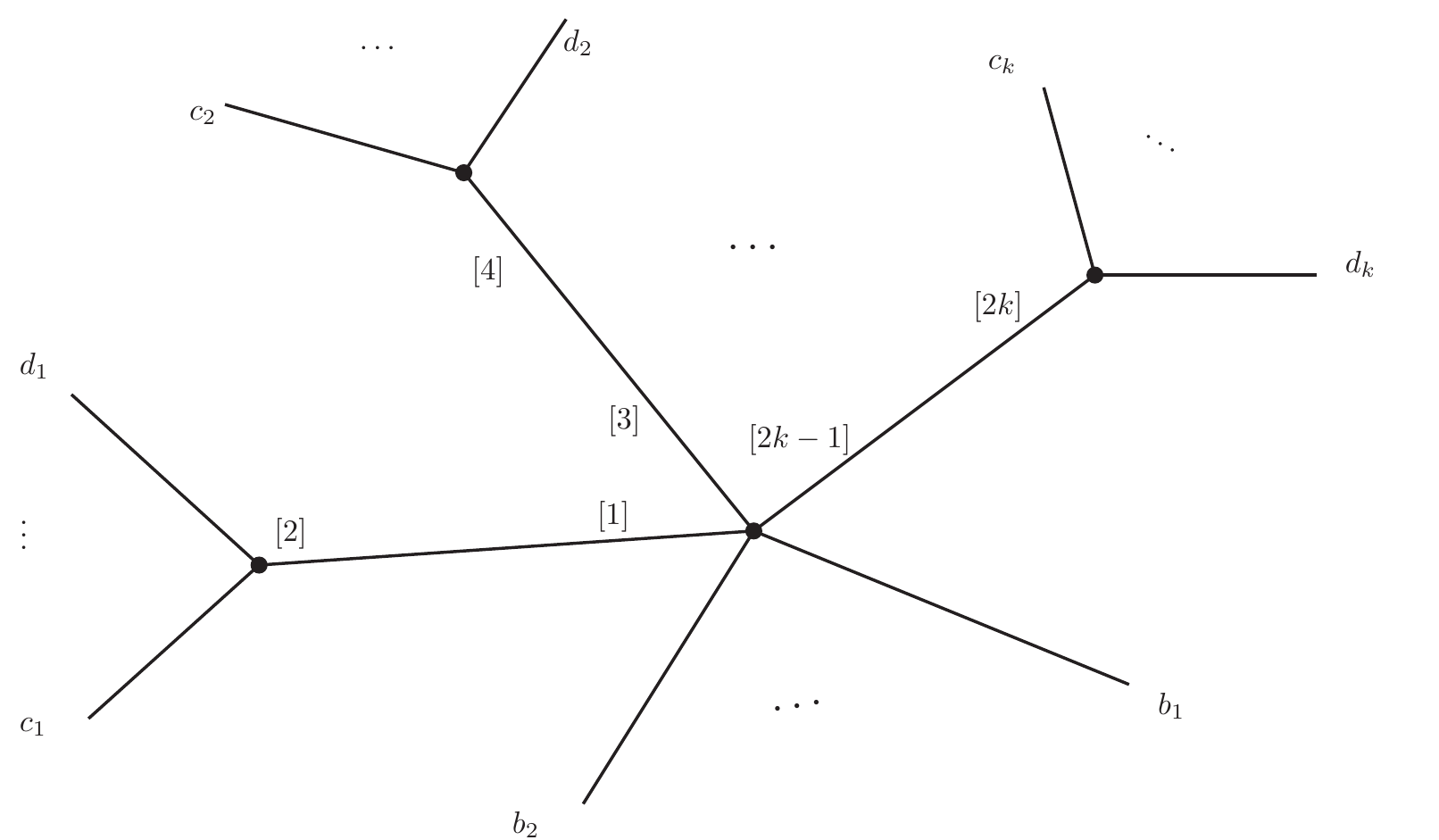}\caption{N$^k$MHV boundary term with $k$ adjacent propagators}\label{kboundary}
\end{figure}

To give the most general case, we compute the shifts at a vertex with 
$k$ adjacent propagators as in Figure \ref{kboundary}. 
We now use \eqref{inverse-soft} to decompose the central vertex into a
product 
\begin{multline*}
\bar{\delta}^{2|4}(b_{2},[1],[3]) \bar{\delta}^{2|4}(b_{2},[3],[5])
\dots \bar{\delta}^{2|4}(b_{2},[2j-1],[2j+1])
\cdots\bar{\delta}^{2|4}(b_{2},[2k-1],b_{1}) \times \\
\V(b_{1},\ldots, b_{2}).  
\end{multline*}
It is clear that we have made a choice in doing this and we could
easily have chosen the opposite orientation or indeed made other
choices.  The factor of $\V(b_{1},\ldots, b_{2})  $ will be left as
part of our
final answer, but we will seek to use the delta functions to integrate
out the $Z_{[2j-1]}$.
Introducing the propagators, the relevant integrals for the
$Z_{[2j-1]}$ are
\be{adjacent-props}
\int \prod_{j=1}^k\D^{3|4}Z_{[2j-1]} \, 
 \bar{\delta}^{2|4}([2j],\rf,[2j-1]) 
\bar{\delta}^{2|4}(b_{2},[2j-1],[2j+1])
\ee
where $Z_{[2k+1]}=Z_{b_2}$.  These can be done inductively starting
from $Z_{[2k-1]}$ and decreasing using the fact that we know that the $Z_{[2j]}$
lie on the lines $L_{c_j,d_j}$.  Performing the $Z_{[2k-1]}$ integral
against the delta functions $ \bar{\delta}^{2|4}([2k],\rf,[2k-1]) 
\bar{\delta}^{2|4}(b_{2},[2k-1],b_1)$ yields $\bar
\delta^{1|4}([2k],\rf, b_1,b_2)$.  Given that $Z_{[2k]}$ is fixed to
lie on the line $L_{c_k,d_k}$,  
$Z_{[2k-1]}$ must be fixed to lie, not
only on $L_{b_1,b_2}$ but also on the plane through $\rf$ and
$L_{c_k,d_k}$.  Thus we can substitute
\be{shiftk}
Z_{[2k-1]}=L_{b_1,b_2}\cap \la \rf,c_k,d_k\ra
\ee
into the remaining formulae.  Now that $Z_{[2k-1]}$ is fixed we can
carry on and integrate $Z_{[2k-3]}$ and so on by induction.  We 
finally obtain for the integral \eqref{adjacent-props}
\be{adjacent-props2}
\prod_{j=1}^k  \bar{\delta}^{1|4}([2j],\rf,b_2,[2j+1]) 
\ee
where now the $Z_{[2j-1]}$ are fixed twistors defined by
\be{shifts}
Z_{[2j-1]}=L_{b_1,b_2}\cap \la \rf,c_j,d_j\ra
\ee
To obtain a product of R-invariants we must now integrate out the
$Z_{[2j]}$ against the $\bar \delta^{2|4}$ obtained from
\eqref{inverse-soft} applied to the vertex on which it lies.  Assuming
no adjacent propagators to these on those vertices we obtain, as before,
for the diagram in Figure \ref{kboundary} the contribution
\begin{equation*}
V(b_{1},\ldots, b_{2})\prod_{j=1}^{k} V(c_{j},\ldots,
d_{j})[[2j-1],b_{1},\rf,c_{j},d_{j}]\, .
\end{equation*}

We have done this calculation for the case where just one end of a
propagator is adjacent to another, but we can state the rules for a
general diagram as follows:    
\begin{itemize}
\item Each vertex in the diagram gives rise to a factor of an MHV vertex in the
answer that depends only on the external legs at that vertex.
\item Each propagator corresponds to an R-invariant $[\widehat{a_1},a_2,\rf,\widehat{b_1},b_2]$ where $a_1$ and $a_2$ are the nearest
  external twistors with $a_1<a_2$ in the cyclic ordering on the
  vertex at one end of the propagator, and similarly for $b_1<b_2$ on
  the vertex at the other end.  Let $p$ be the insertion point on the
  vertex containing $a_1$ and $a_2$.  We have that $\widehat{a_1}$ is
  shifted according to the rule \be{shift-rule}
  Z_{\widehat{a_{1}}}=\left\{
\begin{array}{ll}
Z_{a_{1}} & \mbox{if $p$ is next to $a_1$} \\
L_{a_{1},a_{2}}\cap \langle c,d,\rf\rangle & \mbox{if $p$ is next to the propagator on the $a_1$ side} \\
 & \mbox{that connects to $L_{c,d}$}\, .
\end{array}\right. 
\ee
The rule for $\widehat{b_1}$ follows by $a\leftrightarrow b$.

\end{itemize}

\subsubsection{Boundary-Boundary Terms} 
\begin{figure}
\centering
\includegraphics[width=5 in, height=1.5 in]{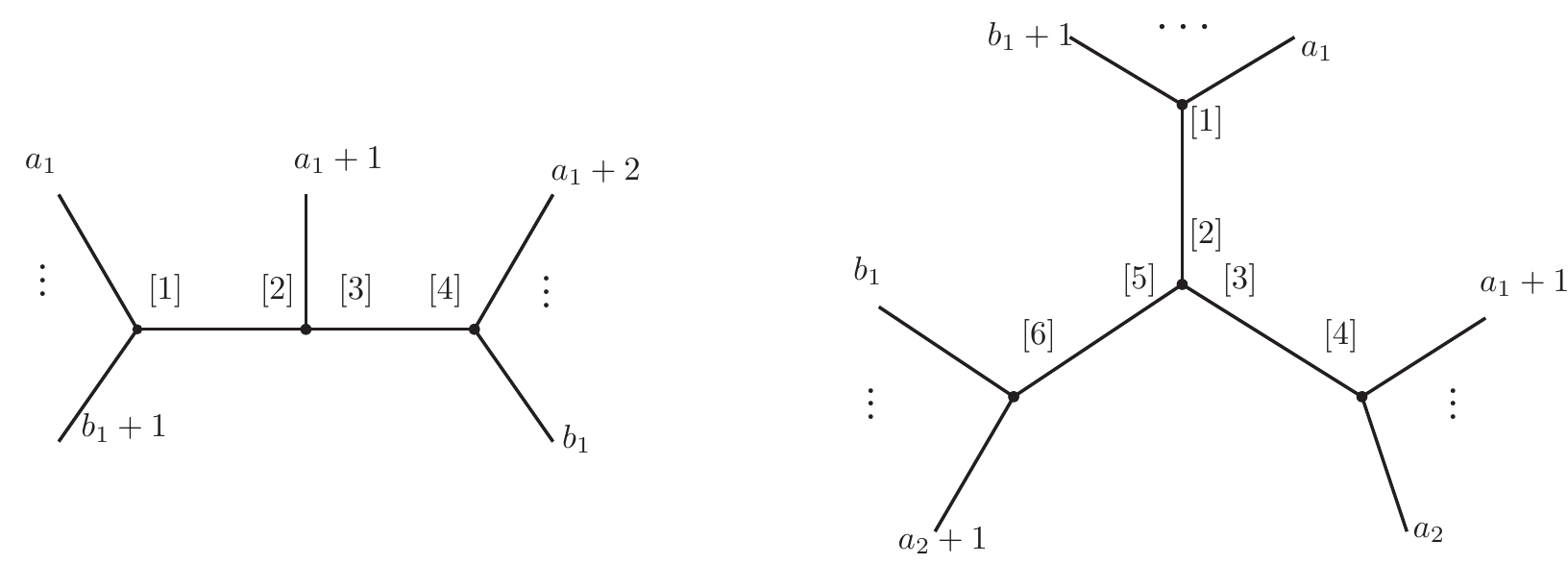}\caption{Boundary-boundary terms at N$^2$MHV and N$^3$MHV}\label{boundary} 
\end{figure}

We now turn to the boundary-boundary contributions when there are
fewer than two external legs on some vertices where the above
prescription breaks down: there will be no line $L_{a_{1},a_{2}}$ to
use in the definition of the shifted $Z_{a}$s so the shifts prescribed by
\eqref{shift-rule} for the boundary terms cannot be defined. See
Figure \ref{boundary} for simple examples of such diagrams; we already
considered the first of these in our discussion of the N$^2$MHV case.
In that case there remained one external leg on the diagram, and we
were left with an integral over one remaining internal twistor in
\eqref{eqn: bb-term}, although in principle, this can be reduced to an
integral over the space of real lines through the given fixed twistor
$Z_{a+1}$ which in Minkowski signature is one-dimensional, and in
Euclidean signature, zero-dimensional.  In general it can be worse than this: we
can have vertices with no external legs and our procedure will leave
us with two remaining twistors to integrate out.  The simplest of
these is the second diagram in Figure \ref{boundary} and we work
through this.

The N$^3$MHV `cartwheel' diagram represents essentially the worst-case scenario for a
boundary-boundary term.  It gives rise to the integral
%\begin{multline*}
%V(b_{1}+1,\ldots, a_{1})V(a_{1}+2,\ldots, b_{1})\int \D^{3|4}Z_{[3]} \bar{\delta}^{1|4}(b_{1}+1,a_{1},N,[2]) V([2],a_{1}+1) \\
% \times \left[[2],a_{1}+1,N,a_{1}+2,b_{1}\right],
%\end{multline*}
%and
\begin{multline*}
V(b_{1}+1,\ldots ,a_{1})V(a_{1}+1,\ldots ,b_{1})V(a_{2}+1,\ldots ,b_{1}) \times \\
\int\D^{3|4}Z_{[2]}\D^{3|4}Z_{[5]}\bar{\delta}^{1|4}(b_{1}+1,a_{1},\rf, [2])\bar{\delta}^{1|4}(a_{2}+1,b_{1},\rf, [5])V([2],[5])\left[a_{1}+1,a_{2},\rf, [2],[5]\right],
\end{multline*}
where $V(\cdot, \cdot)$ is the 2-point MHV amplitude given by \eqref{eqn: 2pt1b-text}.

We emphasize that although we have not been able to reduce
boundary-boundary terms to a simple expression in terms of shifted
twistors, they are still fully described by the twistorial MHV
formalism.  It is possible to reduce these further using the remaining
delta functions, but there is no reason in the case where there are no
external legs on a vertex not to be integrating over the full real four-dimensional space of real lines.  It seems to be impossible to
obtain an expression built only out of R-invariants and MHV
vertices.  However, we again stress that with a choice of real contour these remaining integral could be performed (and do not introduce divergences); this would simply entail the introduction of new signature-dependent machinery which we choose to avoid here.

A full N$^k$MHV tree amplitude is then computed in the MHV formalism on
twistor space by summing the contributions for all non-boundary, boundary,
and boundary-boundary terms for the given specification of external particles
and MHV degree.

\section{Loop diagrams in Twistor Space}
\label{Loop}  

We know that loops are calculated correctly from the MHV formalism in
momentum space at least at 1-loop \cite{Brandhuber:2004yw} and as far
as the loop integrand 
in the planar part of the theory is concerned it has been shown to be
correct to all loops for four-dimensional cut-constructable theories 
\cite{Bullimore:2010dz}.  The status of the MHV formalism at one loop and beyond in non-supersymmetric gauge theories is still speculative, because MHV rules miss the rational contributions to a scattering amplitude.  Furthermore, these loop amplitudes are generically divergent in four dimensions and require regularization.  We give here only a very superficial analysis
and consider only the simplest finite diagrams.  We consider also some of the
diagrams of the MHV 1-loop amplitude but our analysis here will be
inconclusive.  In particular, we will not 
regularize or introduce the Feynman $i\epsilon$ prescription (but see
\S\ref{discuss} for some discussion of this).

\subsection{Finite examples}

In order to start with finite examples, we consider first a non-planar
diagram at MHV and secondly a planar diagram at NMHV to show the
simplicity of the extension of the above ideas to loop amplitudes.

\subsubsection{Non-planar 1-loop MHV diagrams}

\begin{figure}
\centering
\includegraphics[width=3.5 in, height=2.25 in]{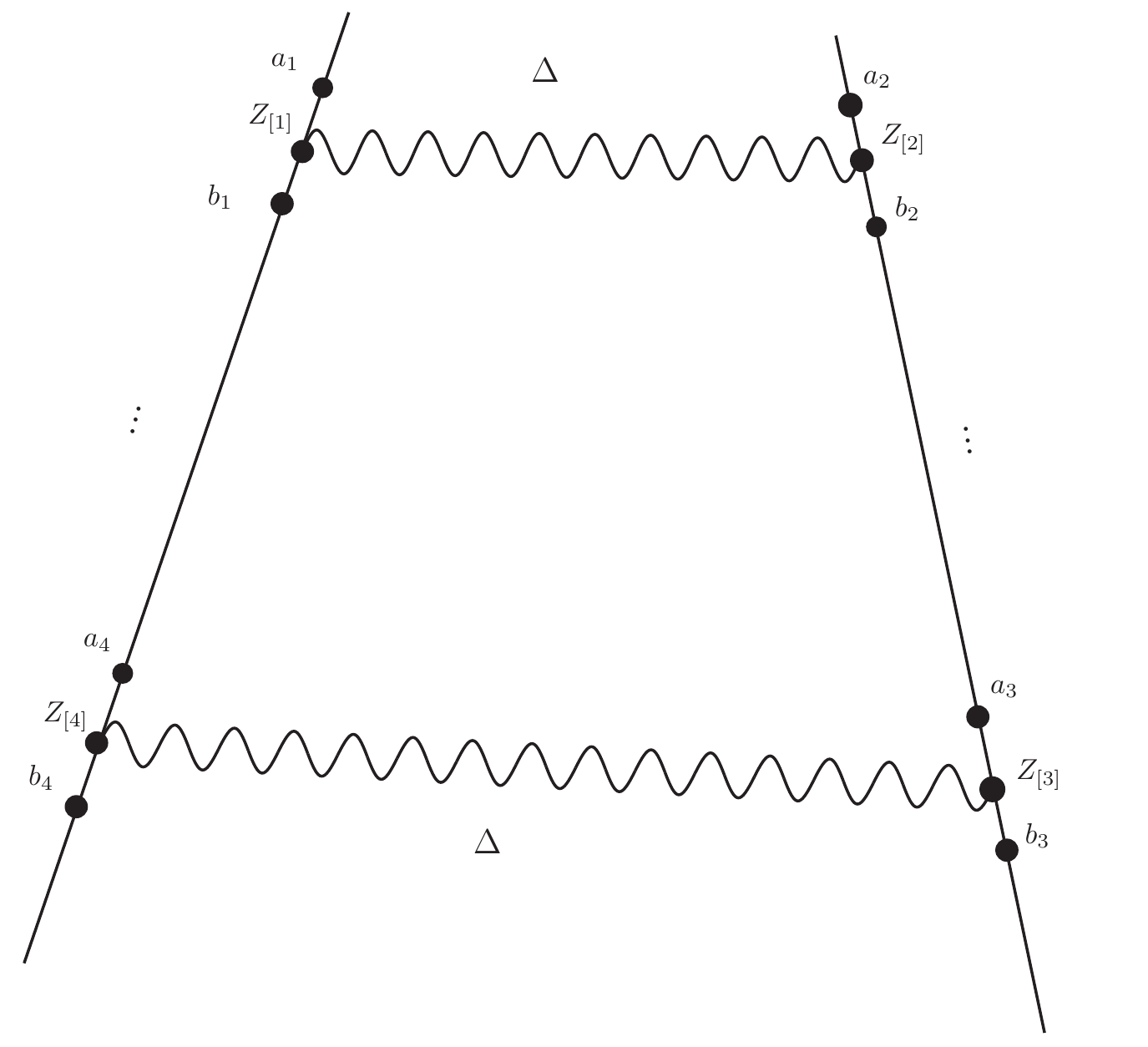}\caption{Non-planar 1-loop MHV diagram in twistor space}\label{NPMHV}
\end{figure}
For 1-loop diagrams at MHV we have two vertices connected by two
propagators.  In the planar case the propagator must be adjacent to
each other on both vertices as in figure \ref{TLoop} and we will see
that these are divergent.  In general
the propagators can have arbitrary separation and we consider these
separated cases first as
shown in Figure \ref{NPMHV}.   Such diagrams are non-planar. 

It is straightforward to see that we can integrate out the
intermediate twistors against delta functions obtained from the
vertices by \eqref{inverse-soft} exactly as at tree-level and indeed
the resulting expression: \be{eqn: NPMHV} V(a_{1},b_{1},\ldots,
a_{4},b_{4},\ldots)\V(a_{2},b_{2},\ldots, a_{3},b_{3},\ldots
)[a_{1},b_{1},\rf, a_{2},b_{2}][a_{3},b_{3},\rf, a_{4},b_{4}] \, .
\ee follows from the rules that we gave for generic tree diagrams.
(We give more details in the calculation for the divergent planar case
below).  However, the geometry underlying this calculation is
instructive.  The twistors that were integrated out to obtain this
formula were constrained to lie on the lines associated to the
vertices.  The propagators also fixed these points so that they lie on
the (unique) line through $\rf$ that is transversal to the lines
associated to the MHV vertices.  Thus the insertion points of the two
propagators at a given MHV vertex end up being the same points.  The
MHV vertex has no singularity when points come together unless they
are adjacent in the colour ordering where they have a pole.  Here in
this generic non-boundary case they are not adjacent.  In the boundary
case one anticipates therefore one degree of divergence as two ends of
the propagators must lie on top of a pole.  In the planar case we will
see a double divergence as both ends of the propagators will be
adjacent.

\subsubsection{Planar NMHV at 1-loop}

\begin{figure}
\centering
\includegraphics[width=4.5 in, height=2.25 in]{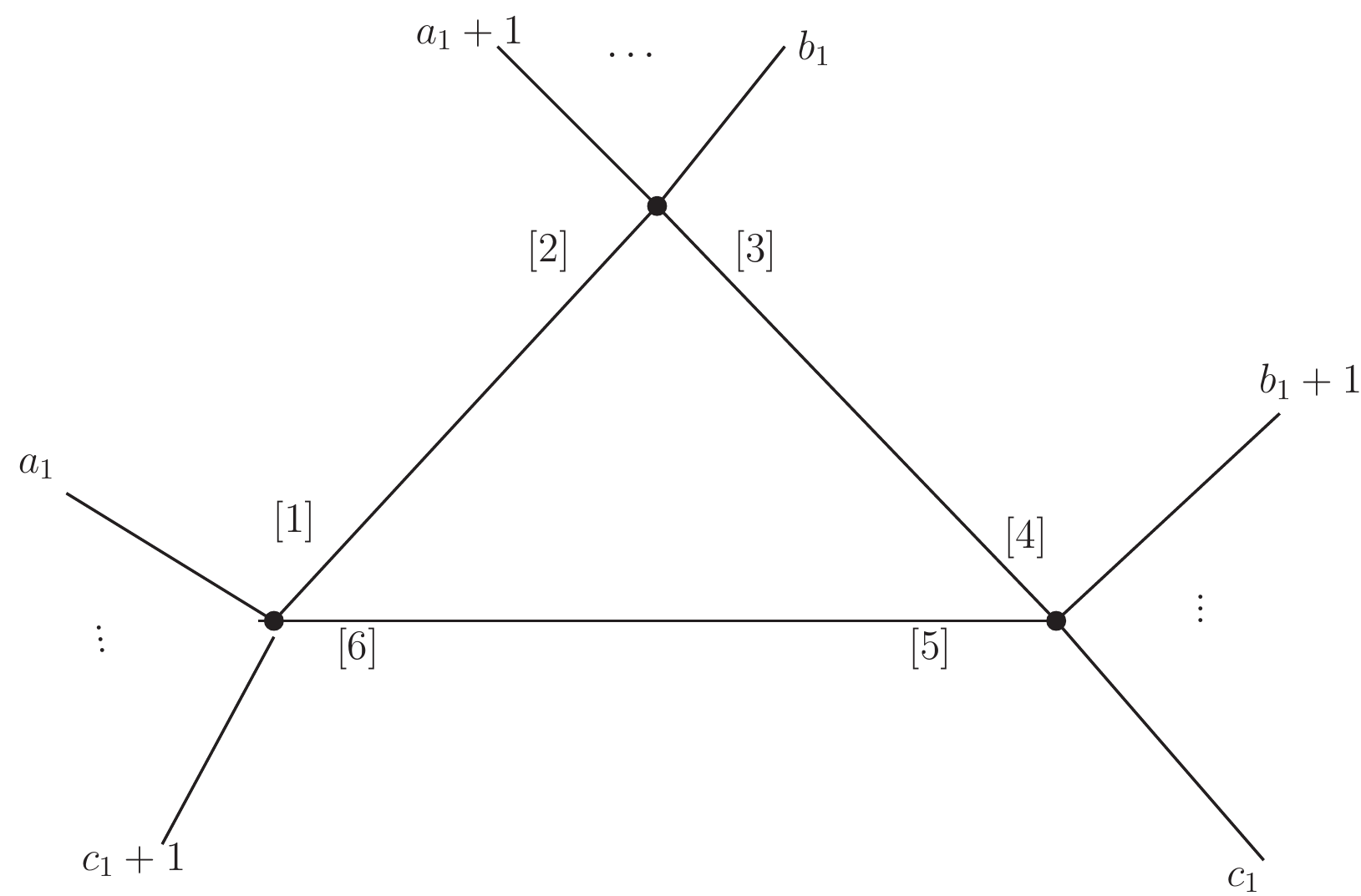}\caption{Triangular 1-loop NMHV diagram}\label{LNMHV}
\end{figure}

Two types of diagram contribute to the NMHV amplitude at 1-loop, one
divergent and one finite. 
The finite cases are as in Figure
\ref{LNMHV} (the divergent ones are the same as the MHV case with an
additional vertex connected by a propagator into one of the vertices).

The computation of the the internal integrals follows identically to
the tree-level cases above.  
Using the numbering scheme for external particles indicated by Figure
\ref{LNMHV}, we can trivially integrate over $\D^{3|4}Z_{[2]}$,
$\D^{3|4}Z_{[4]}$ and $\D^{3|4}Z_{[6]}$ to give: 
\begin{multline*}
\int \D^{3|4}Z_{[1]}\D^{3|4}Z_{[3]}\D^{3|4}Z_{[5]} \bar{\delta}^{2|4}(c_{1}+1,a_{1},[1]) \bar{\delta}^{1|4}([1],\rf, [3],a_{1}+1) \bar{\delta}^{2|4}(a_{1}+1,b_{1},[3]) \\
\bar{\delta}^{1|4}([3],\rf, [5],b_{1}+1) \bar{\delta}^{2|4}(b_{1}+1,c_{1},[5])\bar{\delta}^{1|4}([5],\rf, [1], c_{1}+1) \\
 \V(c_{1}+1,\ldots, a_{1})\V(a_{1}+1,\ldots, b_{1})\V(b_{1}+1,\ldots, c_{1}).
\end{multline*}
The remaining integrals can be performed against the delta functions
yielding the shifted twistors $\hat a, \hat b, \hat c$ for $Z_{[1]}$, $Z_{[3]}$, $Z_{[5]}$ respectively, as before to give:
\begin{multline*}
\V(c_{1}+1,\ldots, a_{1})\V(a_{1}+1,\ldots, b_{1})\V(b_{1}+1,\ldots, c_{1}) \\
\times[c_{1}+1,a_{1},\rf ,\widehat{a},a_{1}+1][a_{1}+1,b_{1},\rf ,\widehat{b},b_{1}+1][b_{1}+1,c_{1},\rf, \widehat{c},c_{1}+1],
\end{multline*}
where
\begin{equation*}
\widehat{a}=L_{a_{1}+1,b_{1}}\cap\langle b_{1}+1,c_{1},\rf\rangle, \qquad \widehat{b}=L_{b_{1}+1,c_{1}}\cap\langle c_{1}+1,a_{1},\rf\rangle, \qquad \widehat{c}=L_{c_{1}+1,a_{1}}\cap\langle a_{1}+1,b_{1},\rf\rangle.
\end{equation*}

\subsection{Planar 1-loop MHV}
\begin{figure}
\centering
\includegraphics[width=5.5 in, height=2 in]{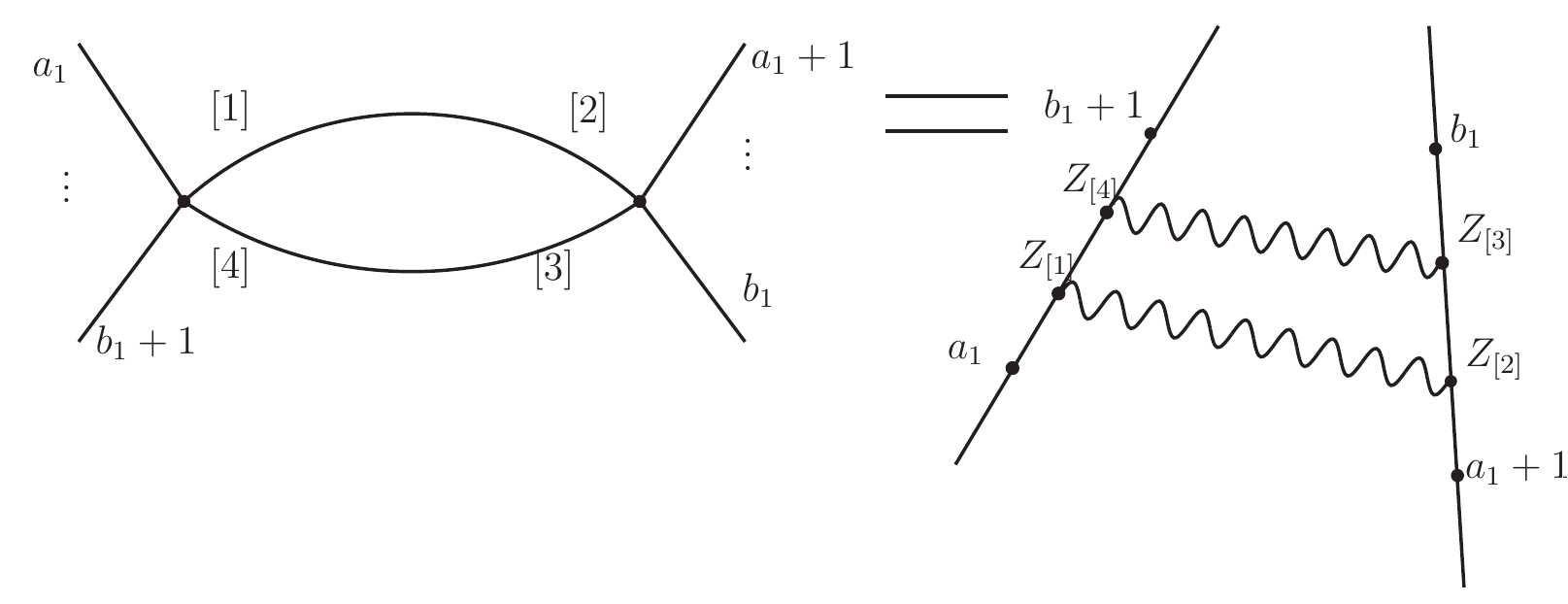}\caption{Twistor support of the 1-loop MHV amplitude}\label{TLoop}
\end{figure}

The diagrams for the planar 1-loop MHV amplitudes all have the same 
form as given by Figure 
\ref{TLoop} (although we do have  boundary terms when there is only
one leg on one of the vertices).

In the generic case, we have enough delta functions to integrate out
the $Z_{[j]}$ as in the tree-level cases above.  However, the geometry
of the relations implied by the delta functions (as in the non-planar
MHV case) forces the point
$Z_{[2]}$ to be coincident with $Z_{[3]}$ and $Z_{[1]}$ to be
coincident with $Z_{[4]}$.  This is because the propagators force both
the pairs $(Z_{[1]},Z_{[2]})$ and $(Z_{[3]},Z_{[4]})$ to lie on the
common transversal to the two lines through $\rf$ as indicated in
Figure \ref{TLoop}).  Given two lines in general position (i.e., those
corresponding to the two MHV vertices) and a point in $\PT$ not on
those lines (here the CSW reference twistor $\rf$), there is a
\emph{unique} transversal connecting the lines and intersecting this
point.  Hence, the lines $L_{[1][2]}$ and $L_{[3][4]}$ must in fact be
the same, which in turn means that $Z_{[1]}=Z_{[4]}$ and
$Z_{[2]}=Z_{[3]}$.  Now, the MHV vertices have simple poles whenever
any two of their arguments coincide, so the geometry evaluates the two
vertices at one of each of their poles.

Using the reduced rules above obtained by naively performing the integrals 
against the delta functions gives
\begin{equation*}
[a_{1},b_{1}+1,\rf,
\widehat{d},a_{1}+1] [b_{1},a_{1}+1,\rf, \widehat{c}, b_{1}+1]
V(a_1+1,\ldots, b_1) V(b_1+1,\ldots, a_1)
\end{equation*}
where
\begin{equation*}
\widehat{c}=L_{a_{1},b_{1}+1}\cap\langle b_{1},a_{1}+1,\rf\rangle, \qquad \widehat{d}=L_{b_{1},a_{1}+1}\cap\langle a_{1},b_{1}+1,\rf\rangle.
\end{equation*}

Considered on its own, each R-invariant can be seen to contain a
divergence arising from the geometry outlined above.
Indeed, recall that 
\begin{equation*}
[b_{1},a_{1}+1,\rf, \widehat{c}, b_{1}+1]=\frac{\delta^{0|4}\left((b_{1},a_{1}+1,\rf, \widehat{c})\chi_{b_{1}+1}+\mathrm{cyclic}\right)}{(b_{1},a_{1}+1,\rf, \widehat{c})\cdots (b_{1}+1,b_{1},a_{1}+1,\rf)},
\end{equation*}
and observe that from its definition, $\widehat{c}$ is co-planar with $b_{1}$, $a_{1}+1$,
and $\rf$ and hence the denominator factor of $(b_{1},a_{1}+1,\rf,
\widehat{c})=0$.  There is a similar zero in the denominator of
$[a_{1},b_{1}+1,\rf,  \widehat{d},a_{1}+1]$, so we obtain a `$1/0$'
from each R-invariant. However, considered together the fermionic
parts of the two R-invariants  
are proportional to each other and so using the nilpotency of these
expressions, these numerator terms vanish \footnote{We thank Mat Bullimore for this observation}

Clearly, the divergence (or non-divergence) properties of this planar
1-loop MHV amplitude are dependent upon a careful treatment of this
`0/0'.  Presumably careful regularization should only be required for
those of these diagrams that are actually divergent, and taking care
of the Fermionic structure first before performing the integrals
should lead to finite answers for the generic finite case of these diagrams.
Genuine divergences only arise in the case when one of the vertices
is the 4 point vertex \cite{Bena:2004xu}.  
To make sense of the genuinely divergent cases, regularization is required and
we discuss this further in \S\ref{discuss}. 

There does not seem to be a correlation between boundary-boundary
terms and divergences as one might have initially feared (although of
course that is not to say that there are not divergent
boundary-boundary diagrams).

\section{Discussion and further  directions} \label{discuss} 

We have seen that it is possible to formulate scattering amplitudes in
twistor space in such a way as to deal with both their
cohomological and invariance properties explicitly if we
regard amplitudes as being in the `topological dual' to the wave
functions; that is, as tensor products of $\Omega^{0,2}$s with compact
support rather than as $H^1$s where the wave functions live.

Using this approach we were able to take the twistor action in the
simplest axial gauge, the CSW gauge, and obtain its Feynman
diagrams. At tree level we discovered that for a large class of diagrams,
as in momentum space, one can perform the internal integrals against
delta functions to give an expression as a product of MHV vertices,
one for each vertex in the diagram but evaluated only on its external
particles, and an R-invariant for each propagator.  This is not quite possible when a line for the vertex is not determined by
external particles on that vertex, which is the case when there are
fewer than two external legs attached.  However, in contradistinction
to momentum space Feynman rules, this integration against delta
functions is still possible for loop diagrams although one then sees
the divergence directly, and we have not yet introduced a suitable
regulation.  This is a major problem that we have not begun to address here.

\subsection*{\textit{Cohomology, crossing symmetry and anomalies}}

As has already been mentioned, wave functions on twistor space are
cohomology classes in $H^1_{\dbar}(U,\cO(n))$, being $(0,1)$-forms
$\phi$ that are $\dbar$-closed ($\dbar \phi=0$) defined modulo the
gauge freedom $\phi\rightarrow \phi+\dbar f$ on a domain $U
\subset \CP^{3|4}$.  Amplitudes are functionals of such wave functions
representing the asymptotic states that are to be scattered.  Crossing
symmetry tells us that it shouldn't matter whether the wave function is
incoming or outgoing (i.e., positive or negative frequency).  The
positive/negative frequency condition is the condition on the set $U$
being $\PT^\pm=\{\pm Z\cdot\bar Z\geq 0\}$.  

In the first instance therefore, the kernel of the
amplitude is in the $n$-fold product of the dual to such $H^1$s. Since
the external wave functions are elements of a Hilbert space, one
usually imagines that one can blur this distinction between the space
of wave functions and its dual as the Hilbert space is dual to itself.
However, for twistor theory in Lorentz signature, this duality is
somewhat involved:  the first step in
defining the Hilbert space structure on a wave function is complex
conjugation.  This maps twistor space to dual twistor space.  The
Hilbert space structure therefore requires the use of the `twistor
transform' from cohomology classes on dual
twistor space to map them back to twistor space to a class with weight
$-n-2$  on $\PT^-$ 
where it is dual to the original class in $H^1(\PT^+,\cO(n-2))$ by
integration over $\PN=\{ Z\cdot \bar Z=0\}$.  Thus, the
duality is highly non-trivial and there is a big difference between
dual descriptions.  Unitarity is not manifest on twistor space.  

Rather than use the nonlocal Hilbert space structure, we have
represented the kernel of an
amplitude using the local duality between $\Omega^{0,1}(\PT,\cO(n-2))$ and
$\Omega^{0,2}_c(\PT,\cO(-n-2))$ where the subscript $c$ denotes compact
support.  Crossing symmetry is then made manifest when this compact support
lies in $\PN$ so that it makes sense when integrated
against external fields of both positive and
negative frequency.   This support was made clear in the definition of
the vertices in which the external twistors were all required to lie
on a line that corresponds to a point of real Minkowski space.   Such
lines automatically lie in $\PN$.

Thus, our amplitudes took values in the $n$-fold tensor product
of $\Omega^{0,2}_c(\PT,\cO(-n-2))$
\begin{equation*}
A(Z_{1},\ldots, Z_{n})\in\Omega^{0,2n}_c(\times_{a=1}^{n}\PT'_{a},\cO).
\end{equation*}
To obtain a functional of wave functions, we use the natural duality between
$(0,2)$- and $(0,1)$-forms on $\PT'$ given by: 
\begin{equation*}
\Omega^{0,2}(\PT',\cO)\otimes\Omega^{0,1}(\PT',\cO)\rightarrow\C,
\qquad (\alpha,\phi)\mapsto
\int_{\PT'}\D^{3|4}Z\wedge\alpha\wedge\phi. 
\end{equation*}
It is defined modulo $\dbar $ of forms with compact support as these
will give zero by integration by parts.
To obtain a formula for the amplitude as a functional of wave
functions of external particles, we simply pair the amplitude
with the external wavefunctions to obtain the normal amplitude.  See
\eqref{eqn: TMHV} above for the example of the MHV amplitude and 
\eqref{v2-exact} where we show that the 2-point vertex is exact so
that it vanishes as an amplitude.

The amplitude ought to be $\dbar$-closed with compact support for the
integral to only depend on the cohomology class of $\phi$.  If so, we
would have that the amplitude take values in the cohomology groups
$H^{2n}_c(\times^n\PT)$.  However, in our
context, we have gauge-fixed and so this is not an absolute
requirement and indeed these groups vanish.  
The MHV amplitude itself seems reasonably canonical, 
and so one might hope that it at least is
$\dbar$-closed.  We find by direct computation along the lines of that
leading to \eqref{dbar-delta-line} \be{dbar-MHV} \dbar \V(1,\dots,n)=
2\pi i \sum_a \left(\bar\delta^{3|4}(a,a+1) \right) \V(1,\dots,\hat
a,\ldots,n) \, .  \ee This is a clear obstruction to realizing the
amplitudes as cohomology, and indeed this equation expresses the
standard infrared singular behaviour of amplitudes under collinear
limits.  One can understand this equation as implying that, when
understood as an $H^2$, the amplitude has a simple pole on the
diagonals.  
See \cite{Atiyah:1981ey} for a discussion of how such
objects can be understood in algebraic geometry.

Thus, the failure of the amplitude to be $\dbar$-closed would seem to
be an anomaly arising from infra-red divergences.  It leads to a
failure of gauge invariance, but the machinery of quantum field theory
has in any case required that we fix our gauge.  A different gauge
fixing would lead to very different formulae for the amplitudes.  The
anomalies are associated to the same poles in the MHV amplitude that
gives rise to anomalies in (super-)conformal invariance as noted by
\cite{Beisert:2010gn} and it seems likely that a proper coholomogical
treatment will require a similar treatment to that given there.

\subsection*{\textit{Comparison to the momentum twistor formulation}}
It is instructive to compare the twistor space version of the MHV
formalism to that in {\em momentum} twistor space. Momentum twistors
were introduced by Andrew Hodges~\cite{Hodges:2009hk} and are based on
{\em dual} conformal invariance.  The dual conformal group is the
conformal group acting on region momentum space, an affine version of
momentum space.  It arises from T-duality in the AdS/CFT
correspondence \cite{Alday:2007hr}, but had already been observed in
the integrands of certain loop amplitudes \cite{Drummond:2006rz} and
was seen to extend to all planar amplitudes in various works
\cite{Drummond:2007cf, Drummond:2008vq, Brandhuber:2008pf,
  Brandhuber:2009xz, ArkaniHamed:2010kv, Boels:2010nw}.  The transform
from momentum space to momentum twistor space is essentially a local
coordinate transformation that uses the twistors for the {\em dual}
conformal group rather than the standard conformal group and makes
manifest that invariance.  The MHV formalism was reformulated in
momentum twistor space in~\cite{Bullimore:2010pj} (and this was the
framework in which the proof of the all-loop integrand for the planar
MHV formalism was obtained \cite{Bullimore:2010dz} by extending the
recursion methods of \cite{Risager:2005vk, Elvang:2008na,
  Elvang:2008vz}).

The correspondence between the two different twistor space MHV
formalisms is relatively simple, but with important differences.
This is despite the fact that momentum twistors and ordinary twistors
are related in a highly non-local way (in split signature by the
half-Fourier transform) reflecting T-duality on space-time.  In 
momentum twistor space, the vertices of the MHV formalism
simply
correspond to 1, and the propagators are given by 
the R-invariants, now for the {\em dual} superconformal group.
If we compare this to the MHV formalism in twistor space obtained in
this paper, in generic diagrams we obtain precisely the same
R-invariants for the propagators, except here they are functions of
twistors, whereas in the momentum twistor version they are functions
of momentum twistors.  However, in twistor space the MHV vertices are
not 1, but are given by the standard twistor MHV formula, i.e., as the product of delta-functions
$\bar\delta^{2|4}$ that ensure collinearity.  However, for the boundary
diagrams, although shifted twistors need to be used in
both versions of the formalism, the geometry of these shifts are
different in the two different twistor versions of the formalism.

At the level of the action, the twistor action was
also used to obtain the momentum twistor version of the MHV diagram
formalism \cite{Mason:2010yk}.  However, in this context it was obtained as a
diagrammatic expansion for the correlation function of a (holomorphic)
Wilson loop in twistor space rather than an amplitude. (This gave the
first proof of the Wilson-loop amplitude correspondence and indeed the
first formulation that extends beyond MHV amplitudes \cite{Bullimore:2011ni}; it also leads to a
definition of a supersymmetric space-time Wilson loop.) However, it is
worth remarking that in this context the Feynman diagrams for the
correlation functions are the planar duals of those for the
amplitudes.  That these can lead to the same formulae is only possible
because, for momentum twistors, the vertices are just given by 1.

\subsection*{\textit{Other axial gauges}}
It is worth noting that the CSW gauge is by no means the only axial
gauge.  One only needs to choose a holomorphic 1-dimensional
distribution $D\subset T^{1,0}\PT$ and require that $\cA|_{\bar D}=0$.
The simplest way to do this is to choose a global holomorphic vector
field $V$ (so that $D$ is the span of $V$) and require that $\bar V
\lrcorner \cA=0$.  The CSW gauge arises when $V$ corresponds to a null
translation, but we can in principle adapt $V$ to any problem we
choose.  

The vertices do not change if we change the gauge, but the propagator
does.  For example, if $V = T^{AA'}\pi_{A'}\p/\p\omega^A$, as arises
from a timelike translation, we obtain for the propagator
\be{time-prop} \Delta(1,2)= \int \frac{\rd s\rd t}{st}
\bar\delta^{4|4}( Z_{1}+ s Z_{2} + t T(Z_1) ) \ee where $T(Z)=
(iT^{AA'}\pi_{A'},0,0)$.  It is conceivable that such a propagator
will give rise to alternative useful formulae.

\subsection*{\textit{Feynman $i\epsilon$ prescription}}
A final issue that we have not explicitly spelt out is how to
ensure that we have incorporated the Feynman $i\epsilon$ prescription
into the propagator.  An automatic way of doing this is to
analytically continue to Euclidean signature; this is
implicit in our form of the propagator. Indeed, we see in
appendices \S\ref{A}--\ref{Prop} that in order to obtain the momentum space propagator
correctly, it is necessary to perform the calculation on the Euclidean
real slice.  However, this also implies that in the definition of the two-point vertex, the `real' contour of Minkowski space must be understood to include an `$i\epsilon$ shift' of the Lorentzian real slice (i.e., one that is topologically equivalent to the Euclidean slice).  This must be
understood in a limiting sense if we also wish to maintain manifest
crossing symmetry which requires us to take the limit back onto the
Lorentzian real slice.

\subsection{Open questions}
\subsubsection*{\textit{Tree amplitudes}}

It would be helpful to have some further analysis of the
boundary-boundary terms.  We have not pushed them further here as they
do not fit into our generic pattern of reducing to MHV vertices with
just external particles multiplied by an R-invariant for each
propagator.  Furthermore, these diagrams do not have a special status
in momentum space (or in momentum twistor space).  There are
nevertheless further delta functions within the two-point vertex that
we haven't exploited.  In the case of a vertex with no remaining
external legs, we will be left with a four dimensional integral that
is unconstrained by delta functions.  With just one external particle,
in Lorentz signature, there is one remaining integral, whereas in
Euclidean signature there are none.  

One unsatisfactory feature of our discussion is our treatment of the
two-point vertex.  We know that it does not contribute from momentum
space arguments, as shown in \S \ref{2pt0}, and it is also easily seen to vanish from the point of
view of the twistor action by evaluating it on off-shell momentum
eigenstates---this was the argument used in \cite{Boels:2007qn}.
However, it is not so obvious why it vanishes within the formalism
used in this paper (see Appendix \ref{2pt} for some further formulae and discussion).  The
puzzle is sharpened by the fact that the 2-point vertex does not
vanish when evaluating correlation functions; in the context of the
holomorphic Wilson loop in twistor space \cite{Mason:2010yk}, it is
the basic ingredient in the MHV 1-loop amplitude.

\subsubsection*{\textit{Loop amplitudes}}
It is striking that for many loop diagrams, it is possible to perform
all the integrals against the delta functions and that their
evaluation is essentially as easy as at tree-level.  This is clearly
quite different from the momentum space (or indeed momentum twistor)
framework in which there are always four remaining integrals at each
loop order.  This leads to the possibility that this could become a
more efficient formalism than that in momentum space.  However, we
need to have a systematic scheme to cope with divergences to make it
useful.  The usual strategy on momentum space is to regularize
divergences using dimensional regularization (c.f.,
\cite{Brandhuber:2004yw}).  This would seem to be awkward to apply in
a twistorial context as twistor-theory does not scale so easily to
higher dimensions.  An easier approach is to use the mass regulation
using the Higgs branch introduced in the context of the AdS/CFT
approach to scattering theory in terms of string theory on
$AdS_{5}\times S^{5}$ \cite{Alday:2009zm}.  In the context of momentum
twistors this leads to local adjustments to the formulae on momentum
twistor space which arise from the same twistor action, and so could
lead to a scheme that is applicable in our context.  Ideally we would
regulate the whole theory in this way and obtain regularized
amplitudes as an output.  It remains to be seen whether such a
regulated theory would be as computable as that described above.

Another approach to regulation is to simply focus on finite terms. It
is a standard fact that divergences are controlled by those at 1-loop
(or indeed from the Wilson loop point of view, from the $U(1)$ theory)
and the cusp anomalous dimension.  Thus one can cancel divergences and
focus on the finite remainder.  A particularly elegant strategy here
would be to directly calculate the finite cross-ratio expressions used
in the OPE approach of \cite{Alday:2010ku}.

An important feature of loop amplitudes is the relationship between
the transcendentality degree of functions (polylogs) of momenta and
the loop order (i.e., at $l$-loops, the functions and coefficients
have transcendentality degree $2l$ and $2l$-logs appear), see
\cite{Goncharov:2010jf,Gaiotto:2011dt} for applications.  Since there
is, at least in a moral sense, a half-Fourier transform between our
relatively accessible expressions and those on momentum space we
shouldn't be expecting to see polylogs directly.  Nevertheless, one
might hope that there should be a direct way of recognising the
symbols of the polylogs that arise.  As a first step, one should
perhaps already be able to see the transcendental coefficients of the
cusp anomalous dimension as one attempts to cancel divergences in
multi-loop diagrams with those in powers of the 1-loop amplitude
according to the definition of the finite parts of the log of the
amplitude that arises in the Wilson-loop point of view.
Perhaps more importantly, we should be able to simply use the
transform to momentum space described above
\eqref{momentum-eigenstate} to obtain the polylogs directly from the
fully integrated loop twistor MHV diagrams.

\subsubsection*{The Grassmannian connection}
In the Grassmannian construction of
\cite{ArkaniHamed:2009dg,Mason:2009qx,Bullimore:2009cb}, tree
amplitudes and the leading singularities of loop amplitudes at
N$^k$MHV are obtained as residues of a contour integral involving
superconformally invariant delta functions of the form used in this
paper over the Grassmannian $G(k+2,n)$ of $k+2$-planes in in $\C^n$.
In \cite{ArkaniHamed:2009sx} the connection with the MHV formalism in
momentum twistor space was established at NMHV using a recursion
argument.  By incorporating this work (and that in
\cite{Bullimore:2010pj} for momentum twistors) we should be able to
obtain a more systematic formulation of the full amplitude in the
Grassmannian derived from Lagrangian principles.  A particular
advantage of the Grassmannian formulation is that global residue
theorems can be used to obtain equivalences between different formulae
for the same amplitude and allow us to obtain improved formulae.  In
particular, it would be a great help in the regularisation problem to
obtain twistor (as opposed to momentum twistor) forms of the `local'
versions of the loop amplitude integrands found in
\cite{ArkaniHamed:2010gh}. 

\begin{acknowledgments}
We are grateful to Fernando Alday, Nima Arkani-Hamed, Freddy Cachazo,
 and Andrew Hodges 
for useful and stimulating conversations and
particularly David Skinner and Mat Bullimore for a number of ideas and
contributions that may not be fully acknowledged in the text.  We
further thank Gabriele Travaglini and Ed Witten for raising issues
that led to the correction of erroneous claims 
in earlier versions of this article. TA is supported by
the NSF Graduate Research Fellowship (USA) and Balliol College; LM is 
supported by the EPSRC grant number EP/F016654.
\end{acknowledgments}

\section*{Appendix} 
\appendix

\section{The 2-Point vertex in twistor space}
\label{2pt}
We showed in \S\ref{2pt0} that the two point vertex vanishes, but we
have not been able to see this directly in twistor space. In this
appendix we give a number of twistor space formulae for the 2-point
vertex showing in particular that it is $\dbar$-exact so that it plays
no role in the Feynman diagram formalism when inserted onto an
external leg.  Recall that the 2-point vertex on twistor space is
given by the expression \be{eqn: 2pt1} \V(Z_{1},Z_{2})=\int_{\M\times
  (\CP^1)^2} \frac{\D^{4|4}Z_{A}\D^{4|4}Z_{B}}{\vol(
  \GL(2,\C))}\frac{\d\sigma_{1}\d\sigma_{2}}{(\sigma_{1}-\sigma_{2})^{2}}\bar{\delta}^{3|4}(Z_{1},Z(\sigma_{1}))\bar{\delta}^{3|4}(Z_{2},Z(\sigma_{2})).
\ee This can be reduced in a number of ways.  Firstly note that the
fermionic part of the integral can be performed directly algebraically
against the fermionic delta functions to yield a factor of
$(\sigma_1-\sigma_2)^4$ leaving an object with no fermionic dependence
\be{eqn: 2pt1a} 
\V(Z_{1},Z_{2})=\int_{\M\times (\CP^1)^2}
\frac{\D^{4}Z_{A}\D^{4}Z_{B}}{\vol(\GL(2,\C))}\,
{\d\sigma_{1}\d\sigma_{2}}{(\sigma_{1}-\sigma_{2})^{2}}\bar{\delta}^{3}_{0,-4}
(Z_{1},Z(\sigma_{1}))\bar{\delta}^{3}_{0,-4}(Z_{2},Z(\sigma_{2})). 
\ee 
Here we define
$$
\bar{\delta}^{3}_{p,-p-4}(Z_1,Z_2)= \int s^{p-1}\rd s \, \bar \delta^4(Z_1+sZ_2)
$$
where the subscripts denote the homogeneity in the first and second
entry respectively.
The $\vol(\GL(2))$ can be taken out by fixing $\sigma_1=0$,
$\sigma_2=\infty$ and reducing the $Z_A$ and $Z_B$ integrals to
projective integrals.  Removing the appropriate Jacobian factor we
obtain 
\be{eqn: 2pt1b} 
\V(Z_{1},Z_{2})
=\int_{\M\times (\CP^1)^2}
{\D^{3}Z_{A}\D^{3}Z_{B}}\, 
\bar{\delta}^{3}_{0,-4} (Z_{1},Z_A)\, \bar{\delta}^{3}_{0,-4}(Z_{2},Z_B),
, \ee where 
the $\CP^1$ factors in the contour are now understood as arising from
integrating $Z_A$ and $Z_B$ over the $\CP_1$ corresponding to $x\in\M$
and then integrating over the real slice $\M$.  This is an integral of
a $12$-form over an $8$-dimensional contour so that we are left with a
$4$-form in $Z^1$ and $Z^2$ (a $(0,2)$-form in each factor).

It is possible to reduce this further, but it is at this point
possible to express this as an exact form using a bosonic version of
\eqref{dbar-delta-line}.  Define \be{bos-delta-line}
\bar\delta^2_{0,0,-4}(Z_1,Z_2,Z_3) =\int \frac{c_3^3
  D^2c}{c_1c_2}\bar\delta^4( c_1 Z_1 + c_2 Z_2 + c_3 Z_3)\, , \qquad
D^2c=c_1\rd c_2\rd c_3+ \mbox{cyclic}\, .  \ee Then as in
\eqref{dbar-delta-line} we have \be{dbar-bos-delta-line} \dbar
\bar\delta^2_{0,0,-4}(Z_1,Z_2,Z_3)=2\pi i(\bar\delta^3_{0,-4}(Z_1,Z_3)+
\bar\delta^3_{0,-4}(Z_2,Z_3)) \, .  \ee (There is no $\bar\delta^3(Z_1,Z_2)$
term as there is no pole in $c_3$ in \eqref {bos-delta-line}.)  Thus
we can write 
\be{v2-exact} V(Z_1,Z_2)=\dbar \left(\frac 1{2\pi
    i}\int_{\M\times (\CP^1)^2} {\D^{3}Z_{A}\D^{3}Z_{B}}\,
  \bar{\delta}^{2}_{0,0,-4} (Z_{1},Z_B,Z_A)\,
  \bar{\delta}^{3}_{0,-4}(Z_{2},Z_B)\right)\, , 
\ee 
as ${\D^{3}Z_{A}\wedge \D^{3}Z_{B}}=0 $ on the support of the
$\bar\delta^3(Z_A,Z_B)$ term.  Thus $V(Z_1,Z_2)$ is an exact form. 

We now present another pair of formulae for the two
point vertex.
Breaking manifest conformal invariance we can write:
\be{eqn: 2ptnc}
V(Z_1,Z_2)= \int_{\M}\d^{4}x \,
[\pi_{1}\pi_{2}]^{2}\, \bar{\delta}^{2}\left(\omega^{A}_{1}-x^{AA'}\pi_{1A'}\right)\bar{\delta}^{2}\left(\omega^{A}_{2}-x^{AA'}\pi_{2A'}\right)
\, .
\ee
We can  maintain conformal invariance,
if we are are prepared to use a Euclidean real slice for $\M$ and this
representation could well be useful for Euclidean signature
correlation function calculations (although less appropriate to the
S-matrix computations of this paper).   In
this case one of the $\CP^1$ bundles over $\M$ can be taken to be the
whole of the twistor space parametrized by $Z_B$ and we can then
integrate out the $Z_A$ delta function in 
 \eqref{eqn: 2pt1b} to obtain
\be{eqn: 2ptc}
\V(Z_{1},Z_{2})=\int_{L_{1}}\D^{3}Z_{B}\, \bar{\delta}^{3}_{0,-4}(Z_{2},Z_{B}).
\ee
The remaining integral is that over the real $\CP^1$ in twistor space
that contains $Z_1$. For a Euclidean signature real slice we must have
$Z_B= \hat Z_1+ tZ_1 $ for the bosonic part where $\hat Z_1$ is 
the Euclidean conjugate.  We can now integrate over the complex 
  $t$-plane to obtain 
\be{euc-2pt}
\V(Z_1,Z_2)=(Z_1,\hat Z_1,\d \hat Z_1,\d\hat Z_1)\int
  t^2\d s \d t \, \bar \delta^4(Z_2+sZ_1+ t\hat Z_1).\ee  
This is the bosonic collinear delta function 
\be{euc-2pt1}
\V(Z_1,Z_2)=(Z_1,\hat Z_1,\d \hat Z_1,\d\hat Z_1) \, \bar
\delta^2_{0,-1,-3}(Z_2,Z_1, \hat Z_1)
\ee 
where the subscripts denote the weights in its arguments.

Although our statement of $\dbar$-exactness for the two point vertex
ensures that it does not contribute to the Feynman diagram formalism
when inserted onto an external particle leg in any diagram (using
integration by parts and the fact that it is paired with an on-shell
external wave-function), it is not so obvious that its contribution
vanishes when it is inserted on an internal (i.e., propagator) leg.
This is because an integration by parts argument moves the
$\dbar$-operator onto a propagator (which is not $\dbar$-closed)
rather than a cohomology class.  Explicitly calculating the
contribution in this case yields an integrand which is exactly the
same as that arising in the computation of the so-called `Kermit'
diagrams in the momentum twistor version of the MHV formalism
\cite{Bullimore:2010pj}.  This contribution plays a crucial
(non-vanishing) role in the momentum twistor framework (essentially
computing the 1-loop MHV contribution) and certainly does not vanish
as an integrand.  However, our formalism is a term-by term
transcription of the momentum space MHV formalism to twistor space,
and the two point contribution does not contribute on momentum space.
Furthermore, as we have mentioned, using the Euclidean space formalism
of \cite{Boels:2007qn,Jiang:2008xw} we can evaluate the two-point
vertex on off-shell momentum eigenstates and see directly that it
vanishes.

This tension between the picture on twistor space presented here
(where the two point contribution must vanish) and the momentum
twistor formalism (where it must not) is clearly a subtle issue that
requires further attention.  As we have mentioned, the two
calculations are identical at the level of the integrand they produce,
the crucial difference manifests itself in the choice of real contour. 

\section{Twistor space and Euclidean space-time signature}
\label{A}

The main exposition of this paper has focused on aspects of the
twistor space MHV formalism which are independent of the choice of
space-time signature.  However, certain choices of signature give rise
to formalisms in which calculations may be performed more explicitly
than before.  This appendix reviews the particulars of twistor space
over Euclidean space-time, \ref{Momsp} provides details for the derivation of the momentum space MHV formalims from the twistor action, and \ref{Prop} demonstrates how the twistor propagator can be calculated directly from space-time representatives.  We denote
four-dimensional Euclidean space as $\E$, and its chiral super-space
extension as $\E^{4|8}$.  Twistor space is related to space-time as a
$\CP^{1}$-bundle over $\E^{4|8}$:
\begin{eqnarray*}
\CP^{1} & \rightarrow & \PT \\
& & \downarrow^{\rho} \\
& & \E^{4|8}
\end{eqnarray*}
Hence, for every $(x,\theta)\in\E^{4|8}$, there is a line $L_{(x,\theta)}\cong\CP^{1}\subset\PT$.  The incidence relationship between $\E^{4|8}$ and $\PT$ is
\be{eqn: incA}
\omega^{A}=x^{AA'}\pi_{A'}, \qquad \chi^{i }=\theta^{i\:A'}\pi_{A'}.
\ee 

In Euclidean signature, $x^{AA'}$ corresponds to a real point $x\in\E$ provided $x^{AA'}=\hat{x}^{AA'}$, where $\hat{\cdot}$ is the quaternionic conjugation which acts on spinors as \cite{Wood-85}:
\begin{equation*}
\alpha^{A}=(a, b)\mapsto \hat{\alpha}^{A}=(\bar{b}, -\bar{\alpha}); \;\;\; \beta^{A'}=(c, d)\mapsto \hat{\beta}^{A'}=(-\bar{d}, \bar{c}).
\end{equation*}
This leads to the reality structure (i.e., complex conjugation) on twistor space given by:
\be{eqn: reality}
Z^{I}\mapsto \hat{Z}^{I}=(\hat{\omega}^{A},\hat{\pi}_{A'}, -\overline{\chi^{2}},\overline{\chi^{1}},-\overline{\chi^{4}},\overline{\chi^{3}}).
\ee
Real points in $\E^{4|8}$ correspond to lines that are invariant under
this conjugation.  This means that we have the
projection $\rho:\PT\rightarrow\E^{4|8}$  where $Z$ is mapped to the
point corresponding to the line through $Z$ and $\hat Z$.  This can be
written explicitly as \cite{Boels:2007qn}
\be{eqn: proj}
Z^{I}\mapsto (x^{AA'}, \theta^{i\;A'})=\left(\frac{\omega^{A}\hat{\pi}^{A'}-\hat{\omega}^{A}\pi^{A'}}{[\pi\hat{\pi}]}, \frac{\chi^{i }\hat{\pi}^{A'}-\hat{\chi}^{i }\pi^{A'}}{[\pi\hat{\pi}]}\right).
\ee

The choice of complex structure on twistor space is that arising from
the complex coordinates $Z$.  The fermionic coordinates $\chi^{i }$
only ever need to be considered holomorphically, so we will just focus
on the definition of the $\dbar$-operator in bosonic directions
$\PT_{b}$.  To write the $\dbar$-operator on $\PT$ explicitly, we
define bases for $\Omega^{0,1}(\PT'_{b})$ and $T^{0,1}\PT'_{b}$
denoted respectively by%\footnote{Choosing such bases is possible in all
%  signatures but the precise form of the $\dbar$-operator will
%  differ among these choices.}  
\be{eqn: basis} \{e^{0},e^{A}\} \qquad
\{\dbar_{0},\dbar_{A}\}.  \ee
where \cite{Mason-05}: 
\be{eqn: 01forms}
e^{0}=\frac{[\hat{\pi}\d\hat{\pi}]}{[\pi\hat{\pi}]^{2}}, \qquad e^{A}=\frac{\hat{\pi}_{A'}\d x^{AA'}}{[\pi\hat{\pi}]}=\frac{\omega^{A}[\hat{\pi}\d\hat{\pi}]}{[\pi\hat{\pi}]^2}-\frac{\d\hat{\omega}^{A}}{[\pi\hat{\pi}]},
\ee
\be{eqn: 01vecs}
\dbar_{0}=[\pi\hat{\pi}]\pi_{A'}\frac{\partial}{\partial\hat{\pi}_{A'}}, \qquad \dbar_{A}=\pi^{A'}\partial_{AA'}=-[\pi\hat{\pi}]\frac{\partial}{\partial\hat{\omega}^{A}},
\ee
and set
\be{eqn: dbare}
\dbar= \d\hat{Z}^{\alpha}\frac{\partial}{\partial\hat{Z}^{\alpha}} =e^{0}\dbar_{0}+e^{A}\dbar_{A}=\frac{\pi_{A'}\hat{\pi}^{B'}\d\hat{\pi}_{B'}}{[\pi\hat{\pi}]}\frac{\partial}{\partial\hat{\pi}_{A'}}+\frac{\pi^{B'}\hat{\pi}_{A'}\d x^{AA'}}{[\pi\hat{\pi}]}\partial_{AB'}.
\ee
so we can set $\dbar:
\Omega^{p,q}(\PT'_{b})\rightarrow\Omega^{p,q+1}(\PT'_{b})$.  We have
$\dbar^{2}=0$ as required for an integrable complex
structure.  An additional structure is provided by the $\dhat$-operator, which represents a holomorphic derivative in an anti-holomorphic direction:
\be{eqn: dhat}
\dhat = \d\hat{Z}^{\alpha}\frac{\partial}{\partial Z^{\alpha}}.
\ee
It is not hard to see that this is also integrable (i.e., $\dhat^{2}=0$) and obeys $\dbar\dhat=-\dhat\dbar$. 

Bosonic twistor space $\PT_{b}$ has the weighted holomorphic volume
form given as 
\begin{equation*}
\Omega=[\pi\d\pi]\wedge \pi_{A'}\pi_{B'}\epsilon_{AB}\d x^{AA'}\wedge\d x^{BB'},
\end{equation*}
a section of $\Omega^{3,0}(\PT'_{b},\cO(4))$.  Supertwistor space
$\PT$ is a Calabi-Yau supermanifold, with the globally
defined holomorphic volume form of weight $0$ on the full supertwistor space
used throughout this paper \cite{Witten:2003nn}: 
\be{eqn: vf}
\D^{3|4}Z=\Omega\wedge\epsilon_{i j  k  l }\d\chi^{i }\wedge\d\chi^{j }\wedge\d\chi^{ k }\wedge\d\chi^{ l }=\Omega\wedge\d^{4}\chi.
\ee

\section{Transformation to momentum space}
\label{Momsp}

This appendix provides the details of the proof of the Fourier
transform of the twistor propagator used to derive the  
momentum space MHV formalism from that in twistor space given in \S
\ref{DMomsp}.  We wish to show that we have the Fourier representation 
of the propagator as 
\be{eqn: momp1-app} 
q^*\Delta(x,\theta,\pi,x'\theta',\pi')= \int\rd^4p\rd^4\eta\,
\e^{i(x-x')\cdot p+ \eta\cdot (\theta-\theta')|p|\hat\iota\ra} \, 
\widetilde \Delta(p,\eta,\pi,\pi')\, ,
\ee 
where $q^{*}\Delta$ is the pullback of the twistor space propagator of
\eqref{eqn: prop2} to the primed spinor bundle, and 
\be{eqn: momp2-app}
\widetilde\Delta(p)=\frac{
  \bar{\delta}^{1}(\langle\hat{\iota}|p|\pi])\,
  \wedge\bar{\delta}^{1}(\langle\hat{\iota}|p|\pi'])}{p^{2}} 
\ee and $\hat{\iota}^{A}$ is related to the original constant spinor
$\iota^{A}$ primary part of $Z_\rf$ by Euclidean complex
conjugation.

The pulled back twistor propagator can be written as 
\be{eqn: momp2}
q^{*}\Delta(x,\theta,\pi,x',\theta',\pi')=\int_{\C^{2}}\frac{\d s}{s}\frac{\d
  t}{t}\bar{\delta}^{2}(s\iota -ix\cdot\pi
-itx'\cdot\pi')\bar{\delta}^{2}(\pi +t\pi')\bar{\delta}^{0|4}(\theta|\pi]
-t\theta'|\pi']) \, . 
\ee
We will abbreviate by writing
$\bar\delta^{2|4}(\pi+t\pi'):=\bar{\delta}^{2}(\pi
+t\pi')\bar{\delta}^{0|4}(\chi -t\chi')$ with $\chi=\theta|\pi]$ etc., 
and use the support of the delta functions to yield
\begin{eqnarray*}
q^{*}\Delta&=&\int
 \frac{\d s \d t}{st}\, \bar{\delta}^{2}\left(s\iota
   -i(x-x')\cdot(\pi)\right) \, 
 \bar{\delta}^{2|4}(\pi +t\pi')   
\end{eqnarray*}
Since this doesnt depend on $x+x'$, we just Fourier transform in
$y=x-x'$ to obtain
\be{}
\widetilde\Delta(p,\chi,\pi,\chi',\pi')
=\int \rd^4y\, \e^{ip\cdot y}\,\frac{\d s \d t}{st}\, 
\bar{\delta}^{2}(s\iota
  -iy\cdot \pi)
\, \bar{\delta}^{2|4}(\pi +t\pi') \, .
\ee
We now evaluate this over a real slice of {\em Euclidean} signature in
the complex space-time.  This will ensure that we obtain the correct
Feynman propagator on continuation back to Minkowski signature. 
Recalling that $\bar{\delta}^{2}(s\iota
  -iy\cdot \pi)$ is four real delta functions multiplied by a $(0,2)$
  form, we see that, in Euclidean signature, the delta function for
  $y$ has the unique 
solution
\begin{equation*}
y^{AA'}=\frac{is\iota^{A}\hat\pi^{A'} - i
  \bar{s}\hat{\iota}^{A}{\pi}^{A'} }{[\hat \pi\, \pi]}\, .
\end{equation*}
(We have taken $\iota$ to be normalized.) In performing the $y$
integral we must also take care of the Jacobian 
factor that will arise and further unpack the the $(0,2)$-form part of
the definition of $\bar{\delta}^{2}(s\iota -iy\cdot \pi)$.  This
$(0,2)$-form will contain a $\rd \bar s$ and $\D\hat \pi$ multiplied by
a further Jacobian factor which gives
\be{}
\widetilde\Delta(p,\chi,\pi,\chi',\pi')
=\int \e^{ip\cdot \left(\frac{is\iota\hat\pi - i
  \bar{s}\hat{\iota}{\pi} }{[\hat \pi\, \pi]}\right)
}\,\frac{\d s \d t}{st}\, 
 \frac{s\rd \bar s \D\hat\pi}{[\hat\pi\, \pi]^2} 
\, \bar{\delta}^{2|4}(\pi +t\pi') \, .
\ee
Because of the cancellation of the $s$ in the numerator with that in
the denominator, the $s$-integral can now be performed introducing a
2-dimensional 
delta-function in the coefficient of $s$ in the exponential.  This
delta-function can be combined with the form $\D\hat\pi$ as one of our
weighted Dolbeault $\bar\delta^1$ forms, at the expense of another
Jacobian factor, to yield
\be{}
\widetilde\Delta(p,\chi,\pi,\chi',\pi')
=\int \frac{\d t}{t}\, \frac{\bar\delta^1_0(p|\hat\iota\ra, 
  \pi)}{|\la\iota|p|\pi]|^2} 
\, \bar{\delta}^{2|4}(\pi +t\pi') \, .
\ee
On the support of the delta function
$|\la\iota|p|\pi]|^2=p^2[\hat\pi\, \pi]$.
We can also use the delta functions to substitute in $p|\hat\iota\ra$
for $\pi$ and  $\pi'$ in $\chi=\theta|\pi]$ and the delta function for
the $\pi+t\pi'$  etc., to find
\be{}
\widetilde\Delta(p,\theta,\pi,\theta',\pi')
= \frac{\bar\delta^1_0(p|\hat\iota\ra,  \pi)\,
  \bar\delta^1_0(p|\hat\iota\ra,  \pi') }{p^2} 
\, {\delta}^{0|4}((\theta -\theta')p|\hat\iota\ra) \, .
\ee
we can finally insert a Fourier representation of the fermionic delta
function to obtain
\be{}
\widetilde\Delta(p,\chi,\pi,\chi',\pi')
= \int\rd^4\eta \, \frac{\bar\delta^1_0(p|\hat\iota\ra,  \pi)\,
  \bar\delta^1_0(p|\hat\iota\ra,  \pi') }{p^2} 
\, \e^{i\eta\cdot((\theta -\theta')p|\hat\iota\ra)} \, .
\ee
This gives \eqref{eqn: momp2-app} and 
we can now Fourier transform back to the spin bundle to obtain the
formula \eqref{eqn: momp1-app} as desired.

We remark here that it was necessary to perform this calculation on
the Euclidean real slice to get the answer in the correct form, see
\S\ref{discuss} for further discussion.

\subsection*{\textit{Off and on-shell momentum eigenstate representatives}}
To prove the correspondence between the twistor and momentum space MHV
formalisms, we inserted momentum eigenstates for the wavefunctions $\cA_{i}$
appearing into the twistorial expression for the MHV vertex,
\eqref{eqn: vert}.  Although it was clear that the representatives we
use evaluate on space-time to give momentum eigenstates, we give here
an alternative derivation by using first 
off-shell momentum eigenstates to give $\cA$ in
the Woodhouse (or harmonic) gauge in terms of the abelian space-time
superconnection:
%\footnote{Considering the non-abelian superconnection
%  adds new terms starting at $O(\theta^{2})$, but for our purposes
%  those terms are not necessary.}  
\be{eqn: supercon1}
\cA=\Gamma_{AA'}\d x^{AA'}+\Gamma_{iA'}\d\theta^{iA'}.
\ee
In this gauge, the multiplet of $\cN=4$ SYM takes the form:
\begin{eqnarray}
A_{AA'}=e^{ip\cdot x}\epsilon_{AA'}, \qquad \Psi_{iA}=e^{ip\cdot
  x}\xi_{A}\eta_{i}, \qquad \Phi_{ij}=\frac{e^{ip\cdot
    x}}{2}\eta_{i}\eta_{j}, \nonumber \\ 
\widetilde{\Psi}^{i}_{A'}=\frac{e^{ip\cdot x}}{3!}
{p}_{A'}\epsilon^{ijkl}\eta_{j}\eta_{k}\eta_{l}, \qquad
G_{A'B'}=\frac{e^{ip\cdot x}}{4!}{p}_{A'}{p}_{B'}\eta^{4}, \label{eqn: multiplet} 
\end{eqnarray}
where the polarization and off-shell momentum spinors are defined in relation to the constant CSW reference spinor $\hat{\iota}^{A}$:
\be{eqn: polarization}
p_{A'}=\hat{\iota}^{A}p_{AA'}, \qquad \hat{\iota}^{A}p^{A'}\epsilon_{AA'}=1, \qquad \hat{\iota}^{A}\xi_{A}=1.
\ee
The superconnection components are given in terms of the multiplet by the expressions:
\begin{multline}\label{eqn: even}
\Gamma_{AA'}=A_{AA'}-\Psi_{iA}\theta^{i}_{A'}+\partial_{AB'}\Phi_{ij}\theta_{A'}^{j}\theta^{iB'}+\frac{\epsilon_{ijlk}}{2}\partial_{AB'}\widetilde{\Psi}^{l}_{C'}\theta_{A'}^{i}\theta^{jB'}\theta^{kC'} \\
+\frac{\epsilon_{ijlk}}{6}\left(\partial_{AA'}\widetilde{\Psi}^{l}_{B'}+\partial_{AB'}\widetilde{\Psi}^{l}_{A'}\right)\theta^{iD'}\theta^{jB'}\theta^{k}_{D'}-\frac{\epsilon_{iklj}}{3}\partial_{AB'}G_{C'D'}\theta^{i}_{A'}\theta^{jB'}\theta^{kC'}\theta^{lD'},
\end{multline}
\be{eqn: odd}
\Gamma_{iA'}=\Phi_{ij}\theta^{j}_{A'}+\epsilon_{ijlk}\widetilde{\Psi}^{l}_{B'}\theta^{k}_{A'}\theta^{jB'}+\epsilon_{jilk}G_{B'C'}\theta^{j}_{A'}\theta^{kB'}\theta^{lC'}.
\ee

To transform to the CSW gauge, one searches for a function $\gamma$ such that $\hat{\iota}^{A}\pi^{A'}\partial_{AA'}\lrcorner(\cA-\d\gamma)=0$.  A calculations reveals that
\be{eqn: gt}
\gamma =-i\frac{e^{ip\cdot x}}{[p\pi]}\left[\langle\hat{\iota} |\epsilon |\pi]+(\eta\cdot\chi)\left(1 +i\frac{(\eta\cdot\tilde{\chi})}{2}-\frac{(\eta\cdot\tilde{\chi})^{2}}{3!}-i\frac{(\eta\cdot\tilde{\chi})^{3}}{4!}\right)\right],
\ee
where $\tilde{\chi}^{i}=\theta^{iA'}p_{A'}$.  Recalling that $e^{0}\dbar_{0}([p\pi]^{-1})=\bar{\delta}^{1}([p\pi])$, it is easy to see that the off-shell multiplet in CSW gauge takes the form:
\begin{multline}\label{eqn: offshell}
\cA_{P}=\bar{\delta}^{1}([p\pi])e^{ip\cdot x}\left[\langle\hat{\iota} |\epsilon |\pi]+(\eta\cdot\chi)\left(1 +i\frac{(\eta\cdot\tilde{\chi})}{2}-\frac{(\eta\cdot\tilde{\chi})^{2}}{3!}-i\frac{(\eta\cdot\tilde{\chi})^{3}}{4!}\right)\right] \\
+\cA_{AA'}\d x^{AA'}+\cA_{iA'}\d\theta^{iA'}.
\end{multline}
In this fully off-shell form, $\cA_{P}$ does not live on twistor space but rather the primed spinor bundle.  On-shell, the terms $\cA_{AA'}\d x^{AA'}+\cA_{i}\d\chi^{i}$ vanish, and what remains descends to twistor space as
\be{eqn: onshell}
\cA_{P}=\int_{\C}\frac{\d s}{s} e^{is(p\cdot x +\eta\cdot\chi)}\bar{\delta}^{2}(s\pi_{A'}-p_{A'}),
\ee
in exact agreement with what was stated in \eqref{momentum-eigenstate}.  This completes the proof.

\section{Derivation of the CSW Propagator from space-time}
\label{Prop}

In this appendix, we provide a derivation of the twistor space
propagator from space-time representatives in Euclidean signature.
This compliments the results of \ref{Momsp}, where the momentum space
propagator was recovered by starting with the CSW gauge in twistor
space and the twistor propagator.  Now, we begin with the CSW gauge on
momentum space and space-time representatives of the propagator. 

We can reduce this task to one on the bosonic twistor space $\PT'_{b}$ by performing the fermionic integrals in the kinetic portion of the action \eqref{eqn: CSWg2} to obtain \cite{Witten:2003nn}: 
\be{eqn: kin}
 \int_{\PT'_{b}}\Omega\wedge\tr\left(g\wedge\dbar a+\chi^{i }\wedge\dbar\lambda^{i }+\frac{\epsilon_{i j  k  l }}{4}\phi^{i j }\wedge\dbar\phi^{ k  l }\right).
\ee
From this, we see that the propagator, $\Delta$, must be a sum of terms (with fermionic coefficients), each of which is a kernel for $\dbar$ on $\PT'_{b}$ taking values in the proper homogeneity configurations.  More formally, we have:
\be{eqn: prop1}
\Delta=(\chi_{2})^{4}\Delta_{0,-4}+\chi_{1}(\chi_{2})^{3}\Delta_{-1,-3}+(\chi_{1})^{2}(\chi_{2})^{2}\Delta_{-2,-2},
\ee
where each bosonic propagator obeys:
\begin{equation*}
\Delta_{i,j}\in H^{0,2}((\PT'_{b}\times\PT'_{b})\setminus\Delta, \cO(i,j)), \qquad \dbar \Delta_{i,j}=(\dbar_{1}+\dbar_{2})\Delta_{i,j}=\bar{\delta}_{\Delta},
\end{equation*}
for $\Delta\subset\PT'_{b}\times\PT'_{b}$ the diagonal in the cartesian product and $\bar{\delta}_{\Delta}$ the anti-holomorphic Dirac current.  

To find an expression for the propagator on bosonic twistor space, one begins with a space-time representative; for the term taking values in $\cO(-2,-2)$, this is just $(x_{1}-x_{2})^{-2}$, and the other homogeneity configurations are appropriate derivatives of this.  We then use Woodhouse's theorems \cite{Wood-85} to construct twistor space representatives in the Woodhouse gauge.  These objects are, by definition, $\dbar$-closed $(0,2)$-forms away from $\Delta$ which are in the Woodhouse gauge (i.e., their restriction to $\CP^{1}$ fibers on either factor of the product $\PT'_{b}\times\PT'_{b}$ are holomorphic $(0,1)$-forms).  We then use the freedom of adding a $\dbar$-exact $(0,2)$ form to these Woodhouse gauge representatives to transform them into objects that obey the CSW gauge on momentum space.  Writing $N^{\alpha}=(\hat{\iota}^{A},0)$, the CSW gauge condition for is:
\be{eqn: CSWgeq}
\hat{\iota}^{A}\pi^{A'}\partial_{AA'}\lrcorner\Delta_{i,j}=N^{\alpha}\frac{\partial}{\partial\hat{Z}^{\alpha}}\lrcorner\Delta_{i,j}=0.
\ee

At the level of space-time representatives, we begin with the photon propagator of QED in the Feynman gauge:
\begin{equation*}
\frac{\epsilon_{AB}\epsilon_{A'B'}}{(x_{1}-x_{2})^{2}}.
\end{equation*}
To obtain a space-time representative of the propagator for each of the three required homogeneity configurations, this expression must be modified properly.  For instance, in the $\cO(0,-4)$ configuration, we are dealing with an ASD potential $a$ and a SD field $g$, so we contract in a derivative with respect to $x_{1}$ and symmetrize over free primed indicies.  Such considerations give us the following space-time representatives:
\be{eqn: st22}
\Delta_{-2,-2}(x_{1},x_{2})=\frac{1}{(x_{1}-x_{2})^{2}},
\ee
\be{eqn: st13}
\Delta_{-1,-3}(x_{1},x_{2})=\frac{(x_{1}-x_{2})_{AB'}}{(x_{1}-x_{2})^4},
\ee
\be{eqn: st04}
\Delta_{0,-4}(x_{1},x_{2})=2\frac{(x_{1}-x_{2})_{B(C'}\epsilon_{A')B'}}{(x_{1}-x_{2})^4}.
\ee
In \eqref{eqn: st13}, the index $A$ is associated with $\PT_{b\;1}$, while $B'$ is associated with $\PT_{b\;2}$; in \eqref{eqn: st04} the indices $B,B'$ are associated with $\PT_{b\;1}$ and $A',C'$ with $\PT_{b\;2}$.  By construction from the photon propagator, it is clear that all three of these objects are zero rest mass fields on $\E\times\E$, away from the diagonal.

Hence, we can use theorems of Woodhouse \cite{Wood-85} to construct $(0,2)$-forms on $\PT'_{b}\times\PT'_{b}$ which are $\dbar$-closed away from the diagonal and are in the Woodhouse gauge \cite{Wood-85}; this is an explicit application of the Penrose transform with the choice of Woodhouse gauge.  Using
\begin{equation*}
(x_{1}-x_{2})^{2}=\frac{(1,\hat{1},2,\hat{2})}{[\pi_{1}\hat{\pi}][\pi_{2}\hat{\pi}_{2}]},
\end{equation*}
(where $(1,\hat{1},2,\hat{2})=\epsilon_{\alpha\beta\gamma\delta}Z^{\alpha}_{1}\hat{Z}^{\beta}_{1}Z^{\gamma}_{2}\hat{Z}^{\delta}_{2}$) along with the Woodhouse theorems, one finds:
\begin{multline}\label{eqn: wg22}
\Delta^{W}_{-2,-2}(Z_{1},Z_{2})=\dhat_{1}\dhat_{2}\left(\frac{1}{(1,\hat{1},2,\hat{2})}\right) \\
=2\frac{(\d\hat{Z}_{1},\hat{1},2,\hat{2})\wedge(1,\hat{1},\d\hat{Z}_{2},\hat{2})}{[(1,\hat{1},2,\hat{2})]^3}-\frac{(\d\hat{Z}_{1},\hat{1},\d\hat{Z}_{2},\hat{2})}{[(1,\hat{1},2,\hat{2})]^2},
\end{multline}
\begin{multline}\label{eqn: wg13}
\Delta^{W}_{-1,-3}(Z_{1},Z_{2})=\dhat_{2}\left(\frac{i[\pi_{1}\hat{\pi}_{1}](x_{1}-x_{2})_{AB'}\hat{\pi}_{2}^{B'}\hat{\pi}_{1\;D'}\d x_{1}^{AD'}}{[(1,\hat{1},2,\hat{2})]^{2}}\right)=\dhat_{2}\left(\frac{-i(\hat{2},1,\hat{1},\d\hat{Z}_{1})}{[(1,\hat{1},2,\hat{2})]^{2}}\right) \\
=-2i\frac{(\hat{2},1,\hat{1},\d\hat{Z}_{1})\wedge(1,\hat{1},\d\hat{Z}_{2},\hat{2})}{[(1,\hat{1},2,\hat{2})]^{3}},
\end{multline}
\begin{multline}\label{eqn: wg04}
\Delta^{W}_{0,-4}(Z_{1},Z_{2})=2\dhat_{2}\left(\frac{[\pi_{1}\hat{\pi}_{1}](x_{1}-x_{2})_{B(C'}\epsilon_{A')B'}\pi^{B'}_{1}\hat{\pi}^{C'}_{2}\hat{\pi}^{A'}_{2}\hat{\pi}_{1\;D'}\d x_{1}^{BD'}}{[(1,\hat{1},2,\hat{2})]^{2}[\pi_{2}\hat{\pi}_{2}]}\right) \\
=-2\dhat_{2}\left(\frac{[\hat{\pi}_{2}\pi_{1}](\hat{2},1,\hat{1},\d\hat{Z}_{1})}{[(1,\hat{1},2,\hat{2})]^{2}[\pi_{2}\hat{\pi}_{2}]}\right) \\
=2\frac{[\hat{\pi}_{2}\pi_{1}](\hat{2},1,\hat{1},\d\hat{Z}_{1})}{[(1,\hat{1},2,\hat{2})]^{2}[\pi_{2}\hat{\pi}_{2}]}\wedge\left(\frac{[\hat{\pi}_{2}\d\hat{\pi}_{2}]}{[\pi_{2}\hat{\pi}_{2}]}-2\frac{(1,\hat{1},\d\hat{Z}_{2},\hat{2})}{(1,\hat{1},2,\hat{2})}\right),
\end{multline}
where $\dhat_{i}$ is the operator defined in \eqref{eqn: dhat} on $\PT_{b\;i}$.  

We must now add a $\dbar$-exact form to our Woodhouse gauge
propagators in order to transform them the CSW gauge on momentum space
defined by $N^{\alpha}$. (This breaks the Woodhouse gauge, but preserves $\dbar$-closure away from the diagonal $\Delta\subset\PT'_{b}\times\PT'_{b}$.)  It is easiest to perform this transformation in homogeneity configuration $\cO(-2,-2)$; we do this rather explcitly below.  In order to obtain $\Delta_{-1,-3}$ and $\Delta_{0,-4}$, one repeats the entire process starting from $\Delta^{W}_{-1,-3}$ and $\Delta^{W}_{0,-4}$ respectively.  We employ the following notation to denote contraction over twistor indicies:
\begin{equation*}
 N^{\alpha}\frac{\partial}{\partial Z_{1}^{\alpha}}=N\cdot\partial_{1},\;\;N^{\alpha}\frac{\partial}{\partial\hat{Z}_{2}^{\alpha}}=N\cdot\dhat_{2},\;\;\mathrm{etc.}
\end{equation*}

The gauge condition on for the propagator reads
\begin{equation*}
 N\cdot\dhat_{i}\lrcorner(\Delta^{W}_{-2,-2}+\dbar f)=0,\;\; i=1,2,
\end{equation*}
where $f$ is now a $(0,1)$-form on $\PT'_{b\;1}\times\PT'_{b\;2}$, which we write as
\begin{equation*}
 f=f_{1}\cdot\d\hat{Z}_{1}+f_{2}\cdot\d\hat{Z}_{2}.
\end{equation*}
Using \eqref{eqn: wg22}, the CSW equation in this homogeneity configuration reads:
\be{eqn: cswt}
N\cdot\dhat_{i}\lrcorner\left(\dhat_{1}\dhat_{2}\left(\frac{1}{(1,\hat{1},2,\hat{2})}\right)+(\dbar_{1}+\dbar_{2})f\right)=0.
\ee
Expanding out this notation leaves us with a set of four differential equations for the components of $f$:
\be{eqn: cswt1}
\begin{array}{c}
 N\cdot\partial_{1}\left(\partial_{2\;\alpha}\left(\frac{1}{(1,\hat{1},2,\hat{2})}\right)\right)+N\cdot\dhat_{1}(f_{2\;\alpha})-\dhat_{2\;\alpha}(f_{1}\cdot N)=0, \\
 N\cdot\partial_{2}\left(\partial_{1\;\alpha}\left(\frac{1}{(1,\hat{1},2,\hat{2})}\right)\right)+\dhat_{1\;\alpha}(f_{2}\cdot N)-N\cdot\dhat_{2}(f_{1\;\alpha})=0,
\end{array}
\ee
\be{eqn: cswt2}
\begin{array}{c}
 N\cdot\dhat_{1}\lrcorner\dbar_{1}(f_{1}\cdot\d\hat{Z}_{1})=0, \\
 N\cdot\dhat_{2}\lrcorner\dbar_{2}(f_{2}\cdot\d\hat{Z}_{2})=0.
\end{array}
\ee

Our methodology will be to first solve \eqref{eqn: cswt1}-\eqref{eqn: cswt2} away from the support of any delta functions by contracting twistors into the free index of the first set of equations.  For instance, by contracting $N^{\alpha}$ into \eqref{eqn: cswt1}, we obtain the equation:
\begin{equation*}
 2\frac{(1,\hat{1},N,\hat{2})(N,\hat{1},2,\hat{2})}{[(1,\hat{1},2,\hat{2})]^3}+N\cdot\dhat_{1}(f_{2}\cdot N)-N\cdot\dhat_{2}(f_{1}\cdot N)=0,
\end{equation*}
from which we deduce
\begin{eqnarray*}
 f_{1}\cdot N & = & -\frac{(2,\hat{2},N,\hat{1})(N,\hat{2},1,\hat{1})}{2[(1,\hat{1},2,\hat{2})]^{2}(2,N,1,\hat{1})}, \\
 f_{2}\cdot N & = & \frac{(1,\hat{1},N,\hat{2})(N,\hat{1},2,\hat{2})}{2[(1,\hat{1},2,\hat{2})]^{2}(1,N,2,\hat{2})},
\end{eqnarray*}
which are unique solutions up to terms annihilated by $N\cdot\dhat_{2}$ and $N\cdot\dhat_{1}$ respectively.  Repeating this procedure of solving scalar equations by contracting with $Z_{1,2}^{\alpha}$ and $\hat{Z}_{1,2}^{\alpha}$ gives candidate solutions:
\be{eqn: f1}
f_{1\;\alpha}=-\frac{(N,\hat{2},1,\hat{1})(2,\hat{2},\alpha,\hat{1})}{2[(1,\hat{1},2,\hat{2})]^{2}(2,N,1,\hat{1})}-\frac{(2,N,\hat{2},1)(N,1,\hat{1},\alpha)}{2(1,\hat{1},2,\hat{2})[(2,N,1,\hat{1})]^2},
\ee
\be{eqn: f2}
f_{2\;\alpha}=\frac{(N,\hat{1},2,\hat{2})(1,\hat{1},\alpha,\hat{2})}{2[(1,\hat{1},2,\hat{2})]^{2}(1,N,2,\hat{2})}+\frac{(1,N,\hat{1},\hat{2})(N,2,\hat{2},\alpha)}{2(1,\hat{1},2,\hat{2})[(1,N,2,\hat{2})]^{2}}.
\ee

However, we must still check that \eqref{eqn: f1} and \eqref{eqn: f2} satisfy the other two CSW equations in \eqref{eqn: cswt2}.  By the symmetry between $\PT'_{b\;1}$ and $\PT'_{b\;2}$ (which is obvious from the equations and our candidate solutions), we only need to check this for one of the equations, so consider the second of these, which is equivalent to $N^{\alpha}\dhat_{2\;[\alpha}f_{2\;\beta]}=0$.  Directly computing the left-hand side of this equation gives (after placing everything over a common denominator of $4[(1,\hat{1},2,\hat{2})]^{2}[(1,N,2,\hat{2})]^{2}$ and applying the Schouten identity twice) a numerator of
\begin{equation*}
 (1,\hat{1},N,\hat{2})\left[(N,2,1,\hat{2})(N,2,\hat{1},\beta)-(N,2,\hat{1},\hat{2})(N,2,1,\beta)-(N,2,1,\hat{1})(N,2,\hat{2},\beta)\right],
\end{equation*}
which is itself equal to zero by the Schouten identity.  Thus, \eqref{eqn: f1} and \eqref{eqn: f2} do indeed satisfy the full set of CSW tranformation equations away from any delta function support.

To obtain a preliminary expression for the propagator (i.e., without any delta function contributions), we must compute $\Delta^{W}_{-2,-2}+\dbar f$ using \eqref{eqn: wg22} and \eqref{eqn: f1}-\eqref{eqn: f2}.  However, a substantial amount of algebra and applications of the Schouten identity reveal that this quantity vanishes, which indicates that only delta function contributions remain.  When these are accounted for, the mixed bi-degree terms (i.e., those which couple to physical fields) in our transformed quantity are
\begin{multline}\label{eqn: ptilde}
 \tilde{\Delta}_{-2,-2}=\frac{(N,\hat{1},2,\hat{2})}{2[(1,\hat{1},2,\hat{2})]^2}\dbar_{1}\left(\frac{1}{(1,\widehat{N},2,\hat{2})}\right)\wedge(1,\hat{1},\d\hat{Z}_{2},\hat{2}) \\ +\frac{(1,N,\hat{1},\hat{2})}{2(1,\hat{1},2,\hat{2})}\dbar_{1}\left(\frac{1}{[(1,\widehat{N},2,\hat{2})]^2}\right)\wedge(N,2,\hat{2},\d\hat{Z}_{2})+\frac{(N,\hat{2},1,\hat{1})}{2[(1,\hat{1},2,\hat{2})]^{2}}(2,\hat{2},\d\hat{Z}_{1},\hat{1})\wedge\dbar_{2}\left(\frac{1}{(2,\widehat{N},1,\hat{1})}\right) \\ +\frac{(2,N,\hat{2},\hat{1})}{2(1,\hat{1},2,\hat{2})}(N,1,\hat{1},\d\hat{Z}_{1})\wedge\dbar_{2}\left(\frac{1}{[(2,\widehat{N},1,\hat{1})]^2}\right).
\end{multline}
Note that when it appears as a delta-function contribution, the twistor $N$ must be conjugated; this is because
\begin{equation*}
\dbar_{1}\left(\frac{1}{(1,\widehat{N},2,\hat{2})}\right)=\delta\left((1,\widehat{N},2,\hat{2})\right)(\dbar_{1}\hat{1},N,\hat{2},2),
\end{equation*}
and to reach the CSW gauge \eqref{eqn: CSWgeq}, the form portion of the delta function must be skewed with $N$.

Though we may be tempted to say that $\tilde{\Delta}_{-2,-2}$ is the propagator in the CSW gauge, it is possible that the inclusion of these delta function-supported terms could spoil the CSW gauge condition.  Indeed,
\begin{eqnarray*}
 N\cdot\dhat_{1}\lrcorner\tilde{\Delta}_{-2,-2} & = & \frac{(N,\hat{2},1,\hat{1})(2,\hat{2},N,\hat{1})}{2[(1,\hat{1},2,\hat{2})]^2}\dbar_{2}\left(\frac{1}{(2,\widehat{N},1,\hat{1})}\right), \\
 N\cdot\dhat_{2}\lrcorner\tilde{\Delta}_{-2,-2} & = & -\frac{(N,\hat{1},2,\hat{2})(1,\hat{1},N,\hat{2})}{2[(1,\hat{1},2,\hat{2})]^2}\dbar_{1}\left(\frac{1}{(1,\widehat{N},2,\hat{2})}\right),
\end{eqnarray*}
indicating that we must modify our solutions \eqref{eqn: f1}-\eqref{eqn: f2} in order to account for these delta function contributions.

The proper modification is easily seen to be:
\be{eqn: f12}
f_{1}\cdot\d\hat{Z}_{1}\rightarrow f_{1}\cdot\d\hat{Z}_{1}-\frac{(N,\hat{1},2,\hat{2})(1,\hat{1},N,\hat{2})}{2(1,\hat{1},2,\hat{2})(2,N,1,\hat{1})}\dbar_{1}\left(\frac{1}{(1,\widehat{N},2,\hat{2})}\right),
\ee
\be{eqn: f23}
f_{2}\cdot\d\hat{Z}_{2}\rightarrow f_{2}\cdot\d\hat{Z}_{2}+\frac{\epsilon(N,\hat{2},1,\hat{1})\epsilon(2,\hat{2},N,\hat{1})}{2\epsilon(1,\hat{1},2,\hat{2})\epsilon(1,N,2,\hat{2})}\dbar_{2}\left(\frac{1}{\epsilon(2,\widehat{N},1,\hat{1})}\right).
\ee
A calculation involving yet more applications of the Schouten identity reveals that the physical portion of the propagator (i.e., the mixed terms) takes the remarkably simple form:
\be{eqn: csw22*}
\Delta_{-2,-2}=\frac{(N,\hat{1},2,\hat{2})(1,\hat{1},N,\hat{2})}{(1,\hat{1},2,\hat{2})}\dbar_{1}\left(\frac{1}{(1,\widehat{N},2,\hat{2})}\right)\wedge\dbar_{2}\left(\frac{1}{(1,\hat{1}, 2,\widehat{N})}\right),
\ee
which is obviously in CSW gauge since the form component is skewed with $N$.

We can perform the same process for propagators taking values in $\cO(-1,-3)$ and $\cO(0,-4)$, using \eqref{eqn: wg13} and \eqref{eqn: wg04}.  In these cases, the CSW transformations equations \eqref{eqn: cswt2} are unchanged, while \eqref{eqn: cswt1} become respectively
\be{eqn: cswt13}
\begin{array}{c}
 i\partial_{2\;\alpha}\left(\frac{(2,1,\hat{1},N)}{[(1,\hat{1},2,\hat{2})]^2}\right)+N\cdot\dhat_{1}(f_{2\;\alpha})-\dhat_{2\;\alpha}(f_{1}\cdot N)=0, \\
 -iN\cdot\partial_{2}\left(\frac{(2,1,\hat{1},\alpha)}{[(1,\hat{1},2,\hat{2})]^2}\right)-\dhat_{1\;\alpha}(f_{2}\cdot N)+N\cdot\dhat_{2}(f_{1\;\alpha})=0,
\end{array}
\ee
\be{eqn: cswt04}
\begin{array}{c}
 2\partial_{2\;\alpha}\left(\frac{[\hat{\pi}_{2}\pi_{1}](\hat{2},1,\hat{1},N)}{[(1,\hat{1},2,\hat{2})]^{2}[\pi_{2}\hat{\pi}_{2}]}\right)+N\cdot\dhat_{1}(f_{2\;\alpha})-\dhat_{2\;\alpha}(f_{1}\cdot N)=0, \\
-2N\cdot\partial_{2}\left(\frac{[\hat{\pi}_{2}\pi_{1}](\hat{2},1,\hat{1},\alpha)}{[(1,\hat{1},2,\hat{2})]^{2}[\pi_{2}\hat{\pi}_{2}]}\right)-\dhat_{1\;\alpha}(f_{2}\cdot N)+N\cdot\dhat_{2}(f_{1\;\alpha})=0.
\end{array}
\ee
The lengthy process of solving these sets of equations again produces remarkably simple results which have only delta function support:
\be{eqn: csw13*}
 \Delta_{-1,-3}=i\frac{[(1,\hat{1},N,\hat{2})]^{2}}{(1,\hat{1},2,\hat{2})}\dbar_{1}\left(\frac{1}{(1,\widehat{N},2,\hat{2})}\right)\wedge\dbar_{2}\left(\frac{1}{(1,\hat{1},2,\widehat{N})}\right),
\ee
\be{eqn: csw04*}
 \Delta_{0,-4}=2\frac{[\hat{\pi}_{2}\pi_{1}][(1,\hat{1},N,\hat{2})]^2}{(1,\hat{1},2,\hat{2}) [\pi_{2}\hat{\pi}_{2}]}\dbar_{1}\left(\frac{1}{(1,\widehat{N},2,\hat{2})}\right)\wedge\dbar_{2}\left(\frac{1}{(1,\hat{1},2,\widehat{N})}\right).
\ee

The formulae \eqref{eqn: csw22*}, \eqref{eqn: csw13*}, and \eqref{eqn: csw04*} define the bosonic components of the propagator, which in turn leads to the fully super-symmetric expression
\begin{equation*}
\Delta=(\chi_{2})^{4}\Delta_{0,-4}+\chi_{1}(\chi_{2})^{3}\Delta_{-1,-3}+(\chi_{1})^{2}(\chi_{2})^{2}\Delta_{-2,-2}.
\end{equation*}
At the level of $\cN=4$ SYM, we note that the homogeneity factors appearing in each bosonic portion are accounted for by fermions, and $\rf=-\widehat{N}$, so each term in the propagator contains a factor:
\begin{equation*}
\dbar_{1}\left(\frac{1}{(1,\rf,2,\hat{2})}\right)\wedge\dbar_{2}\left(\frac{1}{(1,\hat{1},2,\rf)}\right),
\end{equation*}
\begin{figure}
\centering
\includegraphics[width=4 in, height=3 in]{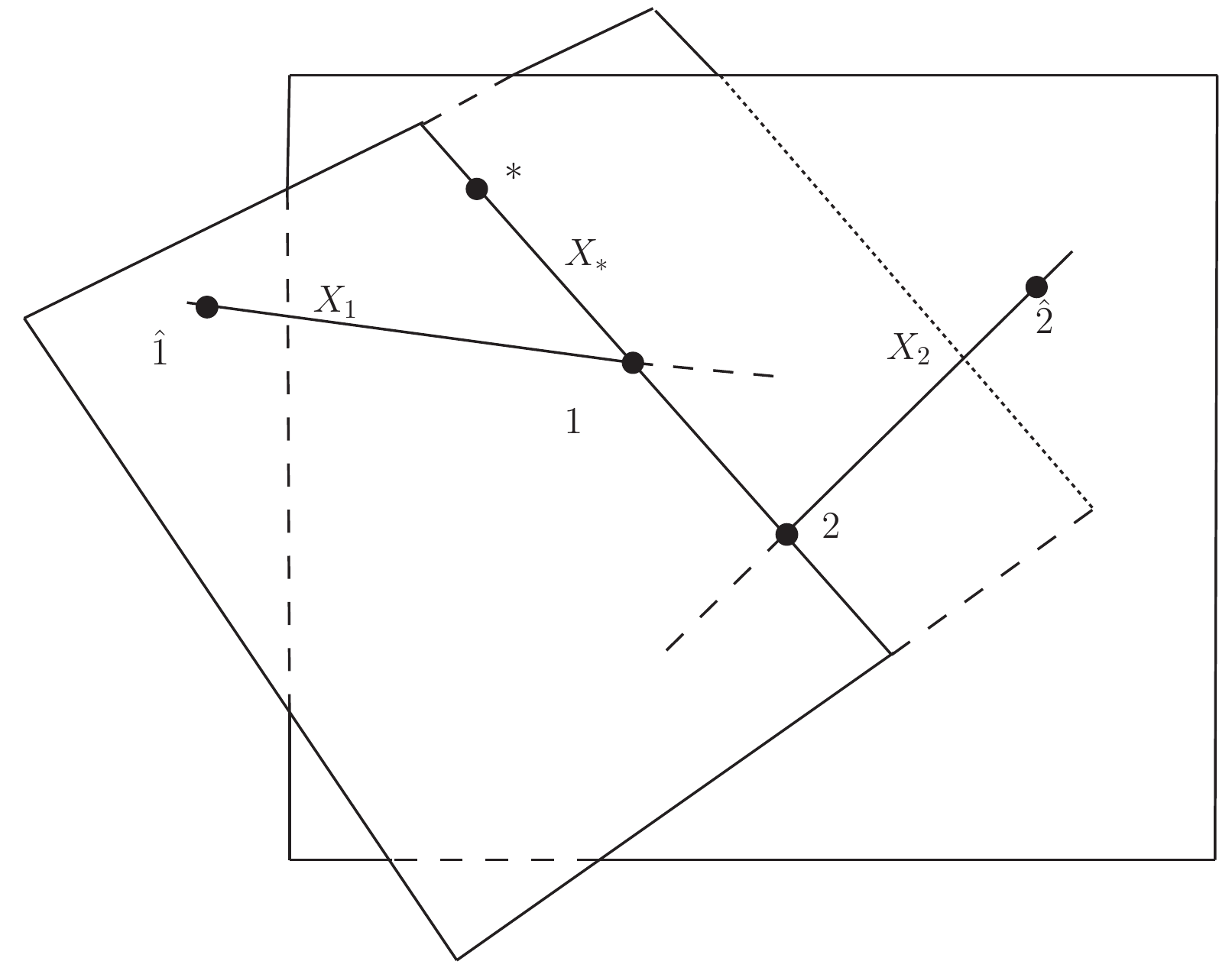}\caption{Twistor space support of the propagator}\label{PropGeo}
\end{figure}
which is supported only on the set $(1,\rf,2,\hat{2})=(1,\hat{1},2,\rf)=0$, as each form in this wedge product represents a delta-function.  The first factor demands that $Z_{1}$ lie in the plane defined by $\rf$ and the line $X_{2}\equiv Z_{2}^{[I}\hat{Z}_{2}^{J]}$; call this plane $\langle\rf ,2,\hat{2}\rangle\subset\PT$.  Identical reasoning tells us that the other factor restricts $Z_{2}$ to lying in the plane $\langle\rf ,1,\hat{1}\rangle\subset\PT$.  Now, it is obvious that the planes $\langle\rf ,2,\hat{2}\rangle$ and $\langle\rf, 1,\hat{1}\rangle$ must intersect in $\PT$ along a line $X_{\rf}$ which contains the point $\rf$.  However, given the support of these two factors, this geometric picture is only possible provided $Z_{1}$ and $Z_{2}$ also lie along $X_{\rf}$.  In other words, the support of the propagator dictates that $Z_{1}$, $Z_{2}$, and $N$ all be co-linear in twistor space (see Figure \ref{PropGeo}).

Hence, modulo some irrelevant numerical factors, our methodology tells
us that the twistor space propagator for $\cN=4$ SYM is: 
\be{eqn: prop2*}
\Delta=\bar{\delta}^{2|4}(Z_{1},\rf, Z_{2}),
\ee
as claimed in the text.  Although our derivation here began in
Euclidean signature, the result for the full propagator is signature
independent and superconformally invariant up to choice of $\rf$.

\bibliography{tsmhv}
\bibliographystyle{JHEP}

\end{document}